\documentclass{aa} 

\usepackage{graphicx}
\usepackage{txfonts}
\usepackage{multirow}
\usepackage{chemformula}
\usepackage{hyperref}
\hypersetup{colorlinks=true, linkcolor=red,citecolor=blue,urlcolor=magenta} 

\usepackage{comment}
\usepackage{booktabs}
\usepackage{tikz}

\newcommand{\met}{CH$_3$OH\xspace}
\newcommand{\metc}{CH$_3$CN\xspace}
\newcommand{\metctr}{CH$_3^{13}$CN\xspace}
\newcommand{\et}{C$_2$H$_5$OH\xspace}

\newcommand{\dme}{CH$_3$OCH$_3$\xspace}
\newcommand{\mf}{CH$_3$OCHO\xspace}

\newcommand{\ad}{CH$_3$CHO\xspace}

\newcommand{\ct}{$^{13}$}
\newcommand{\cd}{$^{12}$}
\newcommand{\ratio}{$^{12}$C\,/\,$^{13}$C\xspace}
\newcommand{\oratio}{$^{16}$O\,/\,$^{18}$O\xspace}
\newcommand{\odh}{$^{18}$}

\newcommand{\kms}{\,km\,s$^{-1}$\xspace}

\begin{document}

   \title{PRODIGE -- envelope to disk with NOEMA} 
   \subtitle{V. Low \ratio ratios for \met and \metc in hot corinos}
%
%
%
   \author{L.~A.~Busch
          \inst{1}
          \and
          J.~E.~Pineda \inst{1}
          \and
          O.~Sipilä \inst{1} 
          \and
          D.~M.~Segura-Cox \inst{2,1}
          \and
          P.~Caselli \inst{1} 
         \and 
          M.~J.~Maureira \inst{1}
          \and
          C.~Gieser \inst{3,1}
          \and 
          T.-H. Hsieh \inst{1,4,5}
          \and
          M.~T.~Valdivia-Mena \inst{6,1}
          \and
          L.~Bouscasse\inst{7}
          \and 
          Th.~Henning \inst{3}
          \and 
          D.~Semenov \inst{3}
          \and 
          A.~Fuente\inst{8}
          \and 
          M.~Tafalla\inst{9}
          \and 
          J.~J.~Miranzo-Pastor\inst{8}
          \and 
          L.~Colzi\inst{8}
          \and
          Y.-R.~Chou\inst{1}
          \and 
          S.~Guilloteau\inst{10}
          }

   \institute{Max-Planck-Institut f\"ur extraterrestrische Physik, Gie\ss enbachstra\ss e 1, 85748 Garching bei M\"unchen, Germany\\
              \email{lbusch@mpe.mpg.de}
        \and
        Department of Physics and Astronomy, University of Rochester, Rochester, NY 14627, USA
        \and
        Max-Planck-Institut f\"{u}r Astronomie, K\"{o}nigstuhl 17, D-69117 Heidelberg, Germany
       \and
       Taiwan Astronomical Research Alliance (TARA), Taiwan
       \and
       Institute of Astronomy and Astrophysics, Academia Sinica, P.O. Box 23-141, Taipei 106, Taiwan
       \and
       European Southern Observatory, Karl-Schwarzschild-Stra\ss e 2, 85748 Garching, Germany
       \and
       Institute de Radioastronomie Millim\'{e}trique (IRAM), 300 rue de la Piscine, F-38406, Saint-Martin d'H\`{e}res, France
       \and
       Centro de Astrobiolog\'{i}a (CAB), CSIC-INTA, Ctra. de Ajalvir Km. 4, 28850, Torrej\'{o}n de Ardoz, Madrid, Spain
       \and
       Observatorio Astron\'{o}mico Nacional (IGN), Alfonso XII 3, E-28014, Madrid, Spain
       \and 
       Laboratoire d'Astrophysique de Bordeaux, Universit\'{e} de Bordeaux, CNRS, B18N, All\'{e}e Geoffroy Saint-Hilaire, F-33615 Pessac, France
        }

   \date{Received ; accepted }

 
  \abstract  
  {The \ratio isotope ratio has been derived towards numerous cold clouds ($\sim$\,20--50\,K) and a couple protoplanetary disks and exoplanet atmospheres.
  However, direct measurements of this ratio in the warm gas ($>$\,100\,K) around young low-mass protostars remain scarce, but are required to study its evolution during star and planet formation.}
   {We aim to derive \ratio ratios from the isotopologues of the complex organic molecules (COMs) \met and \metc in the warm gas towards seven Class 0/I protostellar systems to improve our understanding of the evolution of the \ratio ratios during star and planet formation.} 
   {We used the data that were taken as part of the PRODIGE (PROtostars \& DIsks: Global Evolution) large program with the Northern Extended Millimeter Array (NOEMA) at 1\,mm. The \ct C isotopologue of \met was detected towards seven sources of the sample, the ones of CH$_3$CN towards six. The emission spectra were analysed by deriving synthetic spectra and population diagrams assuming conditions of local thermodynamic equilibrium (LTE).}
   {The emission of \met and \metc is spatially unresolved in the PRODIGE data with a resolution of $\sim$1\arcsec\,\, ($\sim$300\,au) for the seven targeted systems. Rotational temperatures derived from both COMs exceed 100\,K, telling us that they trace the gas of the hot corino, where \metc probes hotter regions than \met on average (290\,K versus 180\,K). The column density ratios between the \cd C and \ct C isotopologues range from 4 to 30, thus, are lower by factors of a few up to an order of magnitude than the expected local ISM isotope ratio of $\sim$68. We conducted astrochemical models to understand the origins of the observed low ratios. We studied potential precursor molecules of \met and \metc, since the model does not include COMs, assuming that the ratio is transferred in reactions from the precursors to the COMs. The model predicts \ratio ratios close to the ISM value for CO and H$_2$CO, precursors of CH$_3$OH, in contrast to our observational results. For the potential precursors of CH$_3$CN (CN, HCN, and HNC), the model predicts low \ratio ratios close to the protostar ($<300$\,au), hence, they may also be expected for \metc.}
   {Our results show that an enrichment in \ct C in COMs at the earliest protostellar stages is likely inherited from the COMs' precursor species, whose \ratio ratios are set during the prestellar stage via isotopic exchange reactions. This also implies that low \ratio ratios observed at later evolutionary stages such as protoplanetary disks and exoplanetary atmospheres could at least partially be inherited. A final conclusion on \ratio ratios in protostellar environments requires observations at higher angular and spectral resolution that simultaneously cover a broad bandwidth, to tackle current observational limitations, and additional modelling efforts.}

    \authorrunning{L.A.~Busch et al.}
    \titlerunning{}
   \maketitle
 
\section{Introduction}\label{s:intro}

The \ratio isotopic ratio is one of the most extensively studied in the interstellar medium (ISM). The main \cd C and the \ct C isotopes are produced in the interior of stars via nucleosynthesis, where \cd C is a mostly primary and \ct C is a primary and secondary product \citep[e.g.][]{Romano2022}.
Therefore, the elemental \ratio ratio is set by the local stellar activity and changes over the lifetime of the galaxy. In the Milky Way, the \ratio ratio increases with Galactocentric distance, where values of $25\,\pm\,7$ are expected for the Galactic centre region and $68\,\pm\,15$ for the Solar neighbourhood ISM \citep[][based on observation of CN, CO, and H\2CO]{Milam2005}. 

\begin{figure}[t] \vspace{-0.5cm}
    \centering
    \includegraphics[width=0.5\textwidth]{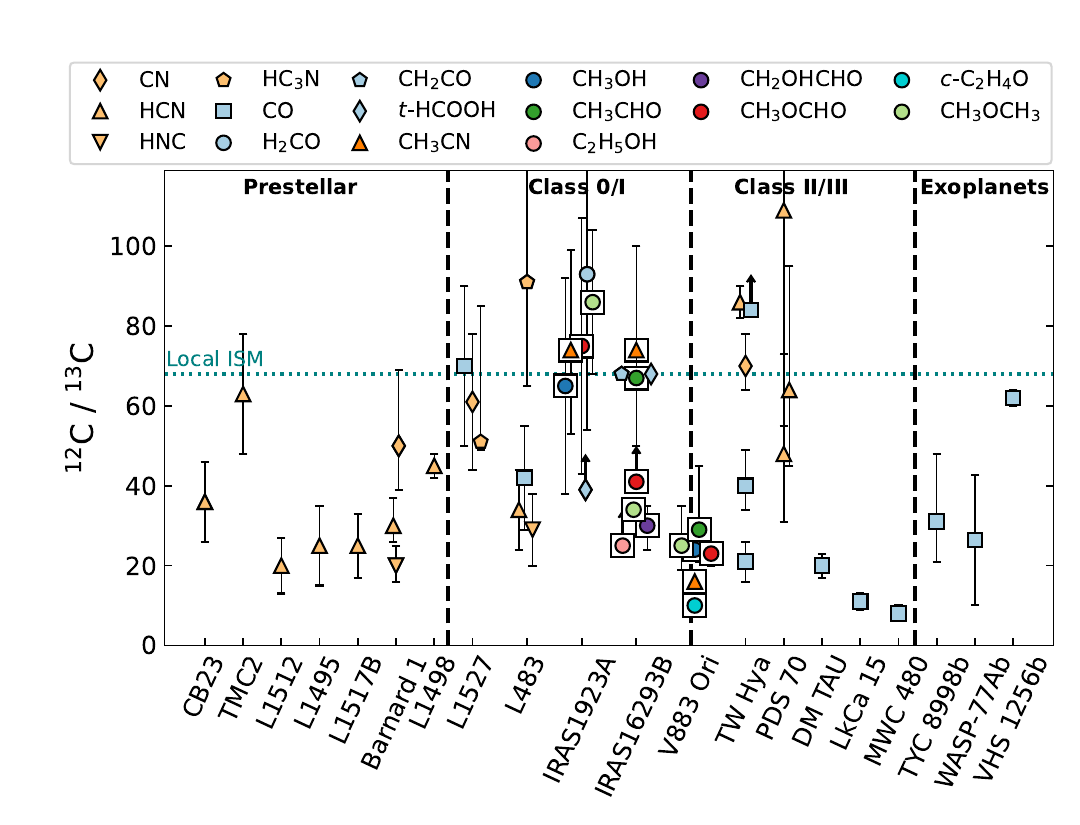}
    \caption{Literature \ratio ratios towards sources at different evolutionary stages using different isotopologues. Simple O- and N-bearing species ($<$\,6 atoms) share a light blue and light yellow colour, respectively, but have different markers. Black rectangles around the orange triangle (N) and coloured circles (O) indicate complex organics ($\geq$\,6 atoms). Arrows indicate lower limits. References to the values can be found in Table\,\ref{tab:refs}. }
    \label{fig:ratio_woP}
\end{figure}

Just as the main isotope, \ct C can react to form molecules. The \ratio isotopologue ratios derived in the Solar system are quite constant, while there can be significant deviations from the local ISM value of 68 (see Fig.\,\ref{fig:ratio_woP}) between nearby star forming regions or between different organic molecules \citep[e.g.][]{Nomura2023}.
The \ct C gets incorporated into CO via the following exchange reaction \citep[e.g.][]{Colzi2020}:
\begin{align}
    \mathrm{^{13}C^+ +\, ^{12}CO} \leftrightarrow \mathrm{^{12}C^+ +\, ^{13}CO + 34.7\,K,} \label{rct:CO}
\end{align}
which preferentially produces \ct CO in the low temperature gas phase \citep{Langer1984}, before it freezes out on dust grains, where it can then be involved in reactions that produce more complex species such as \met. 
Conversely, molecules formed from C$^+$, such as HCN or CN, would rather be depleted in \ct C. 
However, low \ratio ratios have also been measured for the latter species (see Fig.\,\ref{fig:ratio_woP}). New reactions have meanwhile been explored and found to be efficient in enriching these molecules in \ct C \citep[e.g.][]{Roueff2015,Colzi2020,Loison2020}, for example:
\begin{align}
    \mathrm{^{13}C^+ +\, ^{12}CN} &\leftrightarrow \mathrm{^{12}C^+ +\, ^{13}CN + 31.1\,K,}\label{eq2}\\
    \mathrm{^{13}C +\, ^{12}CN} &\leftrightarrow \mathrm{^{12}C +\, ^{13}CN + 31.1\,K,}\\
    \mathrm{^{13}C + H^{12}CN} &\leftrightarrow \mathrm{^{12}C + H^{13}CN + 48\,K.}\label{eq4}
\end{align}
In addition to isotopic fractionation by exchange reactions, there are other processes that can locally alter the isotopic ratio, for example, selective photodissocation \citep[e.g.][]{Visser2009}. 

These local variations in the fractionation of molecules are therefore well suited to shed light onto the formation of the species in general but also on the physical conditions that certain isotopic ratios may be associated with. 
In particular, isotopologue fractionation can be used to address the question of inheritance or in-situ formation of molecules during the various stages of the star-formation process, that is from cold molecular clouds to protostellar environments and, eventually, planets \citep[e.g.][]{Caselli2012,Wampfler2014,Cleeves2014}.
For example, the deuterium fractionation of molecules and the D/H ratio have been studied and evidence for inheritance of deuterated molecules from the prestellar to later stages have been found \citep[e.g.][]{Ceccarelli2014,Drozdovskaya2021}. 

Recently, \ratio ratios have been determined in the atmospheres of exoplanets using CO isotopologues: the super Jupiters VHS\,1256b \citep{Gandhi2023} and TYC\,8998--760--1b \citep[][]{Zhang2021}, and the hot Jupiter WASP--77Ab \citep{Line2021}. 
For VHS\,1256b a \ratio ratio close to the expected local ISM value of 62\,$\pm$\,2 was derived, while the 
ratios for the other two are significantly lower:  $31^{+17}_{-10}$ and 10.2--42.6, respectively (see 
Fig.\,\ref{fig:ratio_woP}). 
Low \ratio ratios for CO isotopologues were also reported for the protoplanetary disks DM TAU (20\,$\pm$\,3), LkCa\,15 (11\,$\pm$\,2), and MWC\,480 (8\,$\pm$\,2) by \citet{Pietu2007} as well as for the disk around TW Hya \citep[$\sim20$ and 40,][]{Zhang2017,Yoshida2022}. The values are summarised in Fig.\,\ref{fig:ratio_woP}.
All together, these observations indicate that exoplanets likely inherit the \ratio of the disk material from which they form. 
In contrast to CO, ratios similar to or even higher than the local ISM value were derived for HCN isotopologues in PDS\,70 \citep{Rampinelli2025} and for HCN and CN in the TW Hya disk \citep{Hily-Blant2019,Yoshida2024}. 
These values were derived within a radius of $\sim$110\,au from the central star. A high \ratio ratio ($>$84) for CO was found at larger radii \citep{Yoshida2022}. All ratios that have been derived for TW Hya are summarised in \citet{Bergin2024}. 
In order to get to the bottom of the apparent presence of two carbon isotopic reservoirs, these latter authors used thermochemical models to trace changes in the \ratio chemistry as a function of time in a disk. 
The observations can be reproduced with a model of a disk at an early evolutionary stage that has C/O$\geq$1, no carbon depletion, and is exposed to cosmic rays (i.e. cosmic rays are not attenuated due to high densities in the disk). This effect of [C/O]$_\mathrm{elem} > 1$ on the \ratio ratios has been confirmed in simulations by \citet{Lee2024}.  
These simulations assume that the two observed carbon isotope reservoirs are the result of in-situ chemistry in the disk. 

However, low \ratio ratios have also been found towards a few low-mass pre- and protostellar sources\footnote{Lower \ratio ratios than expected have also been found towards high-mass analogues \citep[e.g.][]{Palau2017,Beltran2018,Bouscasse2022}. These results are not discussed in the current work, as we focus on low-mass protostellar objects.} (see Fig.\,\ref{fig:ratio_woP}). 
Reliable measurements at these early stages remain scarce as emission from simple species such as CO, HCN, or CN is optically thick, sometimes even for the rarer isotopologues. Instead, complex organic molecules (COMs, C-bearing molecules with $\geq$6 atoms) can be used to shed light on the evolution of the \ratio ratio. This has only been done for a variety of COMs towards the IRAS\,16293-2422 (IRAS\,16293 hereafter) low-mass protostellar system. 
With this work, we aim to increase the number of measurements of \ratio ratios towards young protostars to see whether low ratios are already observed from this earlier evolutionary stage. In addition, we test whether there is a difference between N- or O-bearing molecules in the \ratio ratio that may hint at the presence of two carbon isotope reservoirs, as indicated in TW Hya. 

Besides, it is crucial to know any deviations from the local ISM value at these stages of star formation to accurately derive molecular abundances.  
For example, the simplest and most abundant of the COMs, methanol (\met), is often used to normalise column densities of other COMs to compare molecular inventories of different sources \citep[e.g.][]{Belloche2020,vanGelder2020,Yang2021}. However, the main isotopologue is likely optically thick towards young protostars. Then, its optically thin(ner) \ct C isotopologue can be used to infer the main isotopologue's column density assuming a \ratio ratio \citep[e.g.][]{vanGelder2022}. Therefore, the \ratio ratio can be a crucial parameter in determining and comparing chemical inventories.  
\begin{table*}[h]
    \caption{List of sources from the PRODIGE sample towards which \ct C isotopologues of \met and \metc have been detected.}
    \centering
    \begin{tabular}{cllccccccc}
    \hline\hline \\[-0.3cm]
        ID\tablefootmark{a} & Source & Other names& R.A. & Dec & Class\tablefootmark{b} & $L_\mathrm{bol}$\tablefootmark{c} & $T_\mathrm{bol}$\tablefootmark{d} & $\theta_S$\tablefootmark{e} & $\varv_\mathrm{sys}$\tablefootmark{f} \\
          &     & & (J2000) & (J2000) & & $(L_\odot)$ & (K) & (\arcsec) & (\kms) \\
       \hline\\[-0.2cm]
        008 & IRAS\,2A & Per-emb-27 & 03:28:55.57 & $+$31:14:37.03 & 0/I & $31.0\pm0.7$ & $69.0\pm1.6$ & 0.2\tablefootmark{1} & 7.50 \\ 
        010 & IRAS\,4B & Per-emb-13 & 03:29:12.02 & $+$31:13:08.03 & 0 & $6.50\pm0.5$ & $28.0\pm1.0$ & 0.4\tablefootmark{2} & 7.80 \\ 
        011 & B1-c & Per-emb-29 & 03:33:17.88 & $+$31:09:31.82 & 0 & $6.00\pm0.7$ & $48.0\pm1.0$ & 0.2\tablefootmark{1} & 6.55 \\ 
        012 & L1448-IRS2 & Per-emb-22 & 03:25:22.41 & $+$30:45:13.26 & 0 & $5.90\pm0.8$ & $43.0\pm2.0$ & 0.5\tablefootmark{3} & 4.25 \\ 
        002 & SVS13A & Per-emb-44 & 03:29:03.76 & $+$31:16:03.81 & 0/I & $53.1\pm11.6$ & $188.0\pm9.0$ & 0.3\tablefootmark{2} & 8.00 \\ 
        004 & B1-bS & & 03:33:21.36 & $+$31:07:26.37 & 0 & $1.10\pm0.2$ & $17.7\pm1.0$ & 0.4\tablefootmark{3} & 6.75 \\ 
        005 & L1448C & L1448-mm & 03:25:38.88 & $+$30:44:05.28 & 0 & $13.7\pm2.5$ & $47.0\pm7.0$ & 0.5\tablefootmark{2} & 5.35 \\
        & & Per-emb-26 & & & & & & & \\  
        \hline\hline
    \end{tabular}
    \tablefoot{\tablefoottext{a}{Source ID number within PRODIGE.}\tablefoottext{b}{Young stellar object classification taken from \cite{Tobin2016}.}\tablefoottext{c}{Bolometric luminosity taken from \citet{Tobin2016}, but corrected for the new distance to Perseus of  \citep[230\,pc, ][]{Zucker2018}.}\tablefoottext{d}{Bolometric temperature taken from \citet{Tobin2016}.}\tablefoottext{e}{Assumed size of emitting regions of molecules, taken from (1) \citet{Chen2024}, (2) \citet{Belloche2020}, and (3) \cite{Yang2021}.}\tablefoottext{f}{Approximate systemic velocity stored in the data cubes. Velocities derived from spectral lines in this work are obtained by adding $\varv_\mathrm{off}$ from Tables\,\ref{tab:IRAS2A}--\ref{tab:L1448C}. }}
    \label{tab:Sources}
\end{table*}

We made use of the data taken as part of the Northern Extended Millimeter Array (NOEMA) large program PROtostars \& DIsks: Global Evolution (PRODIGE, PIs: P.\,Caselli and Th.\,Henning) at 1\,mm. The project targets 32 Class 0/I systems in the Perseus molecular cloud, which is at a distance of 294\,$\pm$\,17\,pc \citep{Zucker2018}.
The survey covers a plethora of molecular lines thanks to its broad bandwidth of 16\,GHz. First results using this data set have been published in \citet{Valdivia-Mena2022}, \citet{Hsieh2023,Hsieh2024}, and \citet{Gieser2024}.
We used the isotopologues\footnote{When the mass number of the C atom is not specified, we talk about the main isotopologue $^{12}$C.} of methanol (\met and \ct\met) and methyl cyanide (CH$_3$CN, \ct\metc, and \metctr) to probe their \ratio ratio in the hot gas towards the selected sources. In addition, we include CH$_3^{18}$OH to the analysis to address optical depth effects. 
This article is structured as follows: Sect.\,\ref{s:obs} briefly describes the observations, data reduction and imaging. In Sect.\,\ref{s:method}, we describe the methods of the spectral-line analysis and present the results. The results are compared with studies towards other sources and with predictions of astrochemical models in Sect.\,\ref{s:discussion}. Conclusions are provided in Sect.\,\ref{s:conclusion}.

\section{The PRODIGE data set}\label{s:obs}

The PRODIGE large program is part of the MPG/IRAM Observatory Program (MIOP, Project ID L19MB) observed with NOEMA. A total of 32 Class 0/I protostellar systems were targeted, of which we study seven in this work (see Table\,\ref{tab:Sources} and Sect.\,\ref{s:method}).
The observations are carried out using the Band 3 receiver and the PolyFix correlator, tuned to local-oscillator (LO) frequency of 226.5\,GHz. 
This setup provides a total bandwidth of 16\,GHz with a channel width of 2\,MHz, corresponding to $\sim$2.5--2.8\,\kms, depending on the frequency. These are divided into four sidebands: lower outer (214.7--218.8\,GHz), lower inner (218.8--222.8\,GHz), upper inner (230.2--234.2\,GHz), and upper outer (234.2--238.3\,GHz).
In addition, 39 high spectral resolution windows (62.5\,kHz channel width), each covering a 64\,MHz bandwidth, were placed within the 16\,GHz bandwidth.

Observations were conducted in antenna configurations C and D and finished in 2024. In this work, we only used data taken in 2019 to 2022. 
The maximum recoverable scale (MRS) is 16.9\arcsec at 220\,GHz, which corresponds to approximately 5000\,au at the distance of Perseus.
The data calibration was done using the standard observatory pipeline GILDAS/CLIC package. We used 3C84 and 3C454.3 as bandpass calibrators and 0333$+$321 and 0322$+$222 as phase and amplitude calibrators. The observations for these calibrators were taken every 20\,min. For the flux calibration we used LkH$\alpha$\,101 and MWC\,349. The uncertainty in absolute flux density is 10\%. 
Continuum subtraction and data imaging were done with the GILDAS/MAPPING\footnote{\url{https://www.iram.fr/IRAMFR/GILDAS/}} package using the \texttt{uv\_baseline} and \texttt{clean} tasks, respectively. The imaging of the spectral-line cubes was done with natural weighting to minimise noise, while for the continuum maps we used robust\,=\,1, to improve the angular resolution. 
More details on data reduction and imaging can be found in \citet{Gieser2024}. 
The achieved angular resolution is $\sim$1\arcsec, which corresponds to a spatial resolution of $\sim$300\,au.
The accurate synthesised beams for each source as well as rms noise levels can be found in Table\,\ref{tab:obs}.

\section{Spectral-line LTE analysis and results}\label{s:method}

We detected emission from the rarer \ct C isotopologue of methanol and methyl cyanide towards seven and six protostellar sources of the PRODIGE sample (see Table\,\ref{tab:Sources}), respectively.
In this section we first present the morphology of the molecular emission and the position selection (Sect.\,\ref{ss:morph}), then discuss the line widths and peak velocities (Sect.\,\ref{ss:lprops}). 
The following spectral-line analysis is performed assuming a uniform medium that is in local thermodynamic equilibrium (LTE), which is applicable given the high volume densities \citep{Yang2021}.
The  spectral-line identification and analysis were done using Weeds, which is an extension of the GILDAS/CLASS software \citep{Maret2011}. Weeds produces synthetic spectra by solving the radiative-transfer equation. More details on the modelling with Weeds are in Sect.\,\ref{ss:weeds}. 
The molecular column density as well as the rotational temperature that are obtained from this method are validated by results from a population-diagram analysis, which is described in Sect.\,\ref{ss:PD}. The results are presented in Sects.\,\ref{ss:trot} and \ref{ss:18}.

\begin{figure*}
    \centering
    \includegraphics[width=.89\textwidth]{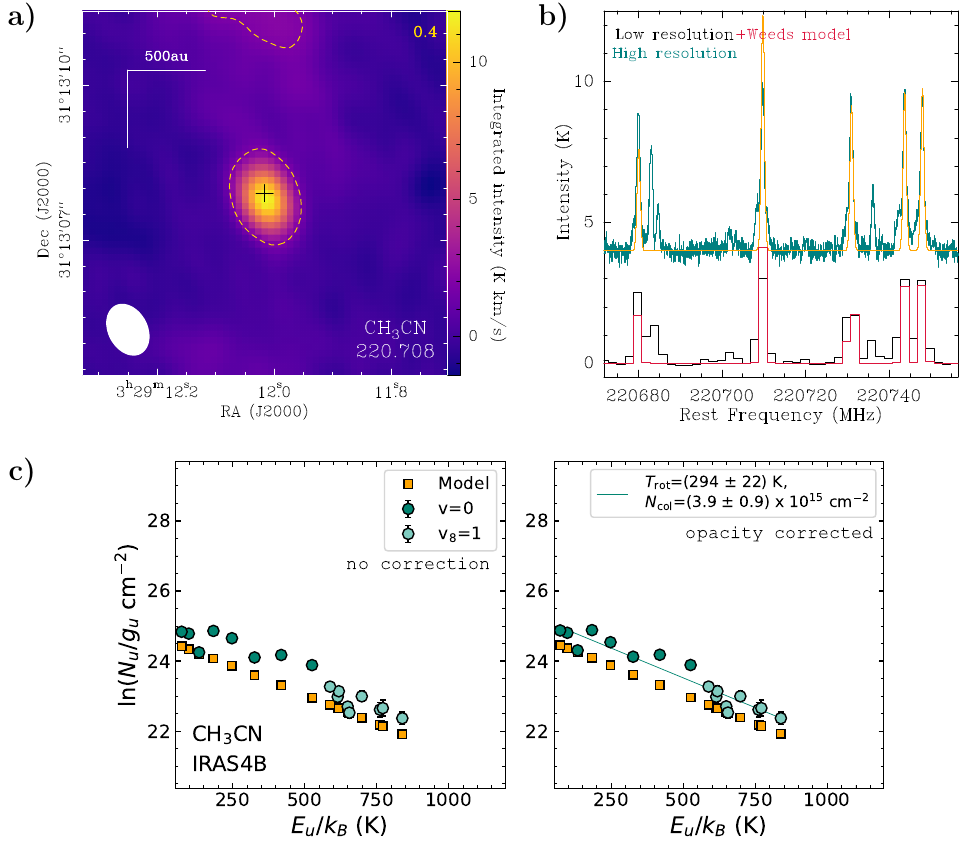}
    \caption{Overview of the PRODIGE data towards IRAS\,4B and presentation of the spectral-line analysis . \textit{Panel a:} Integrated intensity map (in K \kms) for the \metc $12_3-11_3$ transition at 220.708\,GHz. The beam (half power beam width, HPBW) is shown in the bottom left. The dashed yellow contour is at 5$\sigma$, where $\sigma$ is the rms noise level (in K\,\kms) and is written in the top right corner. \textit{Panel b:} Observed spectra towards the peak position, which is at $(\Delta\alpha,\Delta\beta)=(0\arcsec,0\arcsec)$ (black cross in panel a), of \metc at low ($\sim$2\,MHz, black) and high ($\sim$62.5\,kHz, teal) spectral resolution, where the latter was shifted by 8\,K for visualisation. The low spectral resolution data were analysed in this work (see Sect.\,\ref{s:method}). The modelled \metc spectra from Weeds are shown in red (low resolution) and orange (high resolution), where the same input parameters were used for both. \textit{Panel c:} Population diagram for \metc. Teal circles show the observed data points while the modelled data points from Weeds are shown with orange squares (see Sect.\,\ref{ss:caveats} for possible origins of differences between the two). No corrections are applied in the left panel while in the right panel corrections for opacity and contamination by other molecules have been considered (negligible effect in this case) and only transitions with optical depths $<$\,0.5 (i.e. all transitions in this case) are shown. 
    The results of the linear fit to the observed data points are shown in the right panels.}
    \label{fig:overview}
\end{figure*}

\subsection{COM emission morphology and position selection}\label{ss:morph}

As an example, Fig.\,\ref{fig:overview}a shows an integrated intensity map of \metc towards IRAS\,4B.  
Integrated intensity maps of the \cd C and \ct C isotopologues of \met and \metc for all sources are shown in Figs.\,\ref{maps:iras2a}--\ref{maps:b1bs}. 
In most cases, the emission is compact and remains spatially unresolved in the PRODIGE data.
Only for IRAS\,4B, emission outside the compact hot-corino region is identified, which can be associated with its protostellar outflow \citep[e.g.][]{Sakai2012}. 

For the LTE analyses, we extract the spectra at the continuum-peak position at (0\arcsec,0\arcsec), which corresponds to the coordinates in Table\,\ref{tab:Sources}. Although the line data may be affected by continuum optical depth, we do not select a position off-peak \citep[as done, for example, for IRAS\,16293 by][]{Jorgensen2016}. If we did so, we would risk to detect fewer transitions, which might in turn result in larger uncertainties in the LTE analyses as both methods require as many lines as possible.

\subsection{Line widths and peak velocities}\label{ss:lprops}
\begin{figure*}[t]
    \centering
    \includegraphics[width=\textwidth]{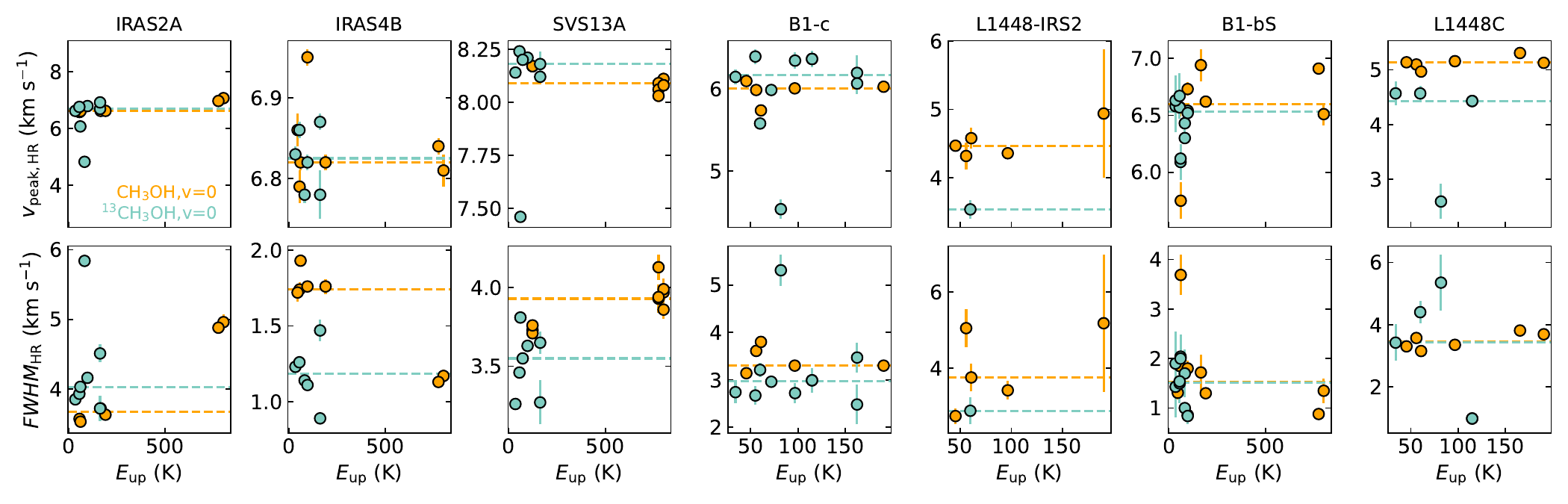}
    \caption{Peak velocity, $\varv_\mathrm{peak}$, and line width, \textit{\small FWHM}, derived from 1D Gaussian fits (including 1$\sigma$ error bars) to the high (HR) spectral resolution PRODIGE data, as a function of upper-level energy, $E_u$ for the vibrational ground states of \met (orange) and \ct\met (teal). The coloured dashed lines indicate the respective average values for $\varv_\mathrm{peak}$ and \textit{\small FWHM}. }
    \label{fig:lineprops}
\end{figure*}
For the spectral-line analysis (Sects.\,\ref{ss:weeds} and \ref{ss:PD}), we made use of the low spectral resolution (LR) data that covers the total 16\,GHz bandwidth, hence, more spectral lines of our molecules of interest.
However, because of the narrow line widths of the spectral lines (full width at half maximum, \textit{\small FWHM}, of $<$\,4\,\kms, see Tables\,\ref{tab:IRAS2A}--\ref{tab:L1448C}) they are only marginally spectrally resolved if at all. To determine the true \textit{\small FWHM} and peak velocity, we used the higher resolution (HR) data.
The HR data only cover transitions of both methanol isotopologues in their vibrational ground state, while the LR data also cover vibrationally excited states. Figure\,\ref{fig:lineprops} shows peak velocities, $\varv_\mathrm{peak}$, and \textit{\small FWHM}, which we derived from 1D Gaussian fits to all available transitions, as a function of upper-level energy, $E_u$. The median values for both species are shown with dashed lines. 
Outliers may appear as a consequence of weak transitions or unidentified contamination by other species, which can lead to shifts. 
In addition, Fig.\,\ref{fig:lineprops} only shows transitions with $\tau<10$, though some \textit{\small FWHM} may be larger due to optical depth effects. For some sources, the presence of outflow wing emission may broaden the lines.
In some cases, there are significant differences between \cd C and \ct C methanol, which may indicate that the two species probe different material along the line of sight. Higher spectral resolution data that cover a larger number of spectral lines per species will help to draw conclusions on this. For the subsequent analysis, we ignore possible kinematic differences between species and use an effective \textit{\small FWHM}, that is the lower median between the two species, and the respective $\varv_\mathrm{peak}$ value.
Because of their unresolved hyperfine structure, methyl cyanide isotopologues are not used to determine line properties. The hyperfine structure (HFS) is however considered later during the column density determination, except for \ct\metc, as the Cologne database of molecular spectroscopy \citep[CDMS,][]{CDMS}, from which all spectroscopic data was taken, does not include the HFS for frequencies beyond 220\,GHz.
%

%

\subsection{Radiative transfer modelling with Weeds}\label{ss:weeds}

We classify a molecule as detected when there are at least five transitions per isotopologue in the observed low spectral resolution spectrum above an intensity threshold of 3$\sigma$, where $\sigma$ is the noise level measured in the 3D data cubes (see Table\,\ref{tab:obs}), and the modelled spectrum reasonably reproduces these  transitions.
In order to derive a synthetic spectrum for a molecule, Weeds requires five input parameters: total column density, rotation temperature, source size, \textit{\small FWHM}, and velocity offset, $\varv_\mathrm{off}=\varv_\mathrm{peak,HR}-\varv_\mathrm{sys}$, of the spectral line from the source systemic velocity, where $\varv_\mathrm{sys}$ is taken from Table\,\ref{tab:Sources}. 
The \textit{\small FWHM} and $\varv_\mathrm{peak,HR}$ present observed values, for which we ignore any unresolved kinematic substructures and velocity gradients, and were determined from 1D Gaussian fits in Sect.\,\ref{ss:lprops}.  For \textit{\small FWHM} that are spectrally unresolved in the observed LR spectra, Weeds smoothes the line width for the given channel width. 
The source size, that is the size (in arcseconds) of the emitting area is needed to compute the beam filling factor. We adopted source sizes from the literature (Table\,\ref{tab:Sources}), which were estimated from COM emission \citep{Belloche2020,Chen2024} or, if this is not available, from continuum emission \citep{Yang2021}.
Because Weeds does not perform a fit to the observed data, the total column density, $N_\mathrm{col}$, and the rotational temperature, $T_\mathrm{rot}$ are adjusted in an iterative way: starting from a first educated guess, population diagrams (see Sect.\,\ref{ss:PD}) are derived and the resulting values for $N_\mathrm{col}$ and $T_\mathrm{rot}$ are fed back into Weeds. 
This is repeated until both Weeds and the population diagrams provide matching results for $N_\mathrm{col}$ and $T_\mathrm{rot}$ within the uncertainties. 
In the process, Weeds computes line optical depths and takes it into account when creating the model spectrum, based on the input parameters. We also use this opacity correction when deriving population diagrams (see Sect.\,\ref{ss:PD}). 
The parameters of all final models  are summarised in Tables\,\ref{tab:IRAS2A}--\ref{tab:L1448C}. 
In Fig.\,\ref{fig:overview}b we show, as an example, a spectrum of \metc towards IRAS\,4B from the low (black) and high (teal) spectral resolution PRODIGE data. The former is used for the spectral-line analysis because a wider bandwidth, hence, a larger number of transitions was covered. The derived Weeds model is shown in red. For comparison, we also show the model of the high-resolution case in orange. 

It happens that the final model over- or underestimates the peak intensity for a few transitions, even if it (and the corresponding population diagram) matches well the rest of the observed spectrum. If the model underestimates the intensity, it likely indicates contamination by another species that adds to the observed intensity. In severe cases, the line is excluded from further analysis. It may also be that the intensity of a transition is not well estimated by the model because the observed emission is not described by a single excitation temperature. This cannot be accounted for in the Weeds models (nor the population-diagram analysis) and can result in an over- or underestimation by the model. Especially, transitions with lower upper-level energies suffer from the latter effect as they can be excited more easily. 
Deviations from Gaussian profiles caused by such an non-uniform medium or optical-depth effects may be worsened due to spectral dilution in the LR data. Such limitations of this analysis and their consequences are taken up on in Sect.\,\ref{ss:caveats}.

\subsection{Population diagrams}\label{ss:PD}

To verify total molecular column densities, $N_\mathrm{tot}$, and rotational temperatures, $T_\mathrm{rot}$, used to derive the synthetic spectra, we derived population diagrams (PDs), which are based on the following formalism \citep[][]{Goldsmith1999,Mangum15}: 
\begin{align}\label{eq:pd}
    \ln\left(\frac{N_u}{g_u}\right) = \ln\left(\frac{8\,\pi\,k_B\,\nu^2 \int J(T_B) \mathrm{d}\varv}{c^3\,h\,A_\mathrm{ul}\,g_u\,f}\right) = \ln\left(\frac{N_\mathrm{tot}}{Q(T_\mathrm{rot})}\right) - \frac{E_u}{k_B T_\mathrm{rot}},
\end{align}
where $N_u$ is the upper-level column density, $g_u$ the upper-level degeneracy, $E_u$ the upper-level energy, $A_\mathrm{ul}$ the Einstein A coefficient, $k_B$ the Boltzmann constant, $c$ the speed of light, $h$ the Planck constant, $f=\frac{\mathrm{source\,size}^2}{\mathrm{source\,size}^2\,+\,\mathrm{beam\,size}^2}$ the beam filling factor, and $Q$ the partition function. Intensities in brightness temperature scale, $J(T_B)$, are integrated over a visually selected velocity range, d$\varv$, in the continuum-subtracted spectra. 
When computing the left-hand term for each transition of a species and plotting it against $E_u$, all data points should ideally follow a linear trend, in which case the level populations can be explained by a single rotational temperature. 

In Fig.\,\ref{fig:overview}c we show one PD of CH$_3$CN obtained for IRAS\,4B as an example (all other PDs can be found in Figs.\,\ref{pd:iras4b}--\ref{pd:18o} in the appendix). Both the observed data points as well as those derived from the Weeds models are shown. If detected, rotational transitions from vibrationally excited states (v=1, v=2 for \met and v$_8$=1 for \metc) are used in addition to transitions from the vibrational ground state (v=0). 
The left panel shows the original data obtained from Eq.\,(\ref{eq:pd}), while two correction factors (i.e. optical depth and contaminating emission, see below) are applied to the modelled and observed data in the right panel (negligible effect in Fig.\,\ref{fig:overview}c). 
One of the correction factors considers line opacities. Besides computing the synthetic spectra for a molecule, Weeds also calculates the optical depth for each transition for the given input parameters. Using the respective peak opacity from Weeds, $\tau_i$, for a given transition $i$, we computed a correction factor, $\frac{\tau_i}{1-\mathrm{e}^{-\tau_i}}$ \citep[e.g.][]{Mangum15}. This factor was then multiplied by the observed and modelled integrated intensities. However, the right-hand panels of the PDs only show transitions for which Weeds computes optical depths of $<$0.5. Plus, the linear fit is only applied to those data points. Since it is possible that we still underestimate optical depth for some lines (see Sect.\,\ref{ss:caveats}), we do not use transitions with higher opacities to ensure that the fit parameters apply to the sufficiently optically thin lines. 

Secondly, contaminating emission from other species is subtracted from the integrated intensities, if identified. In particular, this applies to \metc and CH$_3^{13}$CN, which have overlapping transitions at $\sim$\,220.636\,GHz. 
If the emission spatially coincides, this addition of spectra from different species and a subsequent subtraction of contaminating emission only works for optically thin transitions.
Even with the two correction factors applied, some scatter between the observed and modelled data points remains. In the example in Fig.\,\ref{fig:overview}c, the model clearly underestimates the observations. This is a result of the narrow line widths. As can be seen from the spectra in Fig.\,\ref{fig:overview}b, the spectral lines are rarely broader than one channel. However, for the integration we used at least 2--3 channels, adding contaminating emission that seems to be an additional \metc component in the case of IRAS\,4B, possibly associated with its outflow. Observational limitations such as these are discussed in detail in Sect.\,\ref{ss:caveats}.

The linear fit is applied to the corrected observed data without taking into account their displayed uncertainties, 
since not all of them can be quantified, such as possible contamination from unidentified species. 
The displayed error bars only include the standard deviation from integrating intensities and the 10\% uncertainty in absolute flux density.  
From the linear fit, we obtain $T_\mathrm{rot}$ from the slope and $N_\mathrm{tot}$ from the y-intercept. The latter requires the knowledge of $Q(T_\mathrm{rot})$, which is taken from the CDMS\footnote{Entries for \metc isotopologues use spectroscopic data from \citet{13ch3cn2009,ch3cn2015}, for \met from \citet{Xu2008}, for \ct\met from \cite{Xu199713c}, and for CH$_3^{18}$OH from \citet{Fisher2007.18o}. }, together with other spectroscopic information, such as $A_u$, $E_u$, and the rest frequencies.



\subsection{Rotational temperatures and \ratio isotopologue ratios}\label{ss:trot}

\begin{figure}[t]
    \centering
    \includegraphics[width=0.5\textwidth]{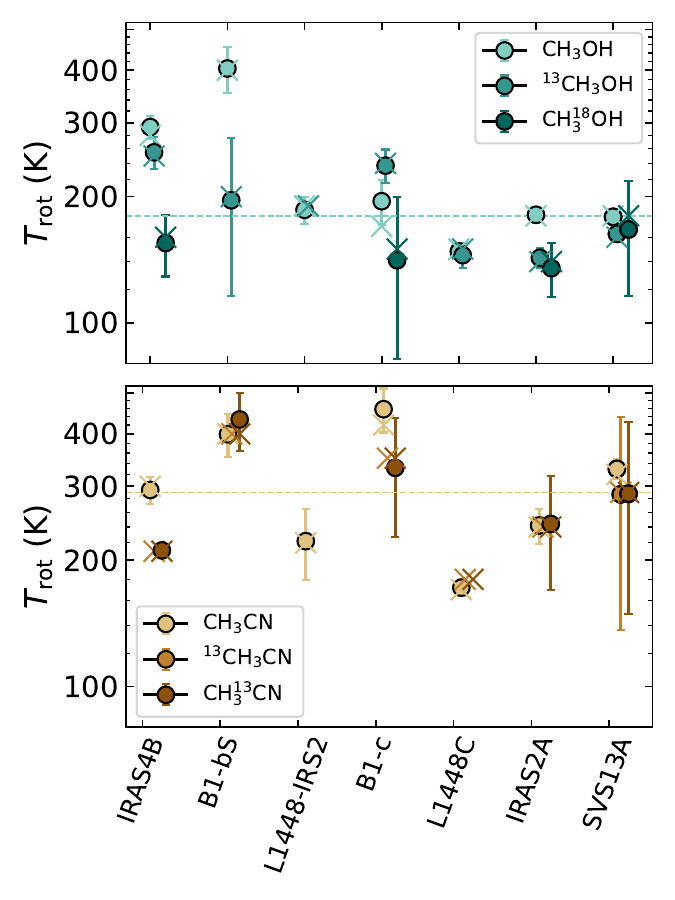}
    \caption{Rotational temperatures used for the Weeds modelling (crosses) and derived from the population diagrams (circles with 1$\sigma$ error bars). The dashed lines indicate the median values considering all isotopologues of a species.} 
    \label{fig:Trot}
\end{figure}
\begin{figure*}[t]
    \includegraphics[width=\textwidth]{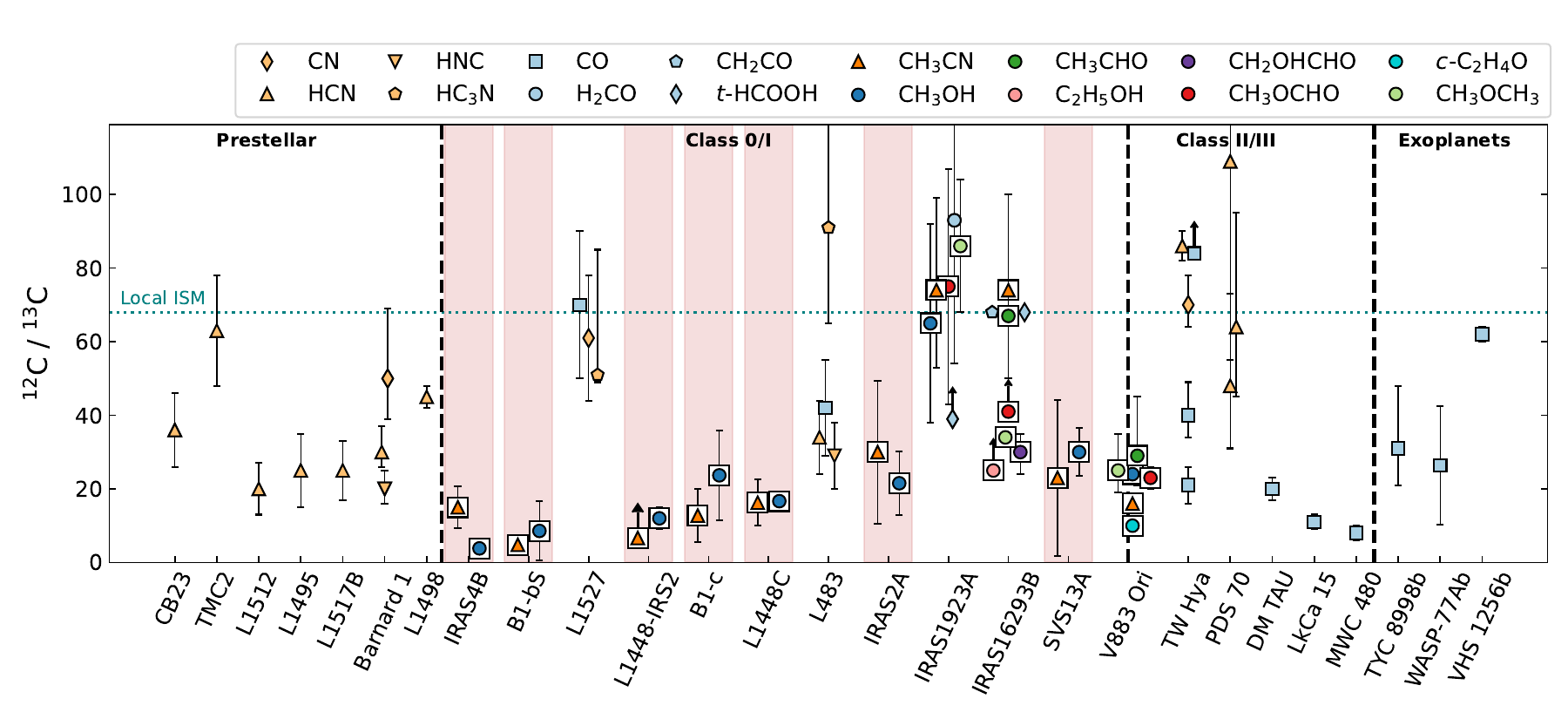}
    \caption{Same as Fig.\,\ref{fig:ratio_woP}, but including the \ratio ratios derived from the PRODIGE data in this work (lightred shaded). Bolometric luminosities of the protostars increase from left to right. Black rectangles around markers indicate complex organic molecules. Arrows indicate lower limits.}
    \label{fig:ratios_all}
\end{figure*}
\begin{table}[t]
    \centering
    \caption{The \ratio column-density ratios for \met and \metc isotopologues derived from the Weeds models.}
    \begin{tabular}{lccc}
        
        \hline\hline \\[-0.3cm] 
        Source & $\frac{N(\mathrm{CH_3OH})}{N(\mathrm{^{13}CH_3OH})}$ & $\frac{N(\mathrm{CH_3CN})}{N(\mathrm{^{13}CH_3CN})}$ & $\frac{N(\mathrm{CH_3CN})}{N(\mathrm{CH_3^{13}CN})}$ \\[0.2cm] \hline \\[-0.3cm]
    
        IRAS\,2A & 26\,$\pm$\,13 & 28\,$\pm$\,23 & 28\,$\pm$\,13 \\
        IRAS\,4B & 4\,$\pm$\,1 & 24\,$\pm$\,8 & 11\,$\pm$\,3 \\
        L1448C & 17\,$\pm$\,2 & 19\,$\pm$\,5 & 14\,$\pm$\,6 \\
        B1-bS & 9\,$\pm$\,8 & 6\,$\pm$\,2 & 4\,$\pm$\,2 \\
        L1448-IRS2 & 12\,$\pm$\,3 & 7\,$\pm$\,1 & 7\,$\pm$\,1 \\
        B1-c & 24\,$\pm$\,12 & 14\,$\pm$\,6 & 12\,$\pm$\,8 \\
        SVS13A & 30\,$\pm$\,6 & 23\,$\pm$\,21 & 23\,$\pm$\,21 \\

    \hline\hline
    \end{tabular}
    \label{tab:ratios}
\end{table}

Figure\,\ref{fig:Trot} compares the rotational temperatures derived from both the Weeds modelling and the PDs for the various isotopologues in each source. 
In all shown cases, the temperatures derived from both methods agree within the error bars. 
Large error bars result from a smaller number of mostly weak transitions that are available to derive the respective PD. 
Some PD temperatures are not shown either because the molecule was not detected (both \ct C \metc isotopologues towards L1448-IRS2) or because the PD fit was unreliable and $T_\mathrm{rot}$ was fixed in the PDs to one from the other detected species (see Tables\,\ref{tab:IRAS2A}--\ref{tab:L1448C}). 

Overall, the results show that for most sources, \metc has higher temperatures than \met. The median rotational temperature derived from methanol is 180\,K (140--400\,K), that from methyl cyanide is 290\,K (180--420\,K).
In some cases, the temperature values for the respective isotopologues are different. 
For example, for B1-bS and IRAS\,2A, \met shows higher temperature than \ct\met. This similarly applies to the \metc isotopologues for IRAS\,4B. This may indicate that the isotopologues probe different excitation conditions. 

Table\,\ref{tab:ratios} summarises the \ratio isotopologue ratios computed from the Weeds column densities of \met and \metc for the PRODIGE sample sources. The ratios are shown in comparison to other sources in Fig.\,\ref{fig:ratios_all}, where we used an average value for the two \ct C isotopologues of methyl cyanide. The error bars are those from the linear PD fits. 
The ratios range from 4 to 30 and are thus all significantly below the local ISM value of 68. There is no significant difference between \met and \metc. 
In some cases, uncertainties are likely larger than depicted (see Sect.\,\ref{ss:caveats}). 

\subsection{CH$_3^{18}$OH}\label{ss:18}

The \odh O methanol isotopologue is detected towards four of the seven sources of our sample: SVS13A, IRAS\,2A, IRAS\,4B, and B1-c.
We applied the same analysis to this molecule: synthetic spectra were derived by fixing all parameters to those of \ct\met, except for the rotational temperature and column density. Population diagrams can be found in Fig.\,\ref{pd:18o}. 

\begin{figure}
    \centering
    \includegraphics[width=.5\textwidth]{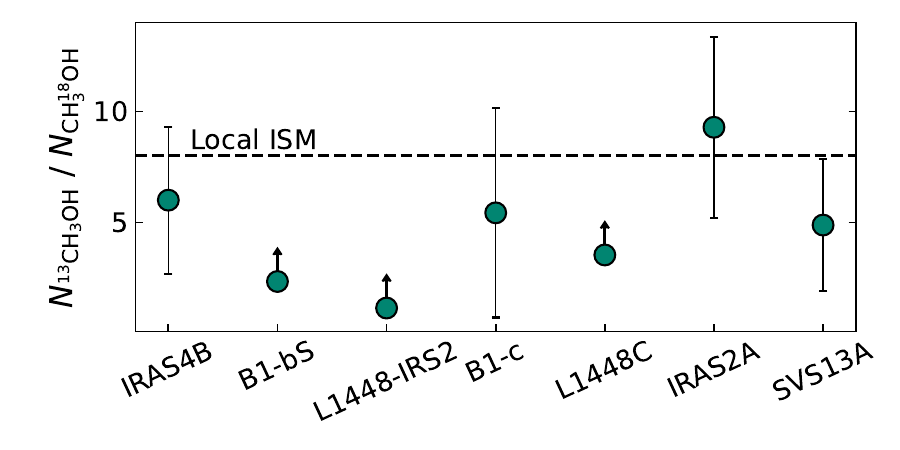}
    \caption{Column density ratio between \ct\met and CH$_3^{18}$OH using the Weeds column densities. The expected local ISM value of $\sim$8 is indicated with a dashed line. Arrows shows lower limits.}
    \label{fig:18o}
\end{figure}

The derived rotational temperatures (Fig.\,\ref{fig:Trot}) remain below 200\,K and are hence either comparable to or lower by a few tens of Kelvin than the other \met isotopologues.
Figure\,\ref{fig:18o} shows the resulting column density ratios between \ct\met and CH$_3^{18}$OH together with lower limits on this ratio for the sources towards which the latter species remains undetected. Based on the local ISM value of $^{16}\mathrm{O}/^{18}\mathrm{O}=557$ \citep[e.g.][]{Nomura2023}, we expect $^{13}\mathrm{CH}_3^{16}\mathrm{OH}/^{12}\mathrm{CH}_3^{18}\mathrm{OH}\sim 8$, which is consistent with our results within the error bars. 

\section{Discussion}\label{s:discussion}

\subsection{Low \ratio COM isotopologue ratios in other protostellar systems}\label{ss:Others}

Low \ratio ratios for (COM) isotopologues towards protostellar systems have been reported in the past. Using the PRODIGE data,
\citet{Hsieh2023} derived a similarly low \ratio ratio ($\sim$15--25) for \metc towards the binary protostellar system SVS13A. These authors tested whether modelling one or two excitation components would lead to a different outcome, which it did not as low ratios were seen in either scenario. 
In a follow-up study at much higher angular resolution with the Atacama Large Millimetre/submillimetre Array (ALMA) the molecular emission is resolved and the low \ratio ratios for \metc are confirmed \citep{Hsieh2025} towards one of the protostars.

Based on data obtained with ALMA at $\sim$0.5\arcsec\,\,angular resolution as part of the Protostellar Interferometric Line Survey (PILS) towards the young protostellar triple system IRAS16293--2422 (IRAS16293 hereafter), \citet{Jorgensen2016} reported \ratio ratios of 24--35 for the glycolaldehyde (CH\2OHCHO) isotopologues for IRAS16293B. 
Based on the same observational data, low \ratio isotopologue ratios were also (tentatively) found for other O-bearing molecules \citep[see Fig.\,\ref{fig:ratios_all} and][]{Jorgensen2018}: \et, \mf, and \dme. 
These three species, together with glycolaldehyde and methanol, likely form in the solid phase during the cold prestellar stage as predicted by astrochemical models \citep[e.g.][]{Garrod2022}. They are formed from CO that was previously enriched with \ct C via the exchange reaction\,(Eq.\,\ref{rct:CO}) in the cold gas phase prior to freeze-out. 
This similarly applies to \ad, which can be formed in the solid phase from CH$_2$CO, which in turn is formed from CO \citep[e.g.][]{Ibrahim2022, Garrod2022}. However, these two species and $t$--HCOOH, whose formation also includes CO, have \ratio ratios close to the local ISM value in IRAS16293B, based on the PILS data. 
At temperatures $>$100\,K, \ad has an efficient formation route in gas phase \citep[C$_2$H$_5$ + O, e.g.][]{Garrod2022}, which might explain why its \ratio ratio is different from the other O-bearing COMs, if the ratio for C$_2$H$_5$ were as high. 
%
%
Again using the PILS data, \citet{Calcutt2018} derived \ratio for the \metc isotopologues of $\sim$70--80 in IRAS16293A and B, which is in contrast to our results. Differences between different star-forming regions in the \ratio ratio of a given species may be linked to the duration of the prestellar phase as indicated by models (see Sect.\,\ref{ss:models}).

The \ratio COM isotopologue ratio has also been investigated within the framework of the ALMA Survey of Orion Planck Galactic Cold Clumps \citep[ALMASOP,][]{Hsu2020,Hsu2022} at an angular resolution of $\sim$0.4\arcsec. Of the 72 targeted clumps (some containing multiple pre- and protostellar sources) 56 Class 0/I protostars have been identified \citep{Dutta2020}, of which eleven show warm methanol emission \citep{Hsu2022}. For another six of these sources, \ct\met has been detected. The resulting \ratio ratios are all below 5, likely because not many methanol transitions are covered in the survey resulting in a strong underestimation of the main isotopologue's optical depth. 

Recently, low \ratio ratios of 20--30 have also been measured for several O-bearing COMs towards the V883\,Ori disk \citep{Yamato2024,Jeong2025}, based on ALMA data at 3\,mm. Similar to the considerations done for IRAS\,16293, the authors argue that one possible reason may be the COMs' formation from CO in ices at early stages, where CO was enriched in \ct C prior to freeze-out, but further investigation is needed.

\subsection{No evidence for an O/N dichotomy based on the \ratio ratio}\label{dss:oVSn} 

We selected \met and \metc for this work not only because they trace the warm gas towards the protostellar sources, but also in order to trace a possible segregation between O- and N-bearing molecules based on their \ratio isotopologue ratios. 
For the disk around the T Tauri star TW Hya \citep[see Sect.\,\ref{s:intro} and, e.g.][]{Bergin2024}, low \ratio ratios were derived for CO inside a 100\,au radius and ratios close to or higher than the local ISM value for C\2H, HCN and CN (see also Fig.\,\ref{fig:ratios_all}). The presence of two different carbon isotopic reservoirs was proposed as the origin for this dichotomy. 
Disk simulations \citep[][]{Lee2024} were able to reproduce the various \ratio isotopologue ratios when considering C/O$>$1. 
In this work, we explored whether any difference between O- and N-bearing molecules in terms of the \ratio ratio can already be observed at the Class\,0 stage.
First evidence for this can be seen from the ratios derived towards IRAS16293B (see Sect.\,\ref{ss:Others} and Fig.\,\ref{fig:ratios_all}), where (some) O-bearing COMs are enhanced in the rarer \ct C isotopologue, while \metc shows ISM-like values. 
Our results do not show significant differences between \met and \metc based on their \ratio ratios (Fig.\,\ref{fig:ratios_all}). It is worth mentioning that for some sources the rotational temperatures that we derived for the \met and \metc isotopologues were different. This may indicate that they probe different conditions. This effect on the \ratio ratio and others that arise from observational limitations (see Sect.\,\ref{ss:caveats}) can be tackled in observations at higher angular and spectral resolution.

\subsection{Implications from the $^{13}$\met / CH$_3^{18}$OH ratio}\label{dss:18o}

The local ISM \oratio isotopic ratio is 557\,$\pm$\,30 \citep{Wilson1999}. Dividing this value by the \ratio ratio of 68, yields a theoretical \ct C$^{16}$O/\cd C\odh O ratio of $\sim$8, which agrees with the values that we derived for \met towards the PRODIGE sources within the error bars (see Fig.\,\ref{fig:18o}). 
Therefore, based on this isotopologue ratio, we do not expect an enrichment in \ct C, which may imply that we are underestimating the column density of the main isotopologue.
As mentioned in Sect.\,\ref{dss:oVSn}, differences in rotational temperatures between \met isotopologues may affect the ratios.

Regarding the robustness of the \ct C$^{16}$O/\cd C\odh O ratios, lower values would mean that the abundance of CH$_3^{18}$OH is enhanced, which has not been reported so far. On the contrary, oxygen fractionation is proposed to be a consequence of isotope-selective photodissociation that depletes CO-related molecules in \odh O \citep[e.g.][]{Nomura2023}.
\citet{Loison2019} modelled the \oratio ratio in dark cloud conditions and compared it with observations. They found that in order for molecules to be enriched in \odh O, the isotope has to be abundant in the gas phase. Consequently, only a small fraction would be available in ices. They focussed on S-bearing species, but also computed the \oratio for CO and \met in the gas phase as a function of time and find no significant deviations from the local ISM value. More simulations are needed to ascertain whether this remains to be the case during the protostellar phase.

The \ct C$^{16}$O/\cd C\odh O ratios have recently been determined for Class II sources, which were observed as part of the PRODIGE -- Planet-forming disks in Taurus with NOEMA \citep[][]{Semenov2024}. They were mostly found to be close to the expected value of 8, in agreement with the ratios we find at earlier stages, although some of these disk systems are rather described by isotopologue ratios deviant from the ISM values, according to simulations \citep{Franceschi2024}.

\subsection{Tentative evidence for an O/N temperature segregation}\label{dss:Tseg} 

Segregation between O- and N-bearing molecules is frequently observed \citep[e.g.][]{Widicus2017,Allen17,Jorgensen2018,Csengeri2019,Pagani2019,Busch2022,Busch2024, Bouscasse2024}, either in the spatial distribution of their emission or in terms of rotational temperature and abundances. 
We derived on average higher rotational temperatures (Fig.\,\ref{fig:Trot}) for the \metc (290\,K) than for the \met isotopologues (180\,K) implying that the two species probe different conditions in the gas, which may trace back to their dominant formation and destruction mechanisms. 
Methanol gas-phase abundances are predicted to peak at $\sim$100--150\,K \citep{Garrod2022}. At this temperature the COM thermally desorbs from the icy dust grains, where it was formed prior to the birth of the protostar via successive hydrogenation of CO. Due to a lack of efficient gas-phase formation routes, the COM was found to be destroyed in the high-temperature gas phase right after its desorption \citep[][]{Garrod2022,Busch2022}. On the other hand, methyl cyanide can efficiently be produced at high temperatures in the gas phase \citep{Garrod2022}. 
Another process that may explain this difference is carbon-grain sublimation \citep[][see also \citeauthor{Nazari2023} 2023 and \citeauthor{vanthoff2024} 2024]{vanthoff2020}. It was proposed to enhance abundances of N-bearing COMs at temperatures $>$300\,K through release of more refractory C- and N-rich species from the grain mantle.

Some of our analysed sources were also covered in the analysis of the spectral-line data obtained as part of the IRAM Plateau de Bure Interferometre (PdBI) large program CALYPSO \citep[Continuum And Lines in Young ProtoStellar Objects,][]{Belloche2020}. This survey resembles the PRODIGE survey in angular and spectral resolution, but covers a smaller bandwidth. 
The authors used the same methods of analysis as in this work and do not find a significant difference in rotational temperature between O- and N-bearing molecules in general. 
Their values are likely more uncertain because they cover fewer transitions and a smaller range of upper-level energies.
The authors do not report column densities for the rarer isotopologues of \met and \metc. Their column densities of the main isotopologues agree however within a factor of 2. Only for SVS13A and IRAS\,2A, we derived methanol column densities that are a factor of 5.5 and 8 higher, respectively, than reported by \citet{Belloche2020}. 
We assumed a source size of $0.2\arcsec$\,\,\citep{Chen2024} for IRAS\,2A, whereas \citet{Belloche2020} used $0.35\arcsec$, resulting in a threefold difference in the column densities. The remaining difference is likely attributed to optical depth.
According to our Weeds models for the two sources, all methanol transitions with $E_u<600$\,K show opacities above 0.5, hence, we did not include them in the PD fits. In contrast, \citet{Belloche2020} use all transitions, likely resulting in an underestimation of the column density.


\subsection{Comparison with astrochemical models}\label{ss:models}

We ran simulations of carbon and nitrogen isotopic fractionation using the gas-grain chemical code {\sl pyRate}, whose basic properties are described in \citet{Sipila15a}. We employed the gas-phase and grain-surface chemical networks presented recently in \citet{Sipila23}, where an extensive set of isotopic exchange reactions for C- and N-bearing species -- also taking into account the possibility of simultaneous isotopic substitution in both elements (e.g. $\rm H^{13}C^{15}N$) -- is considered, though in the present paper we do not provide predictions for $\rm ^{15}N$-bearing species. 
The networks however only include species consisting of up to five atoms, hence, we cannot obtain direct estimates for the fractionation of $\rm CH_3OH$ and $\rm CH_3CN$. Instead, we simulated the abundances of precursor species, and used them to estimate the isotopic ratios of $\rm CH_3OH$ and $\rm CH_3CN$ based on their most likely formation pathways, to constrain the possible origins of the low ratios derived for these COMs from the present observations.

We have not attempted to simulate each observed source individually. Instead, we have adopted as a template the physical model for the protostellar core IRAS16293-2422 presented by \citet{Crimier10}, and carried out chemical simulations to predict the \ratio ratios of various species. Here, we give a brief account of the model setup; an expanded description including the initial conditions is given in Appendix~\ref{a:model}. Following \citet{Brunken14} and \citet{Harju17b}, we have employed a two-stage approach where the chemistry first evolves over a period of $10^6$\,yr in an intermediate-density dark cloud, and then for another period of $10^4$\,yr in the protostellar conditions using the abundances obtained in the first stage as initial abundances for the second stage. For simplicity, the primary cosmic-ray ionisation rate was set to the canonical value of $\zeta_p = 1.3 \times 10^{-17} \, \rm s^{-1}$ across the core\footnote{We tested the effect of a factor of 10 variations in the cosmic-ray ionisation rate, similar to the range reported by \citet{Pineda2024}, but the results were not significantly different to those of the fiducial model, and are hence not shown here.}. The initial elemental \ratio ratio is set to 68 in the first stage \citep{Milam2005}. The model of \citet{Crimier10} is not appropriate for describing the very innermost regions of the protostellar core (inside a few hundred au), but is applicable in the present context where the observations have not been carried out using high angular resolution.

Figure\,\ref{fig:models} shows the \ratio ratios for various $\rm CH_3OH$ and $\rm CH_3CN$ precursors as a function of distance from the center of the protostellar core, taken at the simulation time $t_4=10^4$\,yr in the second (i.e. protostellar) stage of the simulation. We only show this time step because the ratios do not change significantly at later stages, except for H$_2$CO (see below). 
The complex features evident in the various ratios as a function of radius, especially prominent at radii in the range 1000--2000\,au, are a result of complex interactions involving different binding energies between the species, isotopic exchange reactions etc. Quantifying these is outside the scope of the present paper, where we concentrate only on the innermost regions of the core (left of the vertical dotted line in Fig.\,\ref{fig:models}) and analyse the general trends. In what follows, we discuss the results for species associated with methanol and methyl cyanide separately.

\subsubsection{Methanol (precursors)}

Methanol is formed during the prestellar phase in the ice mantles of dust grains through successive hydrogenation of CO. 
Figure\,\ref{fig:darkCloudEvolution} shows the evolution of the \ratio ratio during the prestellar phase, showing that as time advances, the \cd CO/\ct CO ratio approaches the ISM value of 68 in both gas and ice. We would expect the ratio for methanol to follow a similar trend, so that when \met sublimates from the ices, the value should not be too different from what is seen at the end of the prestellar stage.
This is higher than what we observe. 
It may be that the observed protostellar systems spent a shorter time in the prestellar phase than assumed in the model, thereby inherited lower \ratio ratios (Fig.\,\ref{fig:darkCloudEvolution}). Such a time dependence may also explain the differences in the \ratio ratios seen for our Perseus sources and, for example, IRAS16293 (see Sect.\,\ref{ss:Others}), though this only has a significant impact for $t<10^5$\,yr. However, predictions by \citet{Ichimura2024} show that, between 10$^5$--10$^6$\,yr, the \ratio ratio for \met in ices during the prestellar phase can be somewhere between 40--60. Hence, though a shorter prestellar phase may not explain our observed ratios of $<$\,30, it can still affect the observed \ratio ratios during the protostellar stage.
In Section\,\ref{ss:caveats} we discuss that the column densities of especially the main isotopologue of methanol derived from observations can be underestimated due to an underestimation of optical depth, hence the \ratio ratios may be higher.
A final conclusion needs further observational and computational efforts.

Moreover, the \ratio ratios close to the local ISM value for CO as predicted by the current simulations are in agreement with observations of ices in the ISM \citep[][see their Fig.\,10, and references therein]{Brunken2024}.
In addition, we note that our simulation prediction for the \ratio CO ratios is similar to those presented recently by \citet{Bergin2024} and \citet{Ichimura2024}; the latter authors predicted a $\rm CH_3OH$ \ratio ratio close to that of CO (below 68).

\begin{figure}[t]
    \centering
    \includegraphics[width=0.53\textwidth]{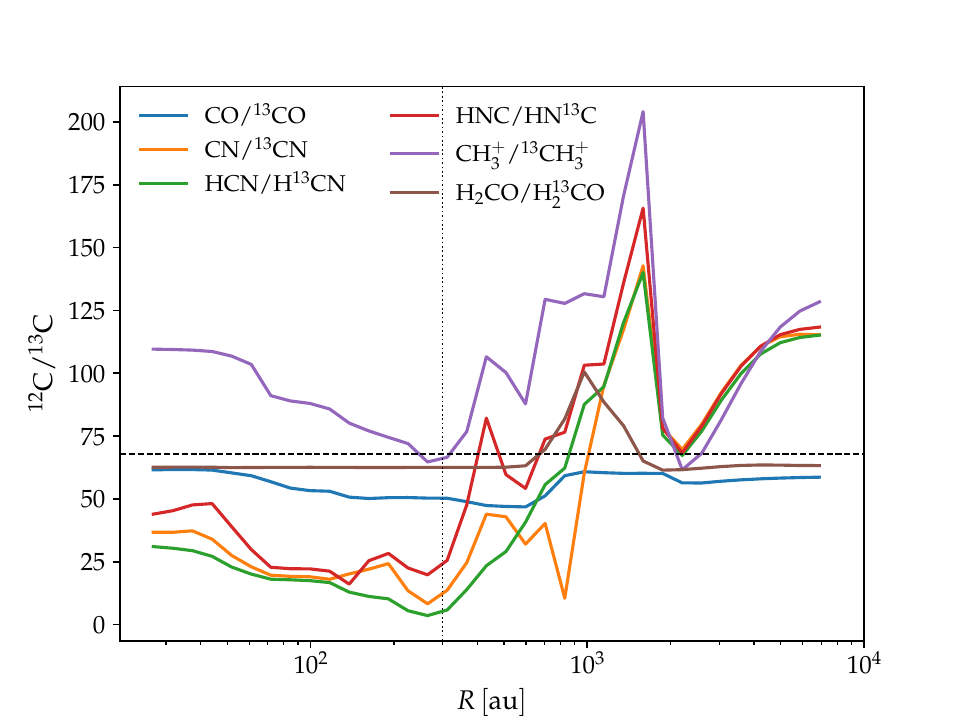}
    \caption{Model predictions for selected gas-phase \ratio ratios as a function of distance from the protostar taken at a simulation time of 10$^4$\,yr. The constant ratio of 68 is indicated by the horizontal dashed black line. The vertical dotted black line separates the spatial scales discussed in this paper ($R<300$\,au) from the larger scales, which are not largely discussed, but shown for completeness. }
    \label{fig:models}
\end{figure}

One intermediate product of the methanol formation in ices is formaldehyde (H$_2$CO), which is included in the model. Its gas-phase \ratio ratio is shown in Fig.\,\ref{fig:models} 
and is $\sim$62 at the scales traced by our observations (i.e. a few 100\,au). The slight difference between the CO and formaldehyde \ratio ratios likely results from competition between the formation of H$_2$CO in the solid phase from CO and the gas phase; 
an in-depth examination of the simulation results shows that the H$_2$CO \ratio ratio rises in the central core at later times (not shown) due to a chain of chemical reactions starting with desorbed $\rm CH_4$. 

\subsubsection{Methyl cyanide (precursors)}

The pathway(s) that dominate the formation of \metc are much less certain. Recently, the gas-phase chemical network of the COM has been revised by \cite{Giani2023}. They used quantum-mechanical calculations to test the relative importance of the various proposed formation routes of \metc. Their results suggest that the COM is dominantly formed from protonated methyl cyanide (CH$_3$CNH$^+$) that recombines with an electron or gives the extra proton to NH$_3$ upon reaction. Protonated \metc is in turn a product of the reactions CH$_3^+$ + HCN and CH$_3$OH$_2^+$ + HNC. Methyl cyanide can also form on dust-grain surfaces via CH$_3$ + CN and hydrogenation of CH$_2$CN \citep[e.g.][]{Garrod2022}.

In Fig.\,\ref{fig:models} we show \ratio ratios for CN, HCN, HNC, and CH$_3^+$. The first three follow a similar trend: the ratio is significantly below the local ISM value at close distances to the protostar and only increases to above the ISM value beyond 500--1000\,au. This is a result of exchange reactions such as Eqs.\,(\ref{eq2}--\ref{eq4}), whose relative importance depends on time and the prevalent physical conditions.  On the other hand, the simulated \ratio ratio of CH$_3^+$ is approximately 100 in the inner core, and therefore if the CH$_3^+$ + HCN formation pathway is dominant, then, we would expect \ct\metc and \metctr to show different fractionation levels. Only for IRAS\,4B, there is tentative evidence for a different ratio (factor 2, see Table\,\ref{tab:ratios}), however, due to the small number of transitions ($<$\,5) available for the rare \metc isotopologues, we cannot be certain.

Therefore, the CH$_3$OH$_2^+$ + HNC formation pathway thus seems more likely, but we cannot quantify this expectation with the present model. 
Similarly, the model predicts a high ($\sim$100) \ratio ratio for $\rm CH_3$ ice and a low ratio ($\sim$30) for CN ice at the end of the cloud stage \textcolor{red}{(Fig.\,\ref{fig:darkCloudEvolution})}, indicating a similar dichotomy between \ct\metc and \metctr if they were predominantly formed on grains via the association of $\rm CH_3$ and CN. An expanded chemical model, taking into account isotopic fractionation of COMs, is required to draw definite conclusions on the \ratio ratios, also keeping in mind that the physical model used in the present work is extremely simple. 

Our simulation predictions for the \ratio ratios for CN, HCN, and HNC are contrary to those presented recently by \citet[][]{Ichimura2024}. In their Fig.\,B5, the authors show that the ratios for these species and \metc remain elevated between 80--90 throughout the protostellar phase. During the prestellar phase, the \cd CN/\ct CN ratio behaves similarly in our and their models. It is not clear whether this discrepancy in the protostellar phase is due to the warm-up scenario considered in their model versus our static approach, or to differences in the chemical network. A detailed investigation is outside the scope of the present paper.  


\subsection{Observational limitations}\label{ss:caveats}

In Sect.\,\ref{ss:PD} we already mentioned that scatter between the observed and synthesised data points in the population diagrams (PDs) shown in Figs.\,\ref{pd:iras4b}--\ref{pd:18o}  
arises from the integration of intensities over at least three channels. Hence, by using the LR data (with a channel width of $\sim$2\,MHz) for the analysis, integrated intensities for lines that are spectrally unresolved (i.e. $FWHM\lesssim$\,channel width) will inevitably contain some contamination or extra noise. 
Contamination may come from unidentified species or unresolved additional velocity components from the same species (e.g. outflow emission, see Fig.\,\ref{fig:overview}b).
In addition, the optical depth of optically thick lines may be underestimated by the model due to the spectral dilution as the true line profiles remain unknown.
This fact together with line optical depths that generally exceed the capability of the Weeds model to correct for them ($\tau\gtrsim2$) are the main reasons for the discrepancy between transitions with low and high upper-level energies for sources with highest molecular column densities (e.g. for \met $E_u=600$\,K for B1-c, IRAS\,2A, and SVS13A).  

An overall underestimation of a molecule's column density may be a consequence of high continuum optical depth. If the continuum was optically thick, it would partially obscure the line emission \citep{deSimone2020}. 
Recent ALMA observations towards IRAS\,2A and B1-c \citep{Chen2024} at 0.9\,mm showed that the continuum is indeed optically thick towards the central position of IRAS\,2A, the effect is however negligible for B1-c. The PRODIGE data cover lower frequencies, hence, continuum opacities are generally lower. We cannot exclude a contribution of this effect to the uncertainties for some (i.e. the brightest) sources, but it does not apply to all. Observations at lower frequencies will help constrain the continuum optical depth at our observed frequencies.

Another source of uncertainty comes from the spatially unresolved emission. An overestimation of the size of the emitting area will lead to an underestimation of the column density. If the emission is optically thick or includes unresolved structures for one but not the other isotopologue, this can also affect the column density ratios. 
Besides, we do not resolve any morphological structures, such as binaries \citep[e.g. SVS13A, \citeauthor{Hsieh2025} \citeyear{Hsieh2025}, and IRAS\,2A, ][]{Jorgensen2022},  streamers \citep[e.g.][]{Valdivia2024}, or outflows \citep[e.g.][see also the HR spectrum in Fig.\,\ref{fig:overview}b, which shows outflow wing emission for IRAS\,4B]{Podio2021}. These may come with different velocity components that we are not able to spectrally resolve, either, but rather smooth over. 
Accordingly, spatial variations in the \ratio ratios can likely be expected. Therefore, we may only see an average ratio using the unresolved data observations, which may not reflect the true conditions at all. 

Lastly, although we are correcting for line opacities and focus the analysis on the optically thin(ner) lines, there is still a chance that we are underestimating the values, as they are model dependent. 
Ultimately, observations at higher angular and spectral resolution that simultaneously cover a broad bandwidth are needed to address these issues and put better constraints on the \ratio ratio.

\section{Conclusions}\label{s:conclusion}

We aimed to increase the number of measurements of \ratio ratios in the warm gas around young embedded protostellar systems using the data obtained as part of the PRODIGE large program. The \ct C isotopologues of \met and \metc were detected towards seven and six sources, respectively, out of in total 32 Class 0/I protostellar systems targeted by the PRODIGE survey. By using isotopologues of these complex organic molecules (COMs), we ensured that the molecular emission stems from the innermost hot gas surrounding the protostar(s). Additionally, we could test whether there is a segregation between O- and N-bearing molecules in terms of the \ratio ratio. 
Our main findings are the following:
\begin{enumerate}
    \item We derived excitation conditions and column densities using optically thin(ner) spectral lines assuming  LTE. \vspace{0.05cm}
    \item The resulting \ratio column density ratios for \met and \metc are in the range 4--30, which are lower than the expected ISM value of 68 by factors of a few up to an order of magnitude. 
    \item There is tentative evidence for an O/N segregation based on rotational temperatures, where \metc isotopologues trace warmer gas on average (290\,K) than \met (180\,K). This may hint at the dominant formation pathways of the two COMs on dust-grain surfaces during the prestellar phase (\met) or in the hot gas phase (\metc).
    \item There is no significant difference between the \ratio ratios from \met and \metc. This is in contrast to the ratios derived in previous works from COMs towards the Class 0 protostellar system IRAS16293-2422. In this source, some O-bearing COMs show \ratio ratios as low as ours, however, the values from \metc isotopologues are around the ISM value. The origin of differences between the various molecules among different star-forming regions is not yet understood and requires further observational and modelling efforts.
    \item Astrochemical models that we conducted for potential precursor molecules of CH$_3$OH and CH$_3$CN predict \ratio ratios for CO and H$_2$CO that are close to or slightly below the local ISM value of 68 (in ice and gas). The ratios do not reach as low as seen in the observational results. One reason may be the assumed duration of the prestellar phase in the model (10$^6$\,yr), but this requires further investigation. The gas-phase \ratio ratios for CH$_3$CN precursors, in particular CN, HCN, and HNC, are predicted to be in the range 20--50 at the spatial scales traced in the observations ($\sim$300\,au), more similar to the observational results.
\end{enumerate}

Our observational and modelling results indicate that there exists  \ct C enrichment at the Class 0 stage, traced by COMs, which likely originates from the prestellar stage. 
There are some observational limitations, which we evaluated in detail, that come with the spatially and (mostly) spectrally unresolved PRODIGE data set and that can addressed in future wideband observations at higher angular and spectral resolution.
Nevertheless, our work opens the avenue to use the \ratio ratio to study the question of chemical inheritance as protostellar and -planetary disks are likely to inherit at least some fraction of the parent cloud material.

In addition, our simulations predict that the \ratio ratios for simpler species, such as H\2CO, HCN, or HNC, greatly vary as a function of distance from the protostar (i.e. density and/or temperature) and depend on the duration of the preceding prestellar phase. An in-depth study of these isotopologues will shed light onto the dependence of the \ratio ratios on time and environment and may help to put constraints on the chemical network of more complex species, for example, \metc.

\begin{acknowledgements}
The authors thank the IRAM staff at the NOEMA observatory for their support in the observations and data calibration. This work is based on observations carried out under project number L19MB with the IRAM NOEMA Interferometer. IRAM is supported by INSU/CNRS (France), MPG (Germany) and IGN (Spain). L.A.B, J.E.P., O.S., P.C., M.J.M, C.G., M.T.V.M, Th.H., D.S., and Y.C. acknowledge the support by the Max Planck Society.
A.F. and J.J.M.P. acknowledge support from ERC grant SUL4LIFE, GA No. 101096293. Funded by the European Union. Views and opinions expressed are however those of the author(s) only and do not necessarily reflect those of the European Union or the European Research Council Executive Agency. Neither the European Union nor the granting authority can be held responsible for them. A.F. also thanks project PID2022-137980NB-I00 funded by the Spanish Ministry of Science and Innovation/State Agency of Research MCIN/AEI/10.13039/501100011033 and by “ERDF A way of making Europe”.  Th.~H. and D.~S. acknowledge support from the European Research Council under the
Horizon 2020 Framework Program via the ERC Advanced Grant Origins 83 24 28
(PI: Th. Henning). L.C. acknowledges support from the grant No. PID2022-136814NB-I00 by the Spanish Ministry of Science, Innovation and Universities/State Agency of Research MICIU/AEI/10.13039/501100011033 and by ERDF, UE.
\end{acknowledgements}

\bibliographystyle{aa} 
\bibliography{refs} 

\onecolumn
\begin{appendix}
\section{Additional tables}

Table\,\ref{tab:obs} lists beam sizes and rms for the seven sources of the PRODIGE sample studied in this work. Tables\,\ref{tab:IRAS2A}--\ref{tab:L1448C} contain the parameters to derive the LTE radiative transfer models for each analysed molecule as well as the results from the population diagram analysis. The \ratio ratios for all additional sources shown in Figs.\,\ref{fig:ratio_woP} and \ref{fig:ratios_all} are listed in Table\,\ref{tab:refs} together with their respective reference and the species that were used to derive them.

\begin{table}[ht]
\caption{Information on the PRODIGE observations.} 
\centering
\begin{tabular}{llcrcc}
\hline\hline \\[-0.3cm] 
Source & sb\tablefootmark{a} & \multicolumn{2}{c}{Synthesised beam} & \multicolumn{2}{c}{$rms$} \\ \cmidrule(lr){3-4} \cmidrule(lr){5-6}
& & $\theta_\mathrm{maj}\times\theta_\mathrm{min}$ & $PA$ & & \\ 
& & $(\,^{\prime\prime}\,\times\,^{\prime\prime}\,)$ & $(^\circ)$ & $\left(\frac{\mathrm{mJy}}{\mathrm{beam}^{-1}}\right)$ & (mK) \\[0.2cm] \hline \\[-0.3cm]
     
IRAS\,2A & li & $1.0\times1.3$ & 21.0 & 2.1 & 40.4 \\
         & lo & $1.0\times1.3$ & 22.2 & 2.2 & 42.8 \\
         & ui & $1.0\times1.2$ & 21.6 & 2.2 & 42.4 \\
         & uo & $0.9\times1.2$ & 22.1 & 2.5 & 48.6 \\
B1-c & li & $1.1\times1.3$ & 27.8 & 2.1 & 35.7 \\
           & lo & $1.1\times1.4$ &     29.3 & 2.2 & 38.1 \\
           & ui & $1.0\times1.3$ &     26.9 & 2.1 & 36.9 \\
           & uo & $1.0\times1.3$ &    27.5 & 2.5 & 43.6 \\
B1-bS & li & $1.0\times1.3$ &     18.6 & 1.9 & 34.9 \\
      & lo & $1.1\times1.3$ &     17.9 & 2.0 & 37.3 \\
      & ui & $1.0\times1.2$ &   $-$7.6 & 2.1 & 37.6 \\
      & uo & $0.9\times1.2$ & 19.1 & 2.4 & 45.9 \\
L1448-IRS2 & li & $0.9\times1.5$ & 10.1 & 2.0 & 35.7 \\
           & lo & $0.9\times1.6$ & 11.6 & 2.0 & 34.6 \\
           & ui & $0.8\times1.5$ & 9.4 & 2.4 & 42.6 \\
           & uo & $0.8\times1.5$ & 10.8 & 2.4 & 43.1 \\
SVS13A & li & $0.7\times1.2$ &     22.5 & 2.4 & 66.9 \\
       & lo & $0.7\times1.3$ & 22.8 & 2.4 & 69.0 \\
       & ui & $0.7\times1.1$ &     22.8 & 2.6 & 76.1 \\
       & uo & $0.7\times1.1$ &     22.9 & 2.9 & 83.3 \\
L1448C & li & $0.9\times1.2$ & 14.8 & 2.3 & 52.2 \\
       & lo & $0.9\times1.3$ & 13.9 & 2.4 & 55.9 \\
       & ui & $0.8\times1.2$ & 12.8 & 2.4 & 55.0 \\ 
       & uo & $0.8\times1.2$ & 13.9 & 2.8 & 65.6 \\
IRAS\,4B & li & $0.9\times1.2$ & 27.7 & 1.8 & 43.3 \\
         & lo & $0.9\times1.2$ & 29.0 & 2.0 & 48.3 \\
         & ui & $0.8\times1.1$ & 27.1 & 2.0 & 47.4 \\
         & uo & $0.8\times1.1$ & 28.1 & 2.3 & 56.7 \\

\hline\hline
\end{tabular}
\tablefoot{\tablefoottext{a}{Lower-inner (li, 218.8--222.8\,GHz), Lower-outer (lo, 214.7--218.8\,GHz), upper-inner (ui, 230.2--234.2\,GHz), and upper-outer (uo, 234.2--238.3\,GHz) sideband of the PolYFiX correlator.}}
\label{tab:obs}
\end{table}
\begin{table*}[ht]
\caption{Weeds parameters and results of the population-diagram analysis for IRAS\,2A.} 
\centering
\begin{tabular}{lcclcccccc}
\hline\hline \\[-0.3cm] 
Molecule\tablefootmark{a} & Size\tablefootmark{b} & $T_{\rm rot,W}$\tablefootmark{c} & $T_{\rm rot,PD}$\tablefootmark{d} & $N_{\rm tot,W}\tablefootmark{e}$ & $N_{\rm tot,PD}\tablefootmark{f}$ & $\Delta\varv$\tablefootmark{g} & $\varv_{\rm off}$\tablefootmark{h} \\ 
 & $(^{\prime\prime})$ & (K) & (K) & (cm$^{-2}$) & (cm$^{-2}$) & (km\,s$^{-1}$) & (km\,s$^{-1}$) \\\hline \\[-0.3cm] 
CH$_3$OH, $v=0$, $v=1$, $v=2$ & $0.2$ & $180$ & $181\pm 8$ & $1.4(19)$ & $(1.9\pm 0.5)(19)$ & $3.6$ & $-0.8$ \\[0.05cm] 
 $^{13}$CH$_3$OH, $v=0$, $v=1$ & $0.2$ & $140$ & $143\pm 8$ & $6.5(17)$ & $(6.9\pm 1.2)(17)$ & $3.6$ & $-0.8$ \\[0.05cm] 
 CH$_3$CN, $v=0$, $v_8=1$ & $0.2$ & $240$ & $242\pm 23$ & $6.0(16)$ & $(8.3\pm 2.8)(16)$ & $3.6$ & $-0.8$ \\[0.05cm] 
 $^{13}$CH$_3$CN, $v=0$ & $0.2$ & $240$ & $240$ & $2.0(15)$ & $(2.4\pm 0.2)(15)$ & $3.6$ & $-0.8$ \\[0.05cm] 
 CH$_3^{13}$CN, $v=0$ & $0.2$ & $240$ & $244\pm 74$ & $2.0(15)$ & $(2.9\pm 1.6)(15)$ & $3.6$ & $-0.8$ \\[0.05cm] 
 CH$_3^{18}$OH, $v=0$ & $0.2$ & $140$ & $135\pm 20$ & $7.0(16)$ & $(9.1\pm 2.8)(16)$ & $3.6$ & $-0.8$ \\[0.05cm] 
 \hline\hline
\end{tabular}
\tablefoot{\tablefoottext{a}{COMs and the vibrational states detected and used to derive population diagrams. }\tablefoottext{b}{Assumed size of the emitting region.}\tablefoottext{c}{Rotational temperature used for the Weeds model.} \tablefoottext{d}{Rotational temperature derived from the population diagram. When a value has no error, it was fixed.}\tablefoottext{e}{Column density used for the Weeds model.} \tablefoottext{f}{Column density derived from the population diagram.}\tablefoottext{g}{$FWHM$ of the spectral lines.}\tablefoottext{h}{Offset from the source systemic velocity (see Table\,\ref{tab:Sources}).}\\ Values in parentheses show the decimal power, where $x(z) = x\times 10^z$ or $(x\pm y)(z) = (x\pm y)\times 10^z$. An upper limit on $N_\mathrm{tot,W}$ indicates that the molecule is not detected. Upper limits were derived from Weeds models, where all parameters, but column density, were fixed and the column density was chosen such that no transition exceeds a 3$\sigma$ threshold, where $\sigma$ is the rms noise level from Table\,\ref{tab:obs}. }
\label{tab:IRAS2A}
\end{table*}

\begin{table*}[ht]
\caption{Same as Table\,\ref{tab:IRAS2A}, but for IRAS4B.} 
\centering
\begin{tabular}{lcclcccccc}
\hline\hline \\[-0.3cm] 
Molecule & Size & $T_{\rm rot,W}$ & $T_{\rm rot,PD}$ & $N_{\rm tot,W}$ & $N_{\rm tot,PD}$ & $\Delta\varv$ & $\varv_{\rm off}$ \\ 
 & $(^{\prime\prime})$ & (K) & (K) & (cm$^{-2}$) & (cm$^{-2}$) & (km\,s$^{-1}$) & (km\,s$^{-1}$) \\\hline \\[-0.3cm] 
CH$_3$OH, $v=0$, $v=1$, $v=2$ & $0.4$ & $280$ & $293\pm 18$ & $2.3(17)$ & $(2.9\pm 0.5)(17)$ & $1.2$ & $-1.0$ \\[0.05cm] 
 $^{13}$CH$_3$OH, $v=0$, $v=1$ & $0.4$ & $250$ & $255\pm 22$ & $6.0(16)$ & $(7.5\pm 1.4)(16)$ & $1.2$ & $-1.0$ \\[0.05cm] 
 CH$_3$CN, $v=0$, $v_8=1$ & $0.4$ & $300$ & $294\pm 22$ & $2.4(15)$ & $(3.9\pm 0.9)(15)$ & $1.2$ & $-1.0$ \\[0.05cm] 
 $^{13}$CH$_3$CN, $v=0$ & $0.4$ & $210$ & $210$ & $1.0(14)$ & $(1.2\pm 0.1)(14)$ & $1.2$ & $-1.0$ \\[0.05cm] 
 CH$_3^{13}$CN, $v=0$ & $0.4$ & $210$ & $211\pm 2$ & $2.2(14)$ & $(3.1\pm 0.1)(14)$ & $1.2$ & $-1.0$ \\[0.05cm] 
 CH$_3^{18}$OH, $v=0$ & $0.4$ & $160$ & $155\pm 26$ & $1.0(16)$ & $(1.4\pm 0.5)(16)$ & $1.2$ & $-1.0$ \\[0.05cm] 
 \hline\hline
\end{tabular}
\label{tab:IRAS4B}
\end{table*}

\begin{table*}[ht]
\caption{Same as Table\,\ref{tab:IRAS2A}, but for SVS13A.} 
\centering
\begin{tabular}{lcclcccccc}
\hline\hline \\[-0.3cm] 
Molecule & Size & $T_{\rm rot,W}$ & $T_{\rm rot,PD}$ & $N_{\rm tot,W}$ & $N_{\rm tot,PD}$ & $\Delta\varv$ & $\varv_{\rm off}$ \\ 
 & $(^{\prime\prime})$ & (K) & (K) & (cm$^{-2}$) & (cm$^{-2}$) & (km\,s$^{-1}$) & (km\,s$^{-1}$) \\\hline \\[-0.3cm] 
CH$_3$OH, $v=0$, $v=1$, $v=2$ & $0.3$ & $180$ & $179\pm 8$ & $1.2(19)$ & $(1.4\pm 0.3)(19)$ & $3.6$ & $0.2$ \\[0.05cm] 
 $^{13}$CH$_3$OH, $v=0$, $v=1$ & $0.3$ & $160$ & $163\pm 5$ & $4.0(17)$ & $(5.1\pm 0.5)(17)$ & $3.6$ & $0.2$ \\[0.05cm] 
 CH$_3$CN, $v=0$, $v_8=1$ & $0.3$ & $320$ & $330\pm 18$ & $4.6(16)$ & $(5.0\pm 0.9)(16)$ & $3.6$ & $0.2$ \\[0.05cm] 
 $^{13}$CH$_3$CN, $v=0$ & $0.3$ & $290$ & $287\pm 151$ & $2.0(15)$ & $(2.1\pm 1.8)(15)$ & $3.6$ & $0.2$ \\[0.05cm] 
 CH$_3^{13}$CN, $v=0$ & $0.3$ & $290$ & $288\pm 139$ & $2.0(15)$ & $(2.1\pm 1.8)(15)$ & $3.6$ & $0.2$ \\[0.05cm] 
 CH$_3^{18}$OH, $v=0$ & $0.3$ & $180$ & $167\pm 51$ & $8.2(16)$ & $(8.4\pm 4.9)(16)$ & $3.6$ & $0.2$ \\[0.05cm] 
 \hline\hline
\end{tabular}
\label{tab:SVS13A}
\end{table*}

\begin{table*}[ht]
\caption{Same as Table\,\ref{tab:IRAS2A}, but for L1448-IRS2.} 
\centering
\begin{tabular}{lcclcccccc}
\hline\hline \\[-0.3cm] 
Molecule & Size & $T_{\rm rot,W}$ & $T_{\rm rot,PD}$ & $N_{\rm tot,W}$ & $N_{\rm tot,PD}$ & $\Delta\varv$ & $\varv_{\rm off}$ \\ 
 & $(^{\prime\prime})$ & (K) & (K) & (cm$^{-2}$) & (cm$^{-2}$) & (km\,s$^{-1}$) & (km\,s$^{-1}$) \\\hline \\[-0.3cm] 
CH$_3$OH, $v=0$, $v=1$ & $0.5$ & $190$ & $186\pm 14$ & $3.0(16)$ & $(3.5\pm 0.8)(16)$ & $3.7$ & $-0.45$ \\[0.05cm] 
 $^{13}$CH$_3$OH, $v=0$ & $0.5$ & $190$ & $190$ & $2.5(15)$ & $(3.0\pm 0.3)(15)$ & $3.7$ & $-0.45$ \\[0.05cm] 
 CH$_3$CN, $v=0$ & $0.5$ & $220$ & $222\pm 43$ & $3.0(14)$ & $(3.2\pm 1.4)(14)$ & $3.7$ & $-0.45$ \\[0.05cm] 
 $^{13}$CH$_3$CN, $v=0$ & $0.5$ & $220$ &  & $<4.5(13)$ &  & $3.7$ & $-0.45$ \\[0.05cm] 
 CH$_3^{13}$CN, $v=0$ & $0.5$ & $220$ &  & $<4.5(13)$ &  & $3.7$ & $-0.45$ \\[0.05cm] 
 CH$_3^{18}$OH, $v=0$ & $0.5$ & $190$ &  & $<2.2(15)$ &  & $3.7$ & $-0.45$ \\[0.05cm] 
 \hline\hline
\end{tabular}
\label{tab:Per-emb-22}
\end{table*}

\begin{table*}[ht]
\caption{Same as Table\,\ref{tab:IRAS2A}, but for B1-c.} 
\centering
\begin{tabular}{lcclcccccc}
\hline\hline \\[-0.3cm] 
Molecule & Size & $T_{\rm rot,W}$ & $T_{\rm rot,PD}$ & $N_{\rm tot,W}$ & $N_{\rm tot,PD}$ & $\Delta\varv$ & $\varv_{\rm off}$ \\ 
 & $(^{\prime\prime})$ & (K) & (K) & (cm$^{-2}$) & (cm$^{-2}$) & (km\,s$^{-1}$) & (km\,s$^{-1}$) \\\hline \\[-0.3cm] 
CH$_3$OH, $v=0$, $v=1$, $v=2$ & $0.2$ & $170$ & $195\pm 24$ & $4.5(18)$ & $(4.3\pm 2.5)(18)$ & $3.0$ & $-0.35$ \\[0.05cm] 
 $^{13}$CH$_3$OH, $v=0$, $v=1$ & $0.2$ & $240$ & $237\pm 22$ & $1.9(17)$ & $(2.2\pm 0.5)(17)$ & $3.0$ & $-0.35$ \\[0.05cm] 
 CH$_3$CN, $v=0$, $v_8=1$ & $0.2$ & $420$ & $458\pm 55$ & $3.0(16)$ & $(3.7\pm 1.3)(16)$ & $3.0$ & $-0.35$ \\[0.05cm] 
 $^{13}$CH$_3$CN, $v=0$ & $0.2$ & $350$ & $350$ & $2.2(15)$ & $(2.1\pm 0.2)(15)$ & $3.0$ & $-0.35$ \\[0.05cm] 
 CH$_3^{13}$CN, $v=0$ & $0.2$ & $350$ & $332\pm 105$ & $2.5(15)$ & $(3.0\pm 1.5)(15)$ & $3.0$ & $-0.35$ \\[0.05cm] 
 CH$_3^{18}$OH, $v=0$ & $0.2$ & $150$ & $141\pm 59$ & $3.5(16)$ & $(3.6\pm 2.9)(16)$ & $3.0$ & $-0.35$ \\[0.05cm] 
 \hline\hline
\end{tabular}
\label{tab:Per-emb-29}
\end{table*}

\begin{table*}[ht]
\caption{Same as Table\,\ref{tab:IRAS2A}, but for B1-bS.} 
\centering
\begin{tabular}{lcclcccccc}
\hline\hline \\[-0.3cm] 
Molecule & Size & $T_{\rm rot,W}$ & $T_{\rm rot,PD}$ & $N_{\rm tot,W}$ & $N_{\rm tot,PD}$ & $\Delta\varv$ & $\varv_{\rm off}$ \\ 
 & $(^{\prime\prime})$ & (K) & (K) & (cm$^{-2}$) & (cm$^{-2}$) & (km\,s$^{-1}$) & (km\,s$^{-1}$) \\\hline \\[-0.3cm] 
CH$_3$OH, $v=0$, $v=1$, $v=2$ & $0.4$ & $400$ & $404\pm 51$ & $6.0(16)$ & $(7.4\pm 2.2)(16)$ & $1.5$ & $-0.25$ \\[0.05cm] 
 $^{13}$CH$_3$OH, $v=0$, $v=1$ & $0.4$ & $200$ & $196\pm 80$ & $7.0(15)$ & $(7.8\pm 6.2)(15)$ & $1.5$ & $-0.25$ \\[0.05cm] 
 CH$_3$CN, $v=0$, $v_8=1$ & $0.4$ & $400$ & $399\pm 46$ & $1.2(15)$ & $(1.7\pm 0.6)(15)$ & $1.5$ & $-0.25$ \\[0.05cm] 
 $^{13}$CH$_3$CN, $v=0$ & $0.4$ & $400$ & $400$ & $2.0(14)$ & $(2.7\pm 0.4)(14)$ & $1.5$ & $-0.25$ \\[0.05cm] 
 CH$_3^{13}$CN, $v=0$ & $0.4$ & $400$ & $433\pm 68$ & $3.0(14)$ & $(3.5\pm 0.8)(14)$ & $1.5$ & $-0.25$ \\[0.05cm] 
 CH$_3^{18}$OH, $v=0$ & $0.4$ & $200$ &  & $<3.0(15)$ &  & $1.5$ & $-0.25$ \\[0.05cm] 
 \hline\hline
\end{tabular}
\label{tab:B1-bS}
\end{table*}

\begin{table*}[ht]
\caption{Same as Table\,\ref{tab:IRAS2A}, but for L1448C.} 
\centering
\begin{tabular}{lcclcccccc}
\hline\hline \\[-0.3cm] 
Molecule & Size & $T_{\rm rot,W}$ & $T_{\rm rot,PD}$ & $N_{\rm tot,W}$ & $N_{\rm tot,PD}$ & $\Delta\varv$ & $\varv_{\rm off}$ \\ 
 & $(^{\prime\prime})$ & (K) & (K) & (cm$^{-2}$) & (cm$^{-2}$) & (km\,s$^{-1}$) & (km\,s$^{-1}$) \\\hline \\[-0.3cm] 
CH$_3$OH, $v=0$, $v=1$ & $0.5$ & $150$ & $148\pm 5$ & $1.0(17)$ & $(1.3\pm 0.2)(17)$ & $1.6$ & $-1.1$ \\[0.05cm] 
 $^{13}$CH$_3$OH, $v=0$ & $0.5$ & $150$ & $145\pm 10$ & $6.0(15)$ & $(1.0\pm 0.1)(16)$ & $1.6$ & $-1.1$ \\[0.05cm] 
 CH$_3$CN, $v=0$, $v_8=1$ & $0.5$ & $170$ & $172\pm 8$ & $1.3(15)$ & $(2.0\pm 0.3)(15)$ & $1.6$ & $-1.1$ \\[0.05cm] 
 $^{13}$CH$_3$CN, $v=0$ & $0.5$ & $180$ & $180$ & $7.0(13)$ & $(7.7\pm 1.0)(13)$ & $1.6$ & $-1.1$ \\[0.05cm] 
 CH$_3^{13}$CN, $v=0$ & $0.5$ & $180$ & $180$ & $9.0(13)$ & $(1.2\pm 0.4)(14)$ & $1.6$ & $-1.1$ \\[0.05cm] 
 CH$_3^{18}$OH, $v=0$ & $0.5$ & $150$ &  & $<1.7(15)$ &  & $1.6$ & $-1.1$ \\[0.05cm] 
 \hline\hline
\end{tabular}
\label{tab:L1448C}
\end{table*}

\begin{table}[ht]
\caption{References for the \ratio isotopologue ratios used in Figs.\,\ref{fig:ratio_woP} and \ref{fig:ratios_all}.} \vspace{-0.2cm}
\centering
\small
\begin{tabular}{llcll}
\hline\hline \\[-0.3cm] 
Source & Species & \ratio & Instrument & References \\[.05cm]\hline\hline \\[-0.3cm] 
CB23 & HCN & 36\,$\pm$\,10 & IRAM 30\,m & \citet{Jensen2024} \\\hline\\[-0.3cm]
TMC2 & HCN & 63\,$\pm$\,15 & IRAM 30\,m & \citet{Jensen2024} \\\hline\\[-0.3cm]
L1512 & HCN & 20\,$\pm$\,7 & IRAM 30\,m & \citet{Jensen2024} \\\hline\\[-0.3cm]
L1495 & HCN & 25\,$\pm$\,10 & IRAM 30\,m & \citet{Jensen2024} \\\hline\\[-0.3cm]
L1517B & HCN & 26\,$\pm$\,8 & IRAM 30\,m & \citet{Jensen2024} \\\hline\\[-0.3cm]
Barnard\,1 & HCN & 30$^{+7}_{-4}$ & \multirow{3}{*}{IRAM 30\,m} & \multirow{3}{*}{\citet{Daniel2013}} \\[0.1cm]
& HNC & 20$^{+5}_{-4}$ & \\[0.1cm]
& CN & 50$^{+19}_{-11}$ & \\[0.1cm]
\hline\\[-.3cm]
L1498 & HCN & 45\,$\pm$\,3 & IRAM 30\,m & \citet{Magalhaes2018} \\\hline\\[-.3cm]
L483 & CO & 42\,$\pm$\,13 &  \multirow{4}{*}{IRAM 30\,m} & \multirow{4}{*}{\citet{Agundez2019}} \\
& HCN & 34\,$\pm$\,10 & \\
& HNC & 29\,$\pm$\,9 & \\[0.1cm]
& HC$_3$N$^a$ & $\sim$91$^{+12}_{-2}$ & \\[0.1cm]\hline\\[-.3cm]
L1527 & CO & 70\,$\pm$\,20 & \multirow{2}{*}{NRO 45\,m} & \multirow{2}{*}{\citet{Yoshida2019}} \\
& CN & 61\,$\pm$\,17 & \\\hline\\[-.3cm]
IRAS16293A & \met & 65\,$\pm$\,27 & \multirow{5}{*}{ALMA} & \multirow{5}{*}{\citet{Manigand2020}} \\
& CH$_3$OCHO & 75\,$\pm$\,32 & \\
& CH$_3$OCH$_3$ & 86\,$\pm$\,18 & \\
& H$_2$CO & 93\,$\pm$\,39 & \\
& $t$-HCOOH & 39\,$\pm$\,7 & \\ \cmidrule(lr){2-5}
& \metc$^b$ & $\sim$74$^{+25}_{-21}$ & ALMA & \citet{Calcutt2018} \\[0.1cm]\hline\\[-.3cm]
IRAS19293B & C$_2$H$_5$OH & $>$\,25 & \multirow{6}{*}{ALMA} & \multirow{6}{*}{\citet{Jorgensen2018}}\\
& CH$_3$OCHO & $>$\,41 & \\
& CH$_3$OCH$_3$ & 34 & \\
&  CH$_3$CHO & 67 & \\
& CH$_2$CO & 68$^c$ & \\
& $t$-HCOOH & 68$^c$ & \\[-0.1cm]\cmidrule(lr){2-5}\vspace{-0.05cm}
& CH$_2$OHCHO & 27 & ALMA & \citet{Jorgensen2016} \\\cmidrule(lr){2-5}
& \metc$^b$ & $\sim$74$^{+26}_{-24}$ & ALMA & \citet{Calcutt2018} \\[0.1cm]\hline\\[-.3cm]
TW Hya$^d$ & CO & 40$^{+9}_{-6}$ & ALMA & \cite{Zhang2017} \\ \cmidrule(lr){2-5}
& & 21\,$\pm$\,5 & \multirow{2}{*}{ALMA} & \multirow{2}{*}{\citet{Yoshida2022}} \\
& & $>$\,84 &  &  \\\cmidrule(lr){2-5}
& HCN & 86\,$\pm$\,4 & ALMA & \citet{Hily-Blant2019} \\\cmidrule(lr){2-5}
& CN & 70$^{+8}_{-6}$ & ALMA & \cite{Yoshida2024} \\[0.1cm]\hline\\[-.3cm]
PDS 70 & HCN (50\,au) & 48$^{+25}_{-17}$ & \multirow{3}{*}{ALMA} & \multirow{3}{*}{\citet{Rampinelli2025}$^f$} \\[0.1cm]
& HCN (75\,au) & 64$^{+31}_{-19}$ \\[0.1cm]
& HCN (100\,au) & 109$^{+100}_{-54}$ \\[0.1cm]\hline\\[-.3cm]
V883 Ori\tablefootmark{*} & CH$_3$OCHO & 23\,$\pm$\,3 & \multirow{3}{*}{ALMA} & \multirow{3}{*}{\citet{Yamato2024}} \\[0.1cm]
& CH$_3$OCH$_3$ & 25$^{+10}_{-6}$ & \\[0.1cm]
&  CH$_3$CHO & 29$^{+16}_{-8}$ & \\[0.05cm]\cmidrule(lr){2-5}
& \met & $\sim$\,24 & \multirow{3}{*}{ALMA} & \multirow{3}{*}{\citet{Jeong2025}$^e$} \\
& $c-$C$_2$H$_4$O & $\sim$\,10 & & \\
& CH$_3$CN & $\sim$\,16 & & \\\hline\\[-.3cm]
DM\,TAU & CO & 20\,$\pm$\,3 & PdBI & \citet{Pietu2007} \\[0.1cm]\hline\\[-.3cm]
LkCa\,15 & CO & 11\,$\pm$\,2 & PdBI & \citet{Pietu2007} \\[0.1cm]\hline\\[-.3cm]
MWC 480 & CO & 8\,$\pm$\,2 & PdBI & \citet{Pietu2007} \\[0.1cm]\hline\\[-.3cm]
TYC\,8998b & CO & 31$^{+17}_{-10}$ & VLT/CRIRES+ & \citet{Zhang2021} \\[0.1cm]\hline\\[-.3cm]
WASP-77Ab & CO & 26\,$\pm$\,16 & GS Observatory & \citet{Line2021} \\\hline\\[-.3cm]
VHS\,1256b & CO & 62\,$\pm$\,2 & JWST & \citet{Gandhi2023} \\
 \hline\hline
\end{tabular}\vspace{-0.2cm}
\tablefoot{\tablefoottext{a,b}{Median (a) and mean (b) value of the ratios derived from all \ct C isotoplogues of the respective molecule. The error bars indicate the range of possible values including uncertainties.}\tablefoottext{c}{The value is not directly measured but estimated from the analysis.} \tablefoottext{d}{The ratios were derived at different distances from the protostar in the disk \citep[see also][]{Bergin2024}.} \tablefoottext{e}{Estimated ratios read off their Fig.\,11. The value for \metc is assumed for both \ct C isotoplogues.}\tablefoottext{f}{Estimated ratios and uncertainties read off their Fig.\,7. Ratios were derived at different radii in the disk (50, 75, 100\,au).} \tablefoottext{*}{A classification of V883 Ori is difficult as it undergoes an outburst. It is likely in a stage between Class I and II \citep{Jeong2025}.}}
\label{tab:refs}
\end{table}
\clearpage
\section{Additional figures: Integrated intensity maps}

\begin{figure}[h]
    \centering
    \includegraphics[width=\textwidth]{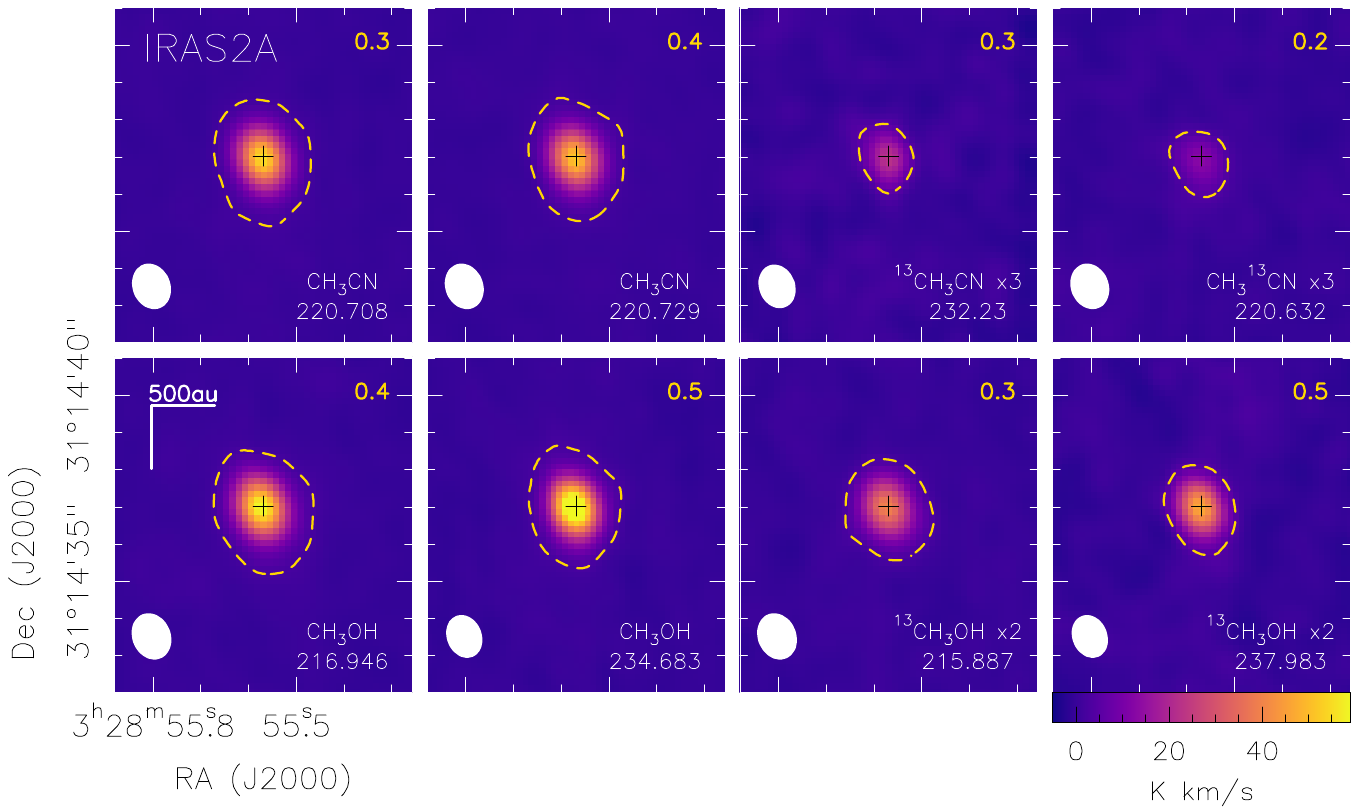}
    \caption{Integrated intensity maps (in K \kms) using two transitions for both isotopologues of \met, the main \metc isotopologue, and one transition for each \ct C isotopologue towards IRAS\,2A. Respective rest frequencies in GHz are shown in the bottom right corner, the beam (half power beam width, HPBW) on the bottom left. Multiplication factors (e.g. $\times$3, next to the species name) were applied to some species and transitions to be able to use the same intensity scale for all panels. The yellow dashed contour is at 5$\sigma$, where $\sigma$ is the rms noise level (in K\,\kms) and is written in the top right corner.}
    \label{maps:iras2a}
\end{figure}

\begin{figure}[h]
    \centering
    \includegraphics[width=.96\textwidth]{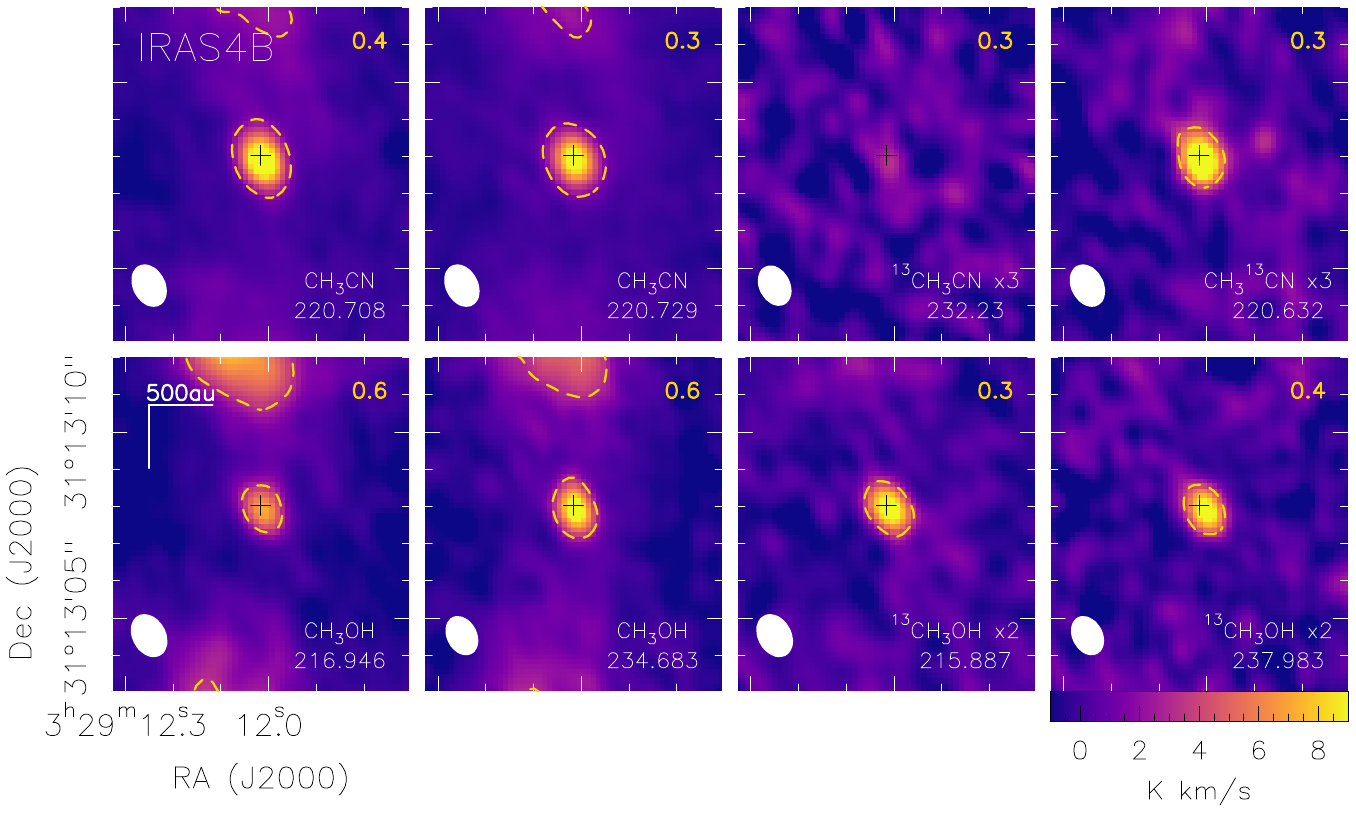}
    \caption{Same as Fig.\,\ref{maps:iras2a}, but for IRAS\,4B.}
    \label{maps:iras4b}
\end{figure}

\begin{figure}[h]
    \centering
    \includegraphics[width=.98\textwidth]{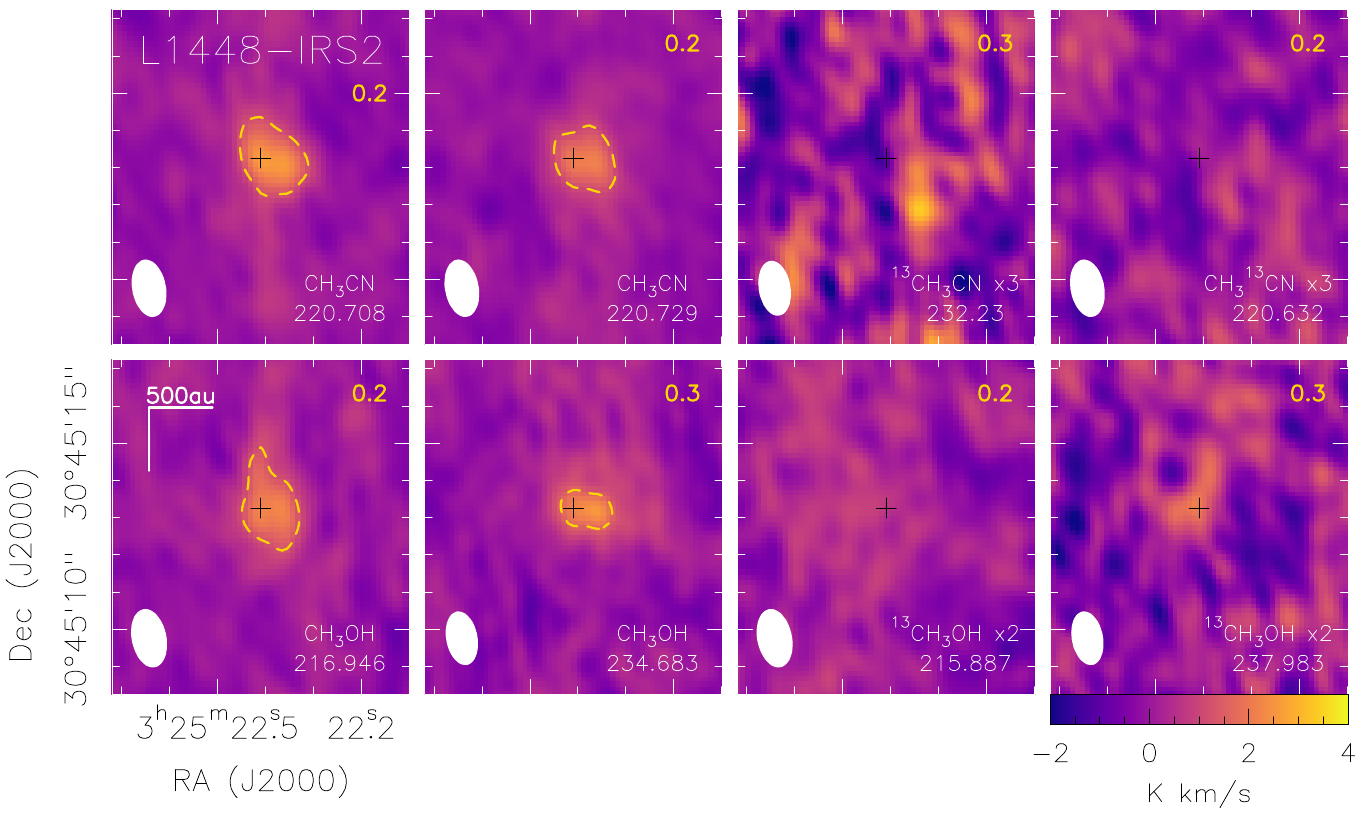}
    \caption{Same as Fig.\,\ref{maps:iras2a}, but for Per-emb-22.}
    \label{maps:peremb22}
\end{figure}

\begin{figure}[h]
    \centering
    \includegraphics[width=.96\textwidth]{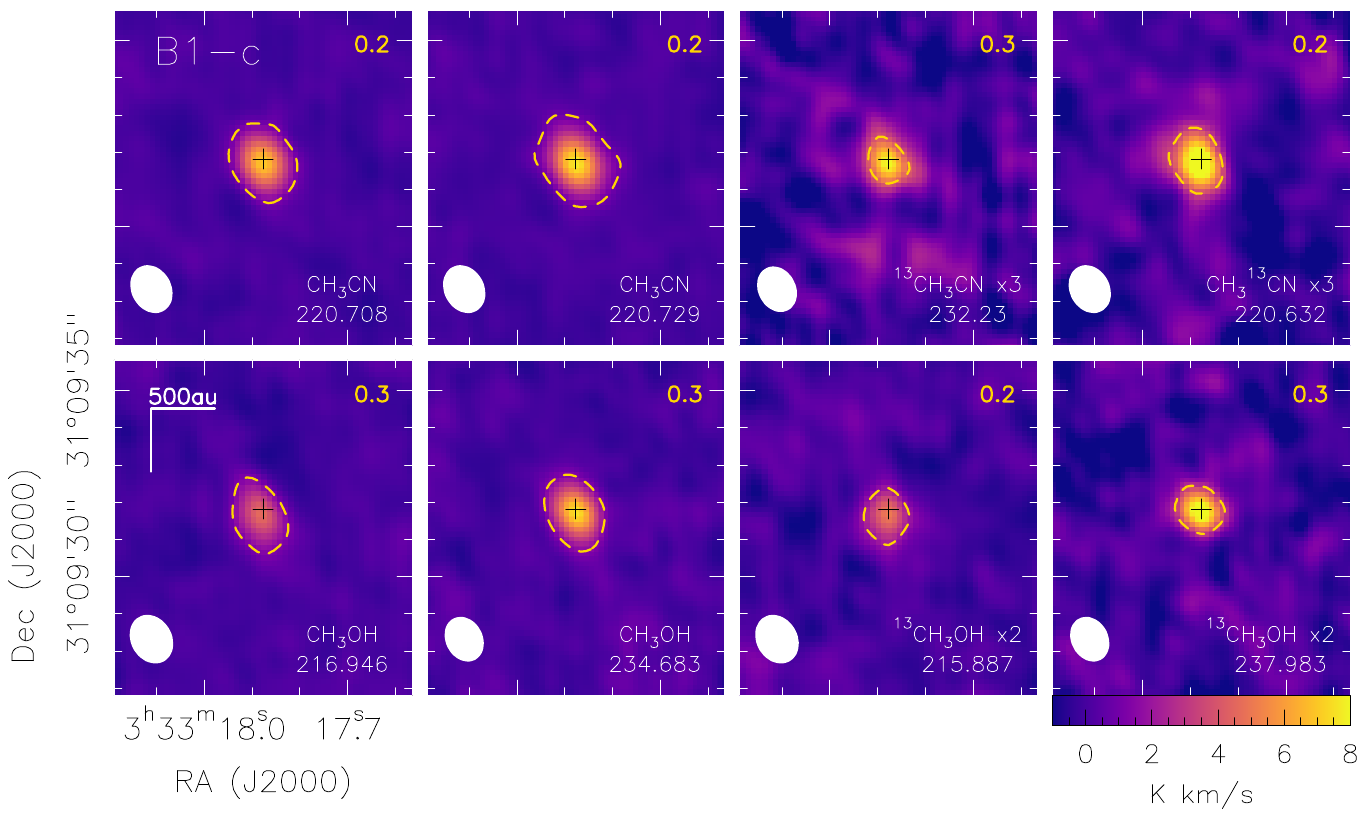}
    \caption{Same as Fig.\,\ref{maps:iras2a}, but for Per-emb-29.}
    \label{maps:peremb29}
\end{figure}

\begin{figure}[h]
    \centering
    \includegraphics[width=.98\textwidth]{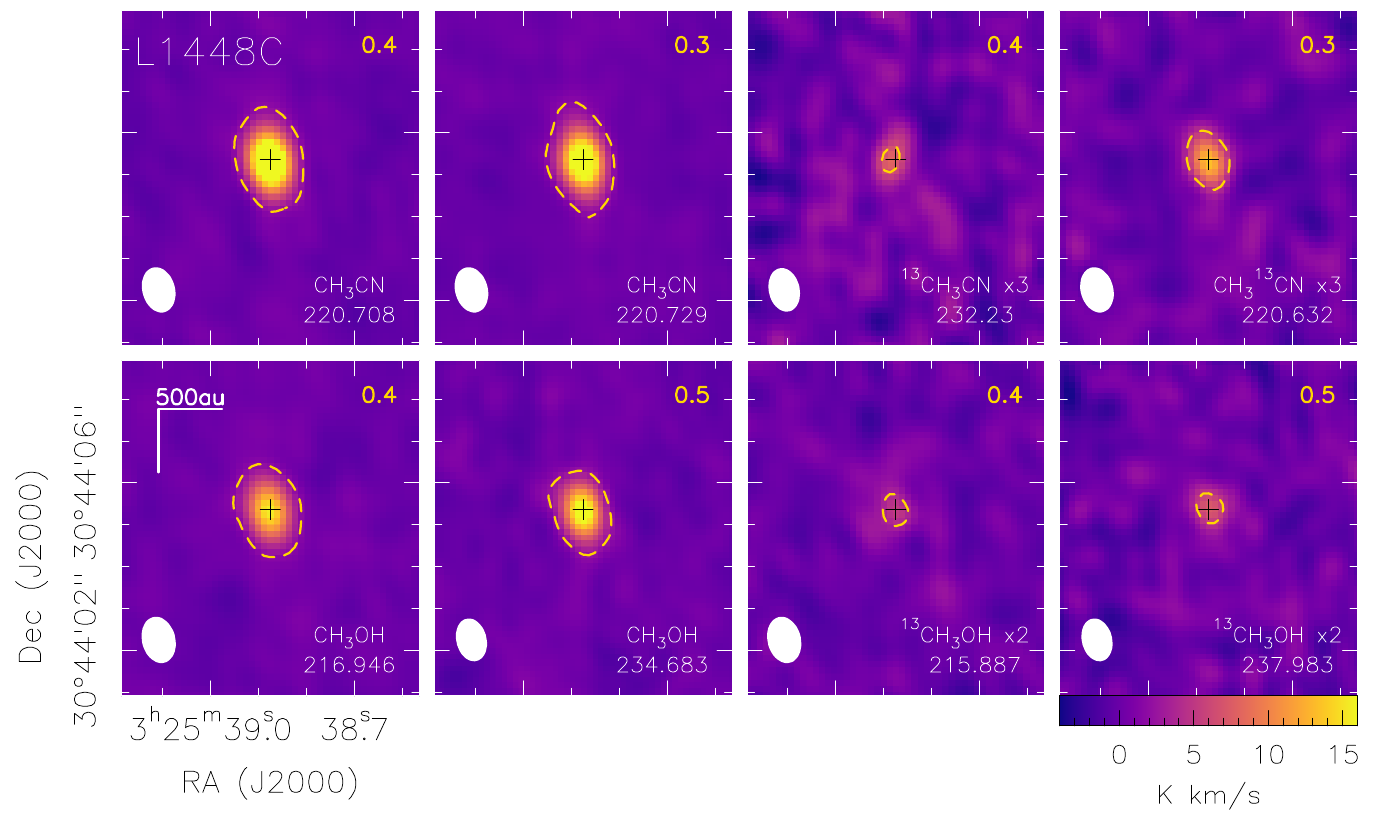}
    \caption{Same as Fig.\,\ref{maps:iras2a}, but for L1448C.}
    \label{maps:l1448c}
\end{figure}

\begin{figure}[h]
    \centering
    \includegraphics[width=.98\textwidth]{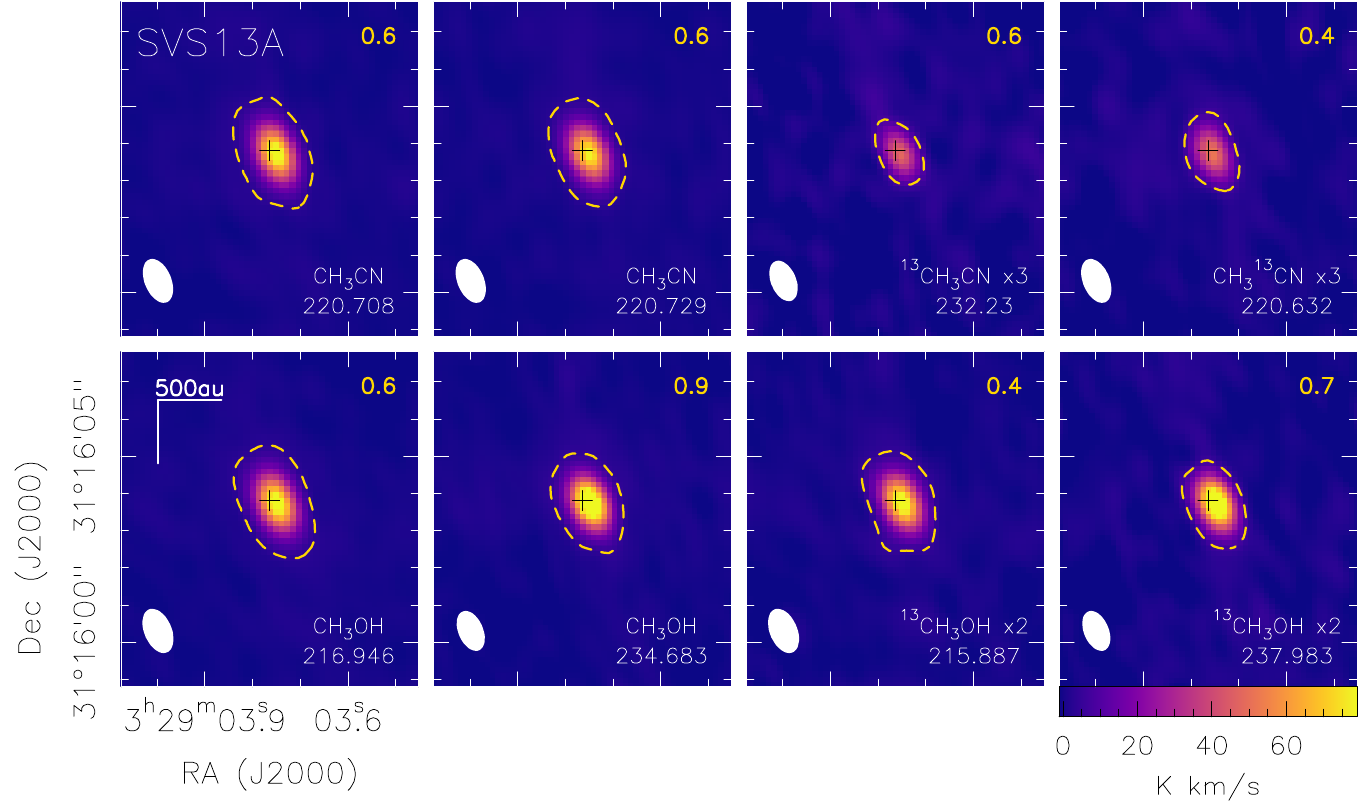}
    \caption{Same as Fig.\,\ref{maps:iras2a}, but for SVS13A.}
    \label{maps:svs13a}
\end{figure}

\begin{figure}[h]
    \centering
    \includegraphics[width=\textwidth]{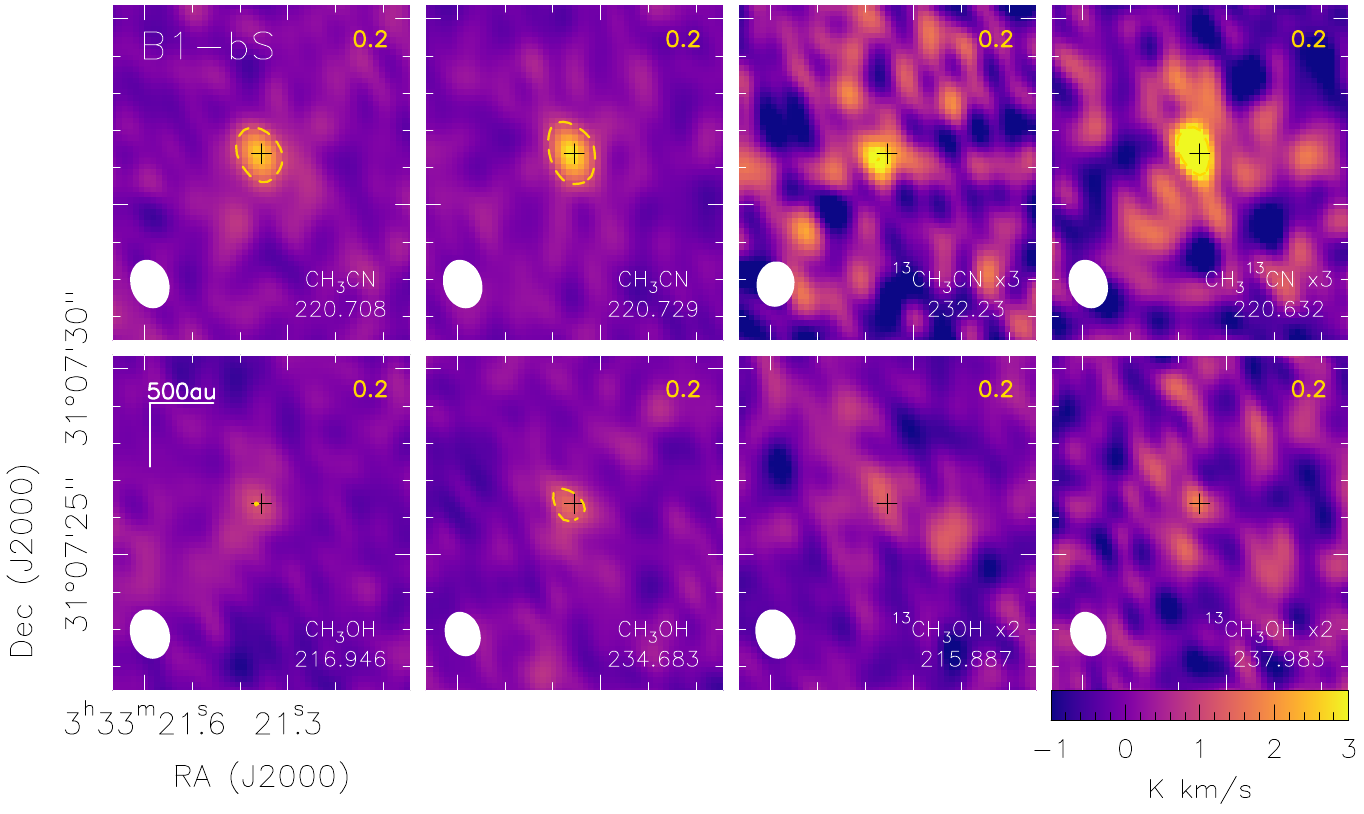}
    \caption{Same as Fig.\,\ref{maps:iras2a}, but for B1-bS.}
    \label{maps:b1bs}
\end{figure}

Figures\,\ref{maps:iras2a}--\ref{maps:b1bs} show integrated intensity maps for selected transitions of  the main and rarer isotopologues towards the seven sources of the PRODIGE sample that were analysed in this work.  The transitions were selected because they are bright and unblended.


\clearpage
\section{Additional figures: Population diagrams}

\begin{figure}[ht]
    \centering
    \includegraphics[width=0.48\textwidth]{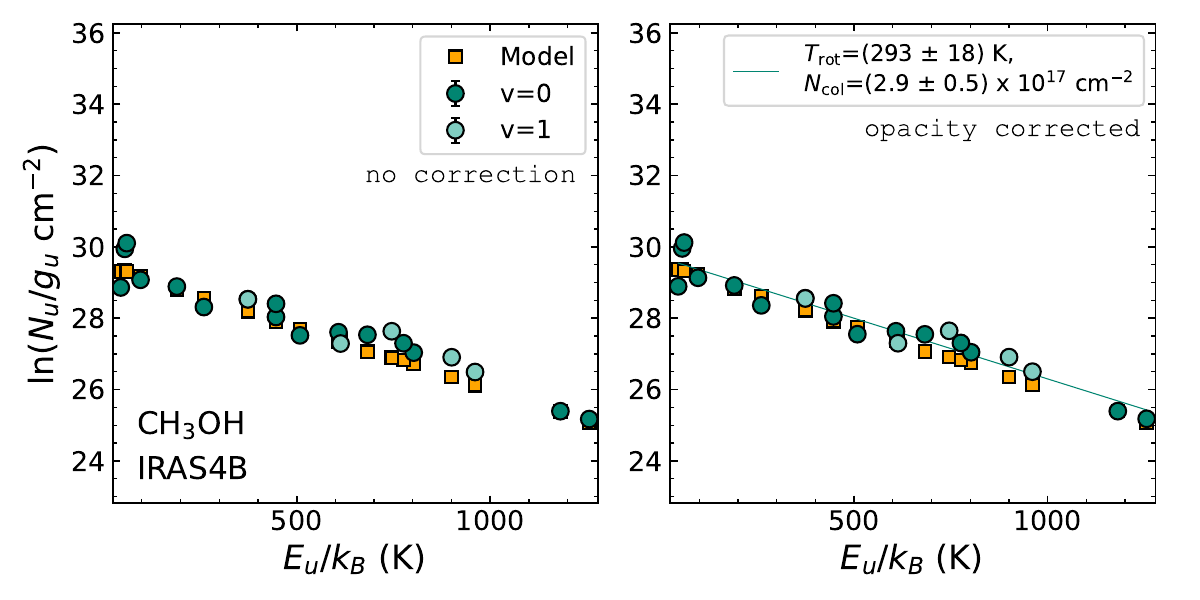}
    \includegraphics[width=0.49\textwidth]{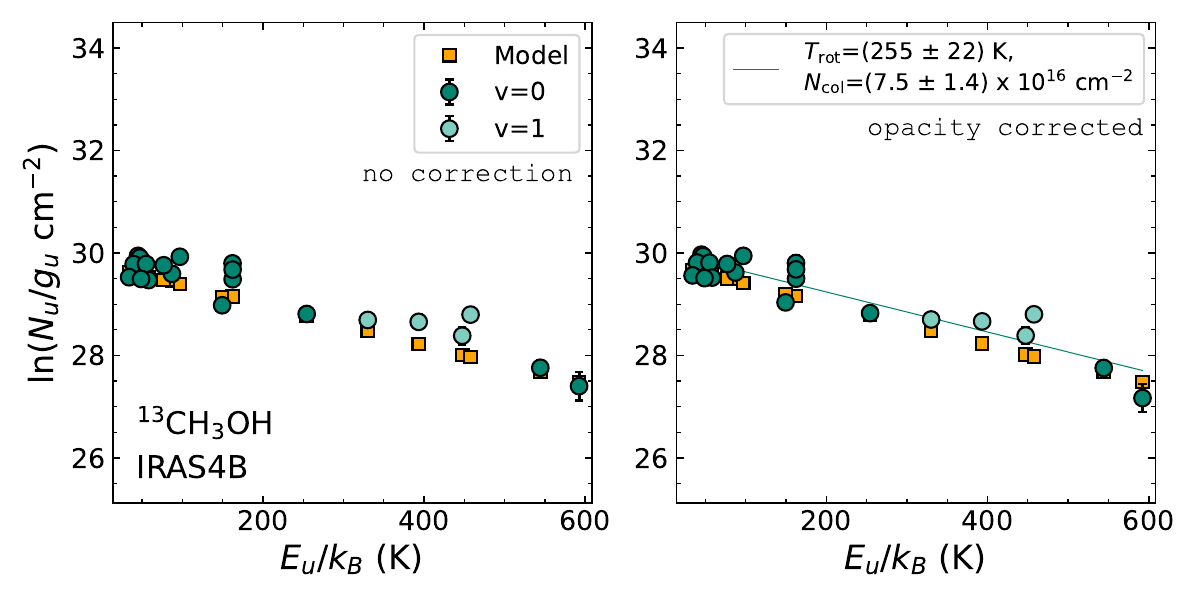}
    \includegraphics[width=0.49\textwidth]{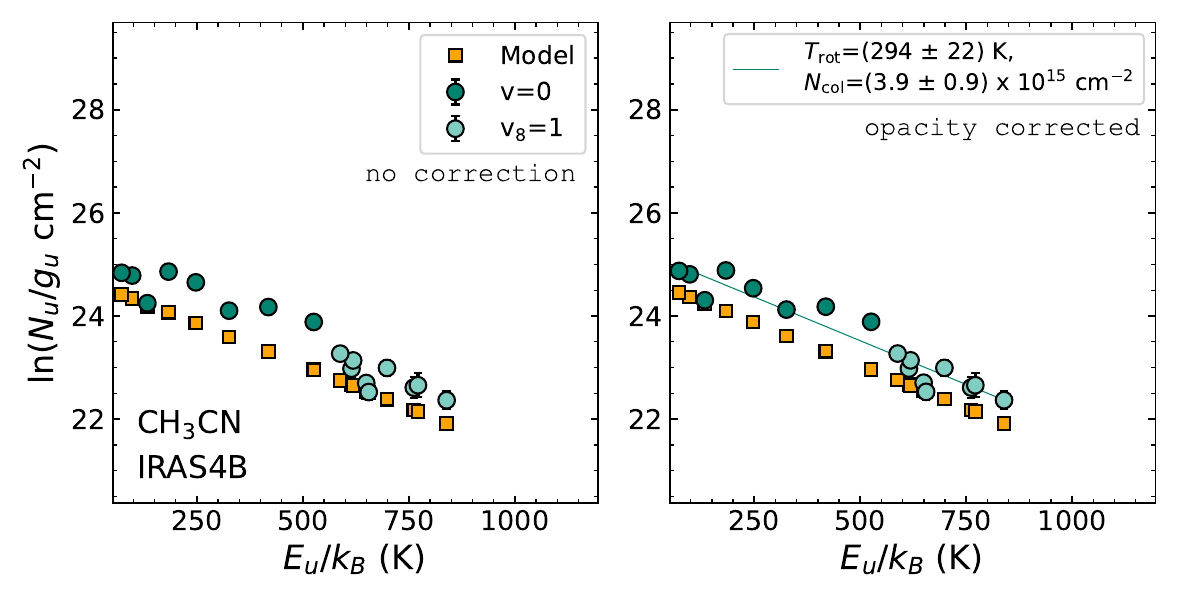}
    \includegraphics[width=0.49\textwidth]{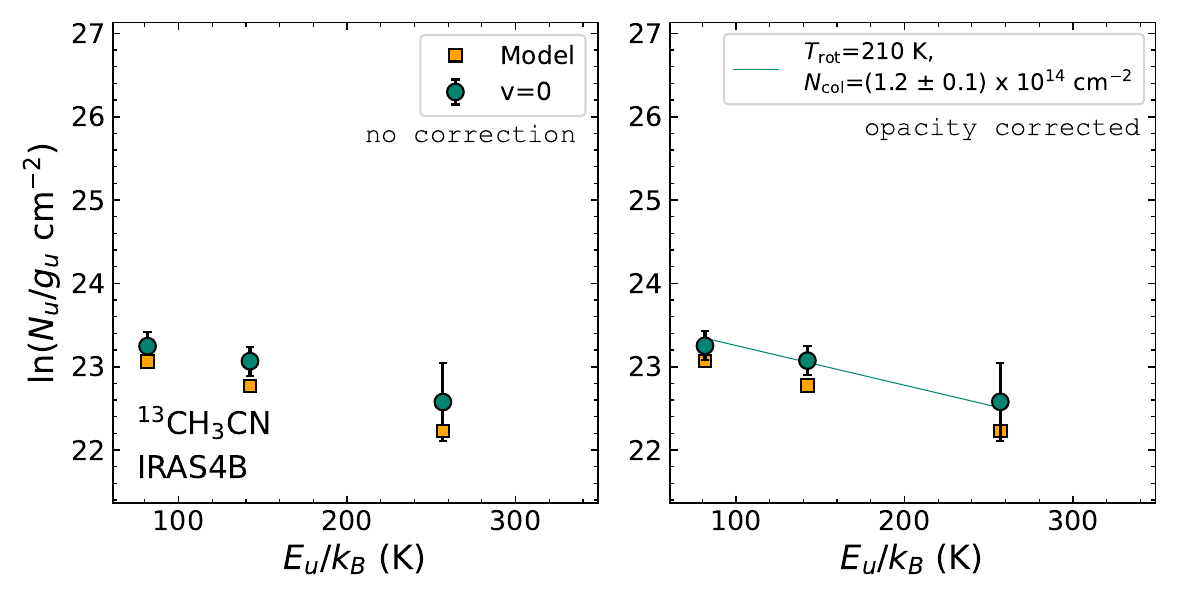}
    \includegraphics[width=0.49\textwidth]{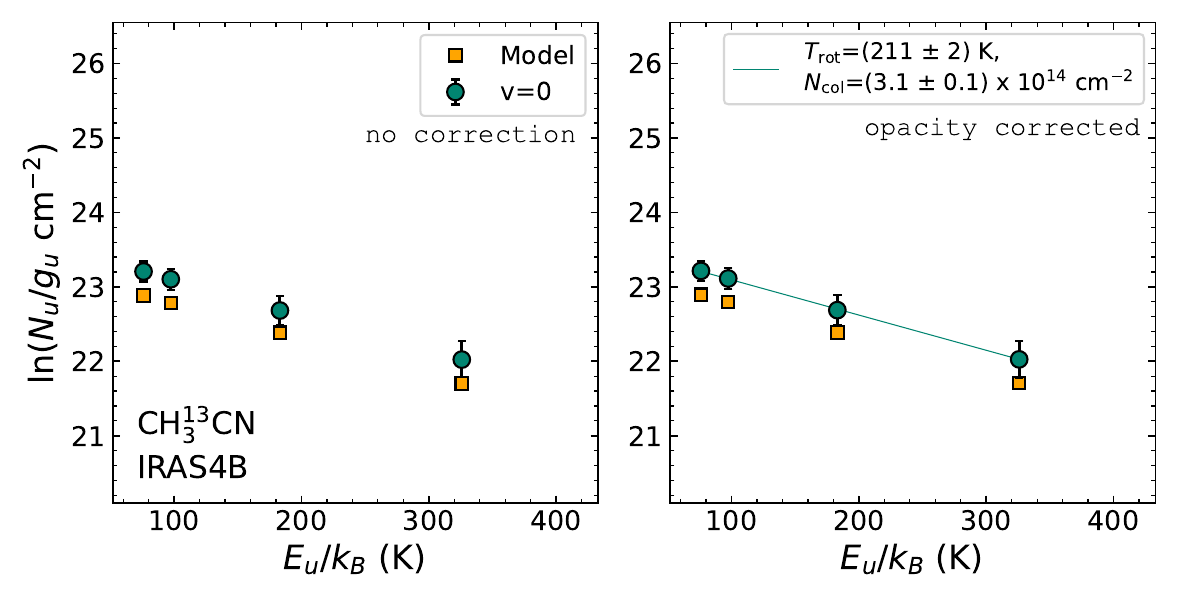}
    \caption{Population diagrams for \met, \ct\met, \metc, \ct\metc, and \metctr towards IRAS\,4B. Observed data points are shown with teal circles as indicated in the top right corner of the left panel while the modelled data points from Weeds are shown orange squares. No corrections are applied in the left panel while in the right panel corrections for opacity and contamination by other molecules have been considered for the observed and modelled populations. 
    The results of the linear fit to the observed data points are shown in the right panels. }
    \label{pd:iras4b}
\end{figure}

\begin{figure}[ht]
    \centering
    \includegraphics[width=0.48\textwidth]{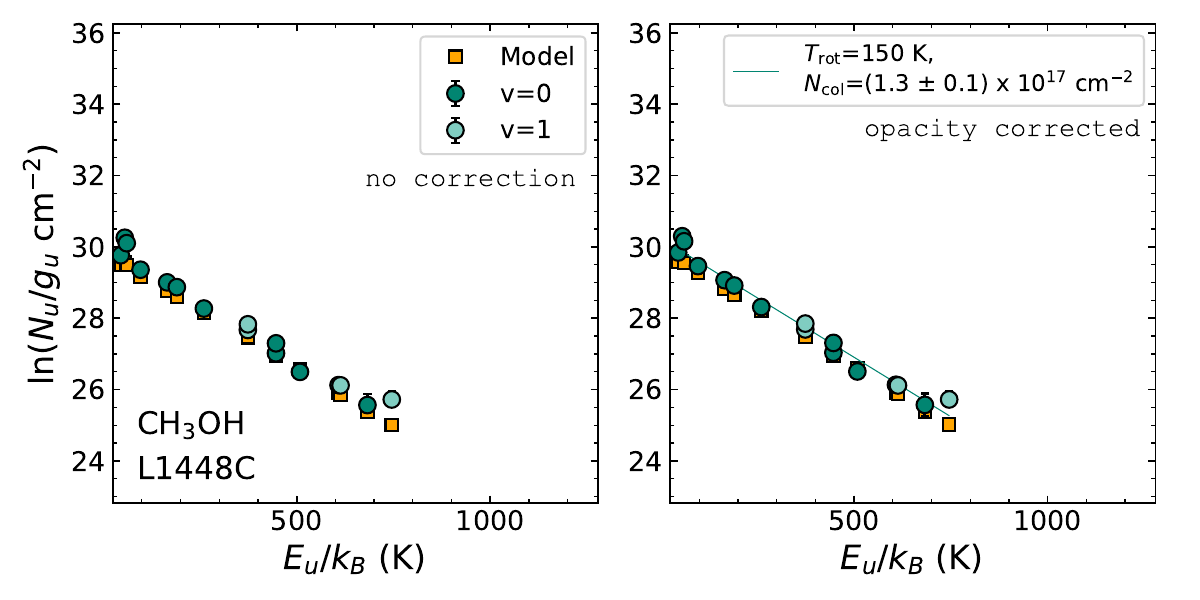}
    \includegraphics[width=0.49\textwidth]{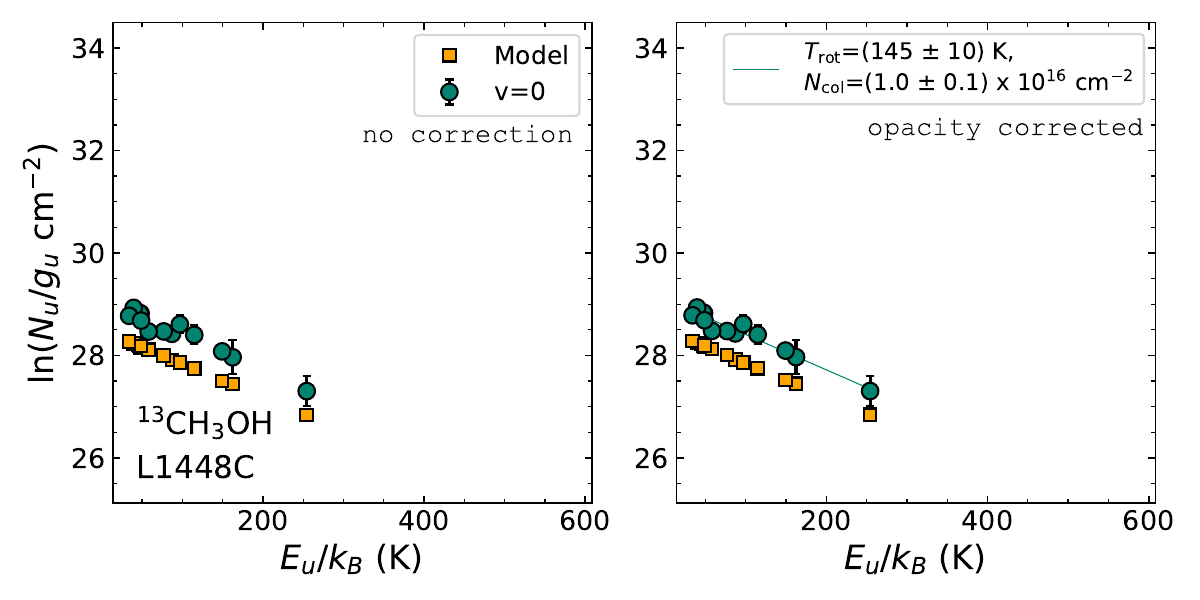}
    \includegraphics[width=0.49\textwidth]{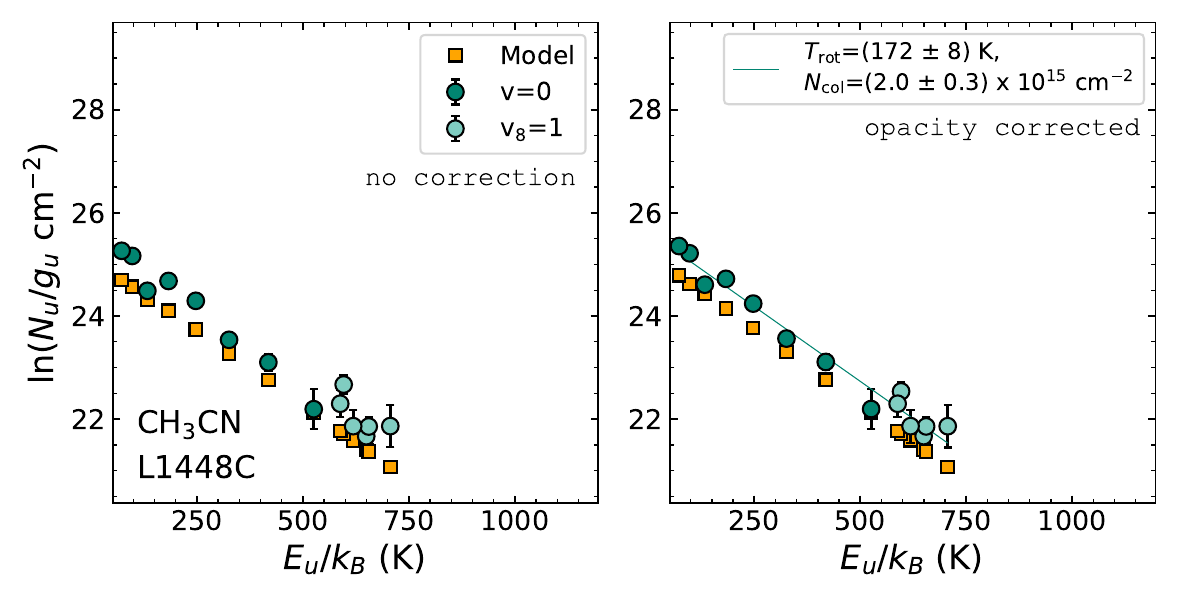}
    \includegraphics[width=0.49\textwidth]{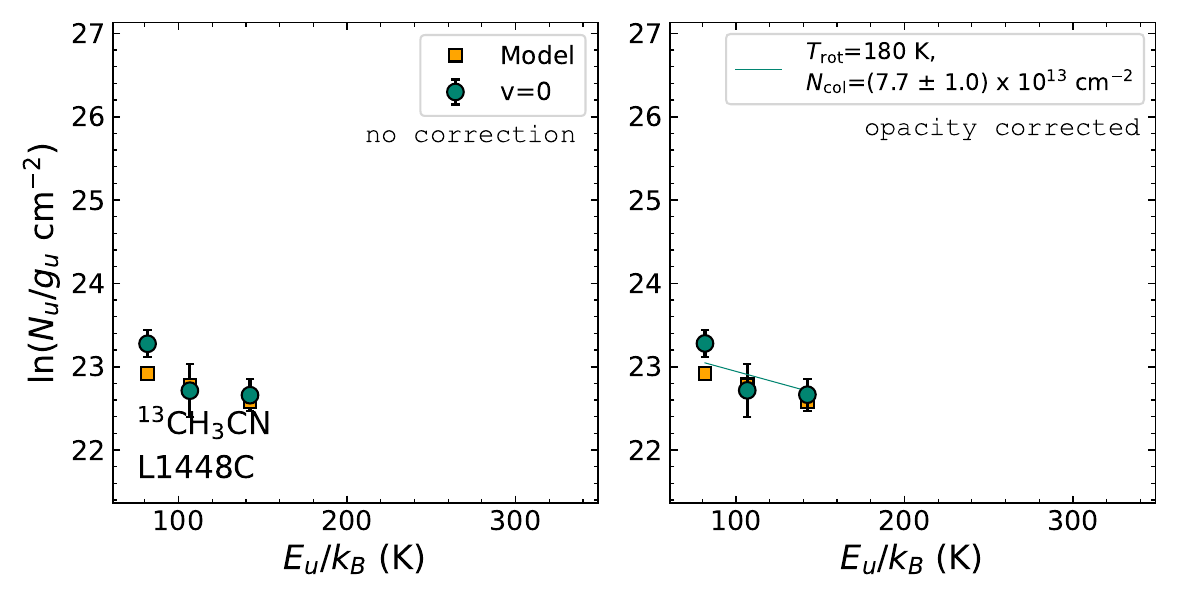}
    \includegraphics[width=0.49\textwidth]{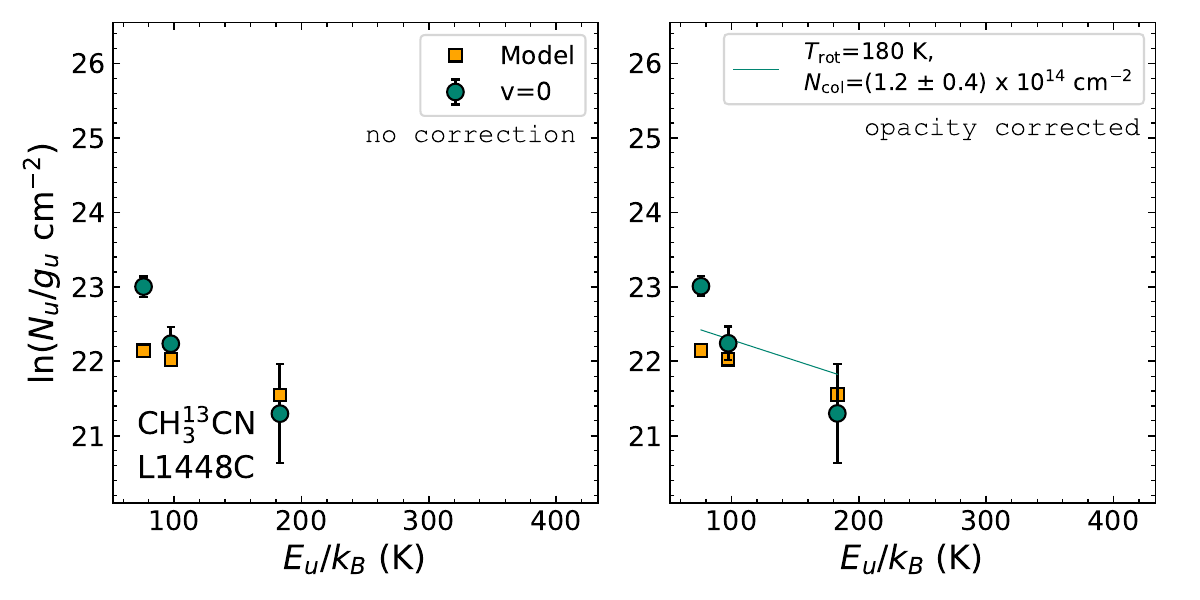}
    \caption{Same as \ref{pd:iras4b}, but for L1448C. }
    \label{pd:l1448c}
\end{figure}

\begin{figure}[ht]
    \includegraphics[width=0.48\textwidth]{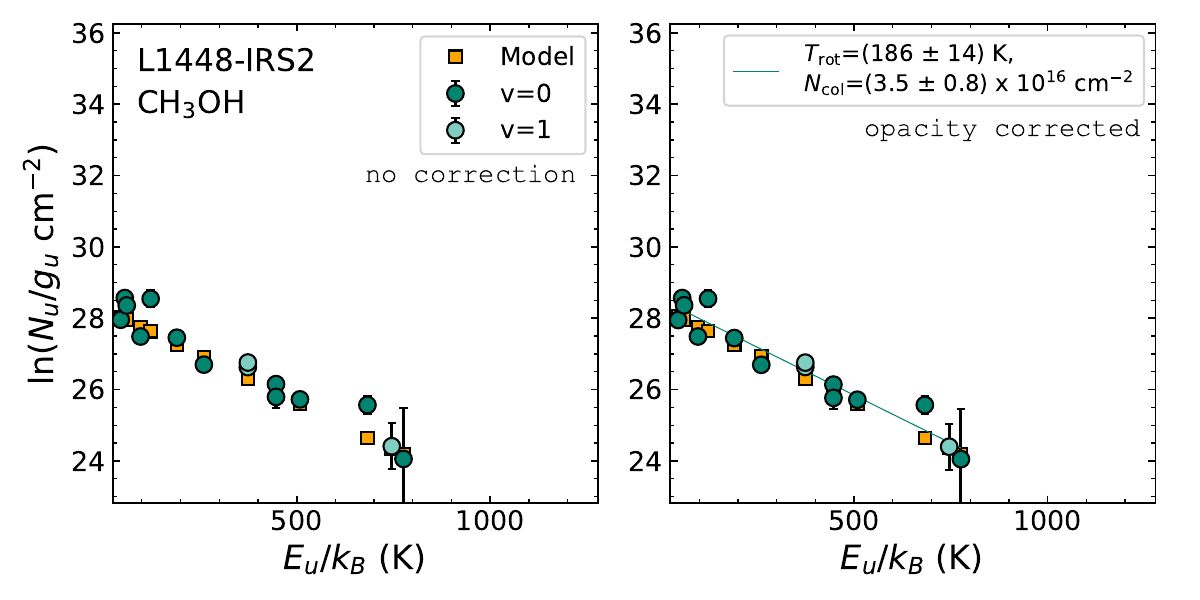}
    \includegraphics[width=0.49\textwidth]{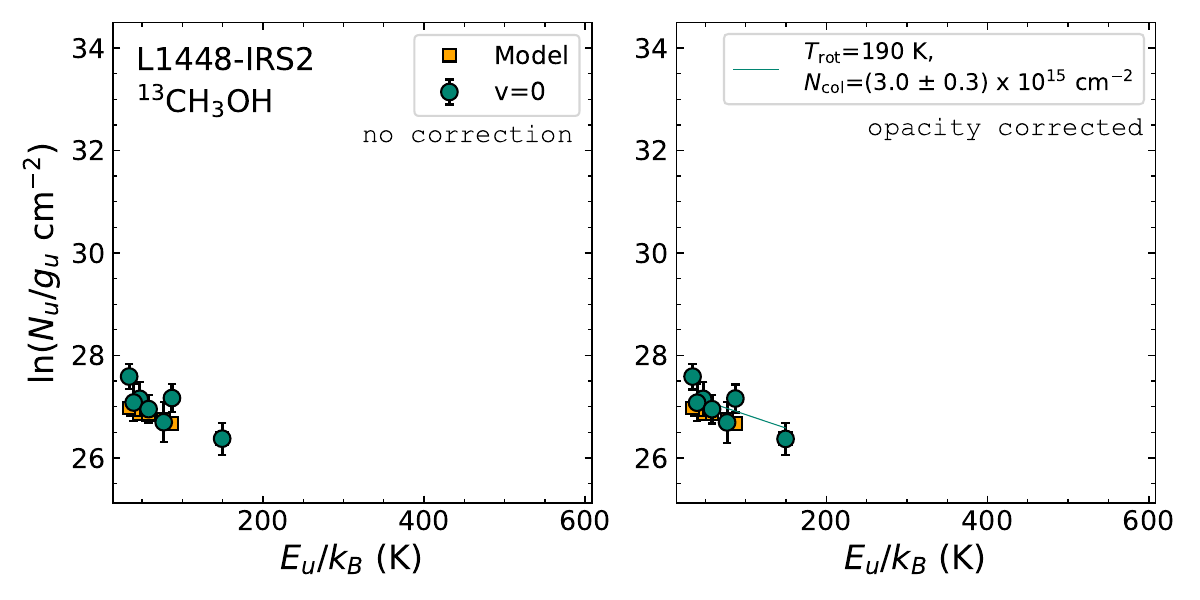}
    \includegraphics[width=0.49\textwidth]{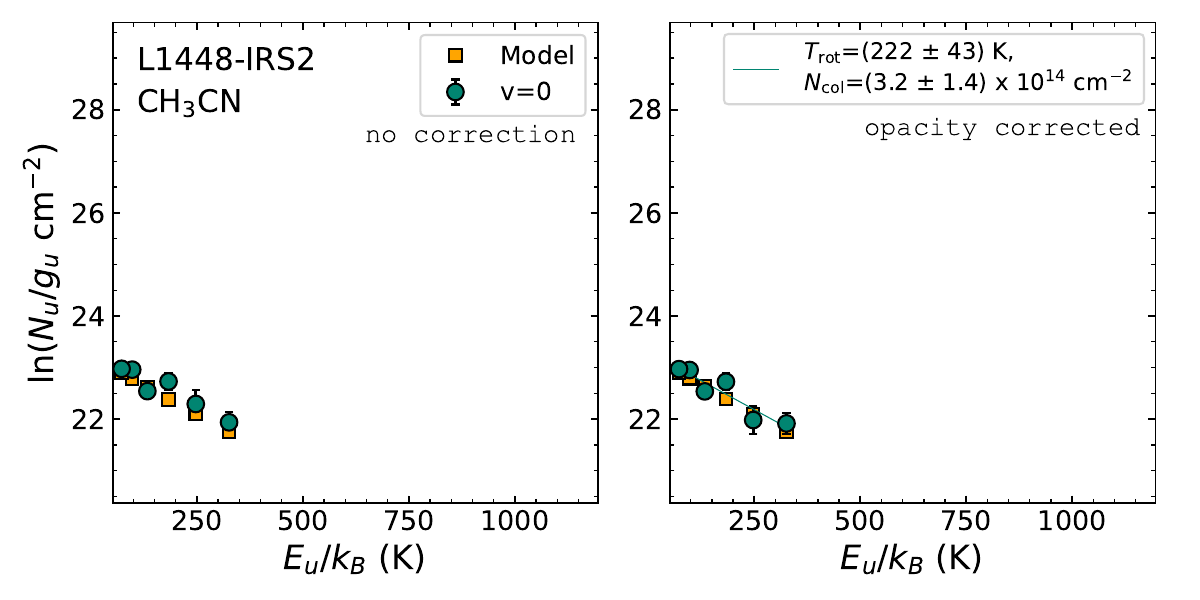}
    \caption{Same as \ref{pd:iras4b}, but for L1448-IRS2. }
    \label{pd:peremb22}
\end{figure}

\begin{figure}[h]
    \centering
    \includegraphics[width=0.48\textwidth]{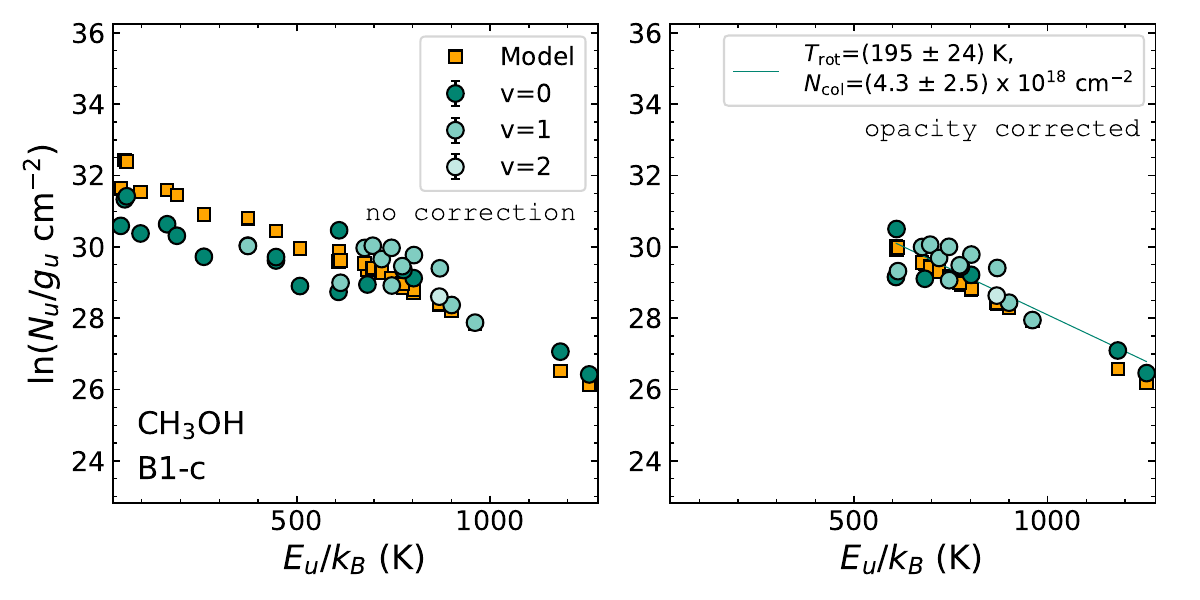}
    \includegraphics[width=0.49\textwidth]{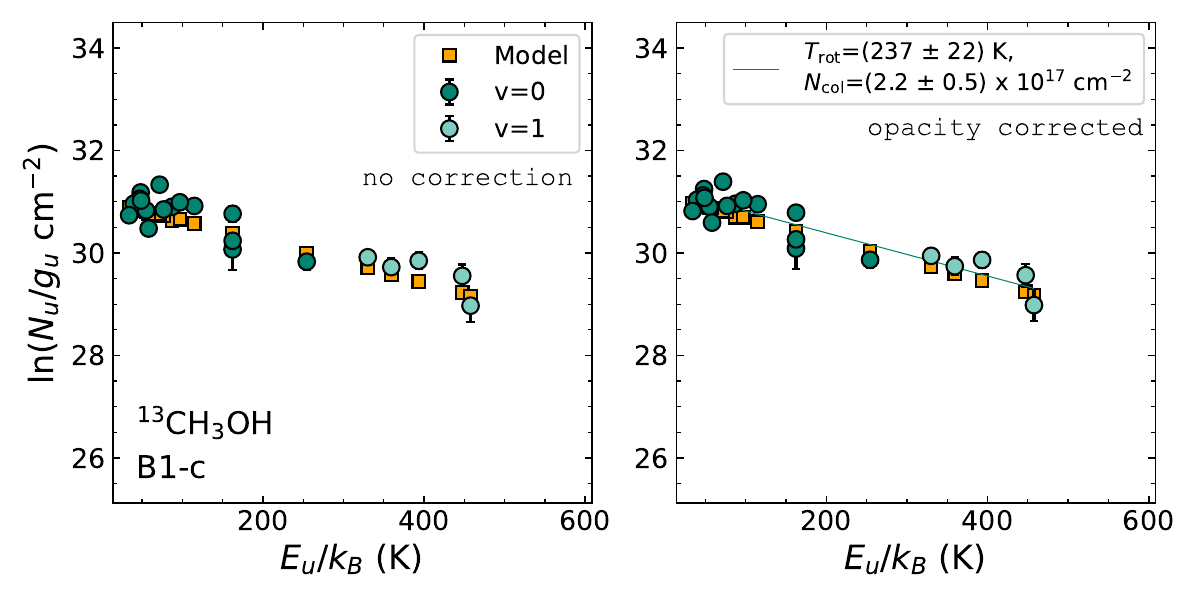}
    \includegraphics[width=0.49\textwidth]{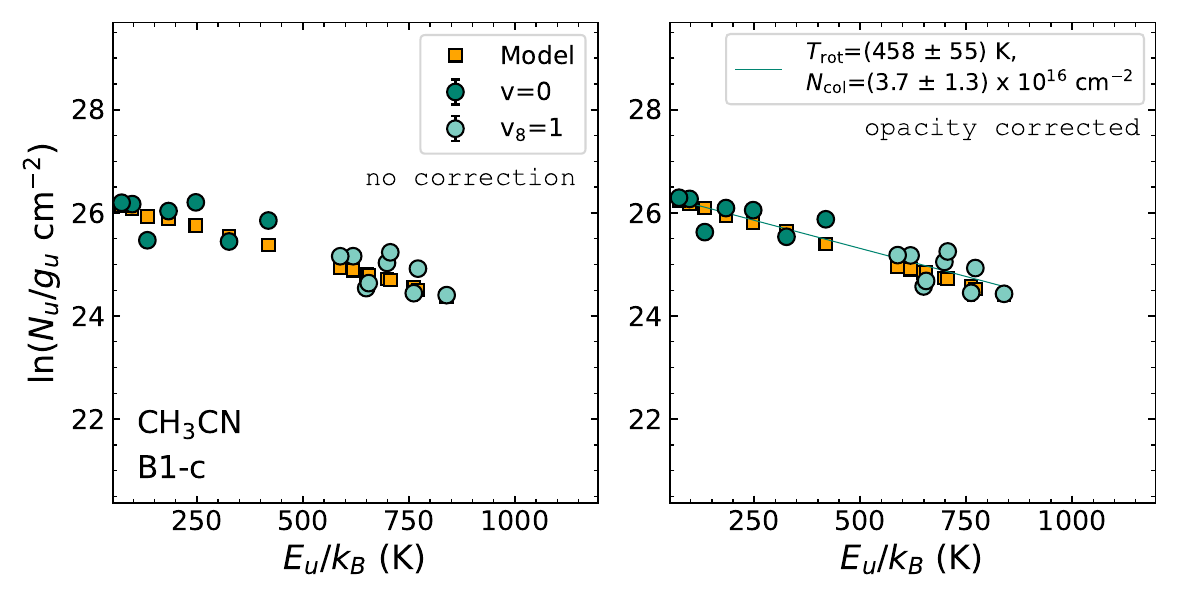}
    \includegraphics[width=0.49\textwidth]{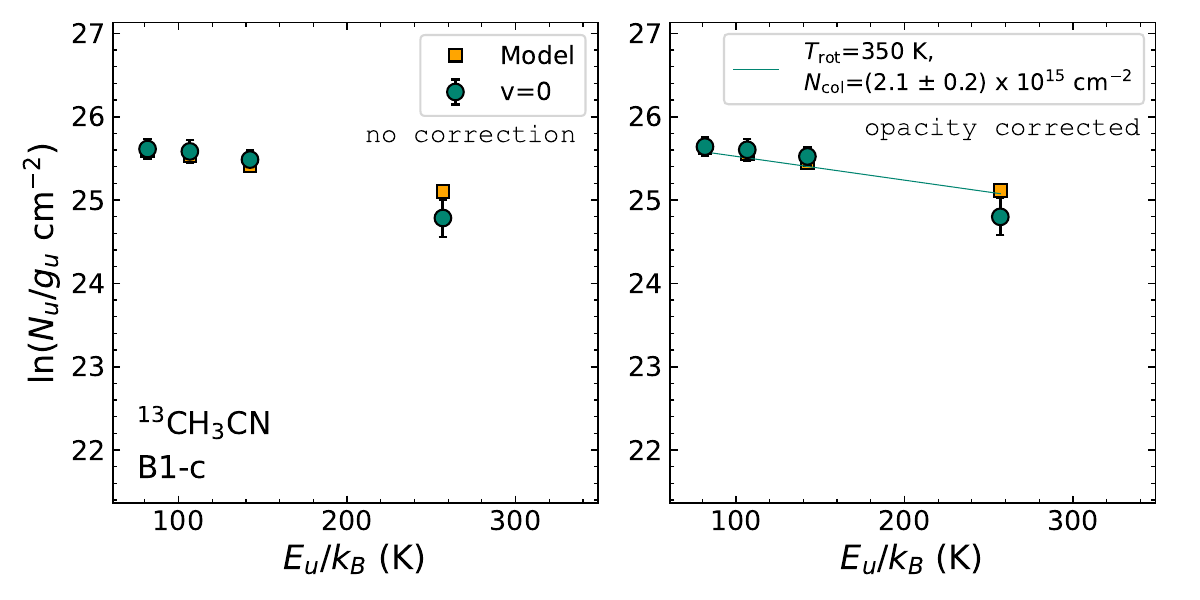}
    \includegraphics[width=0.49\textwidth]{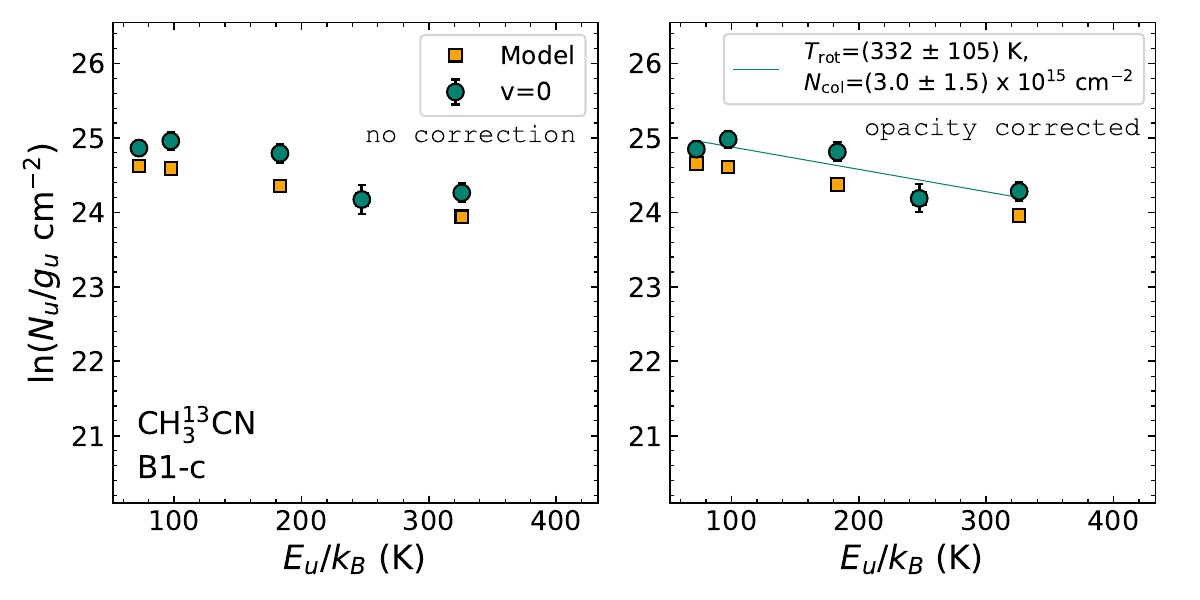}
    \caption{Same as \ref{pd:iras4b}, but for B1-c. }
    \label{pd:peremb29}
\end{figure}

\begin{figure}[h]
    \centering
    \includegraphics[width=0.48\textwidth]{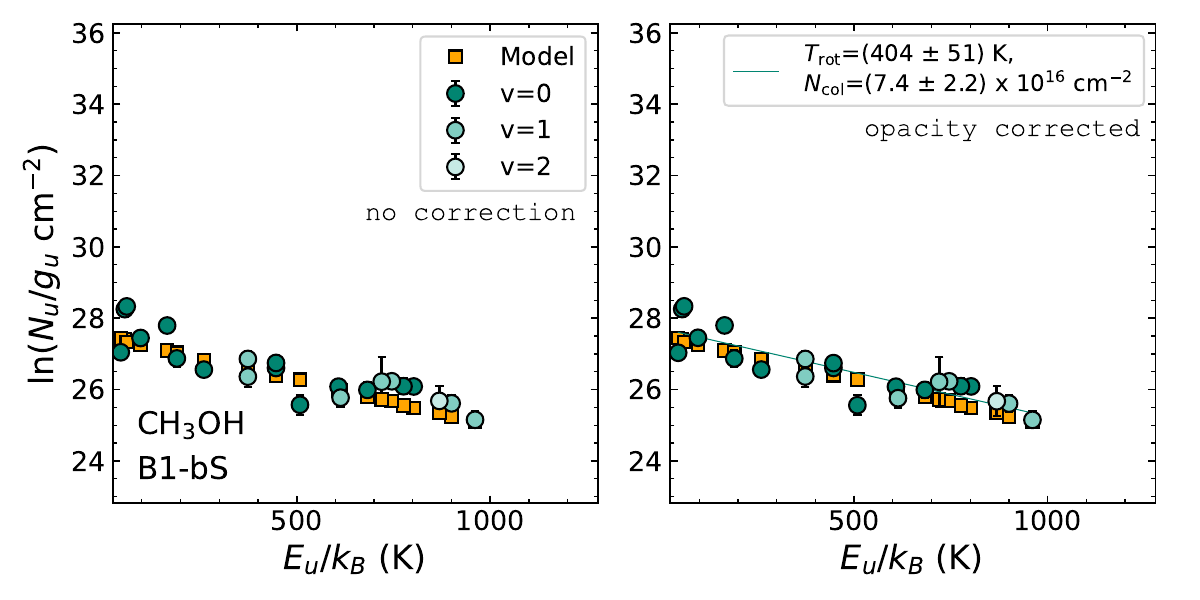}
    \includegraphics[width=0.49\textwidth]{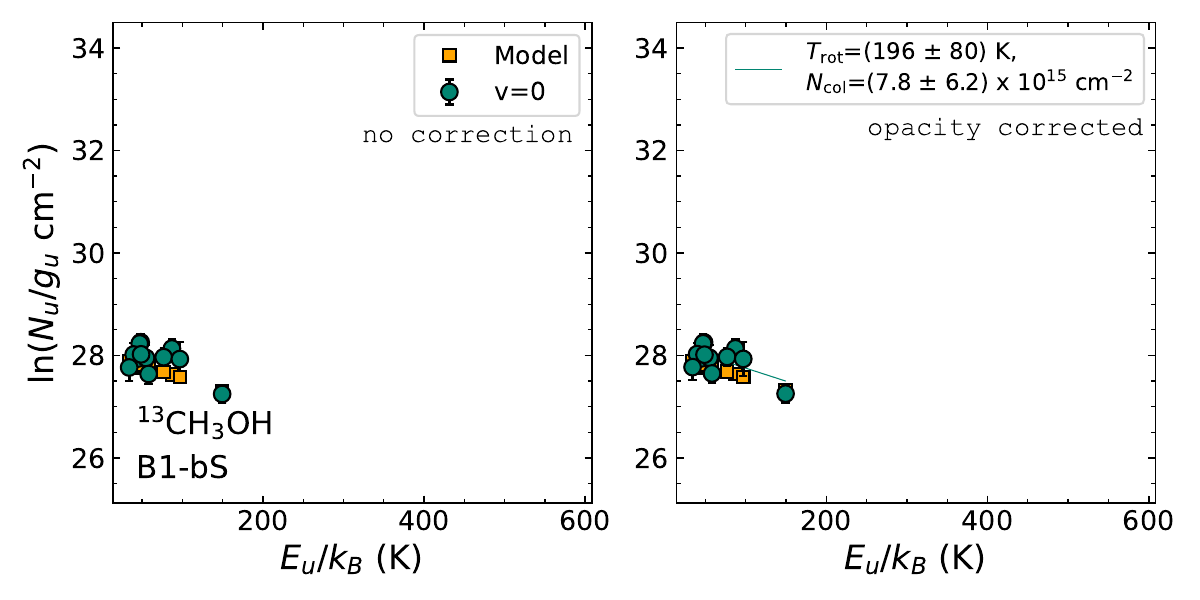}
    \includegraphics[width=0.49\textwidth]{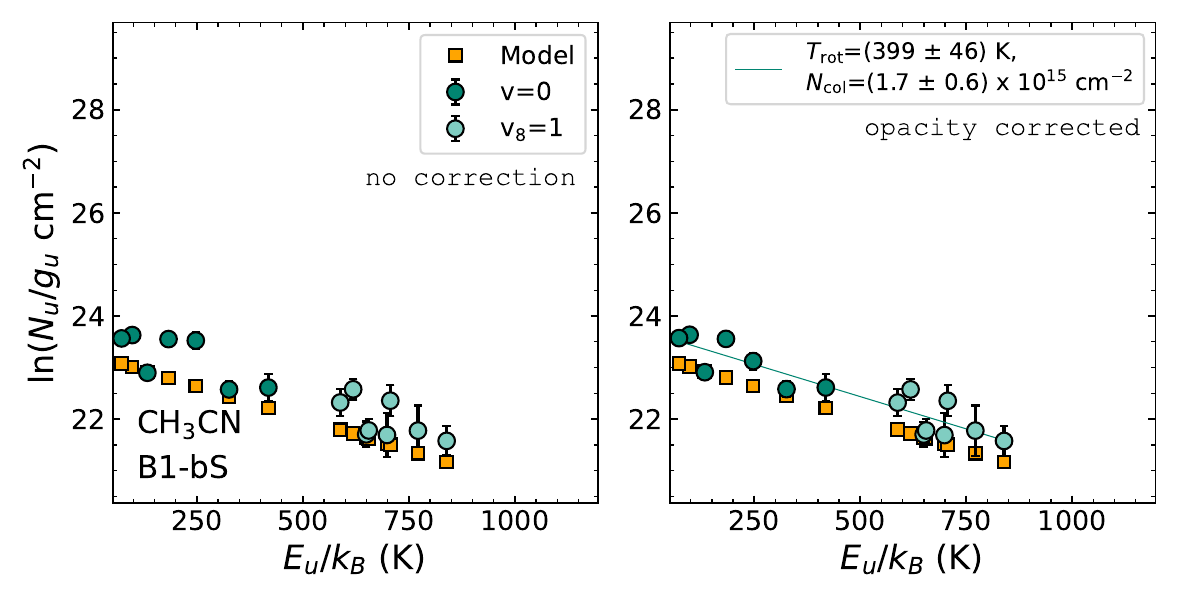}
    \includegraphics[width=0.49\textwidth]{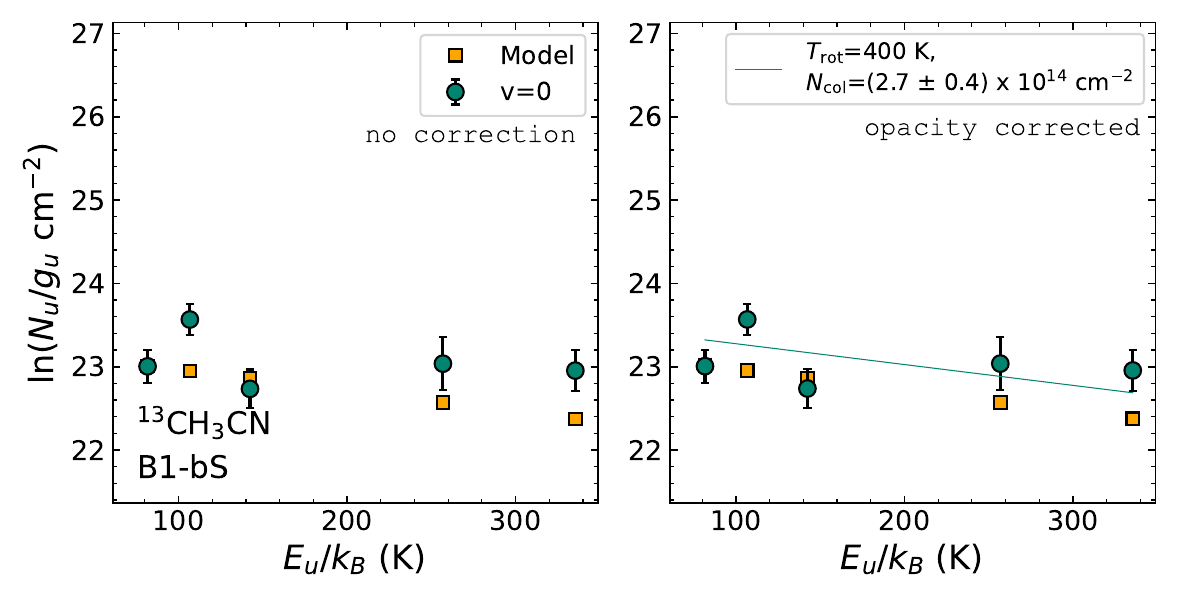}
    \includegraphics[width=0.49\textwidth]{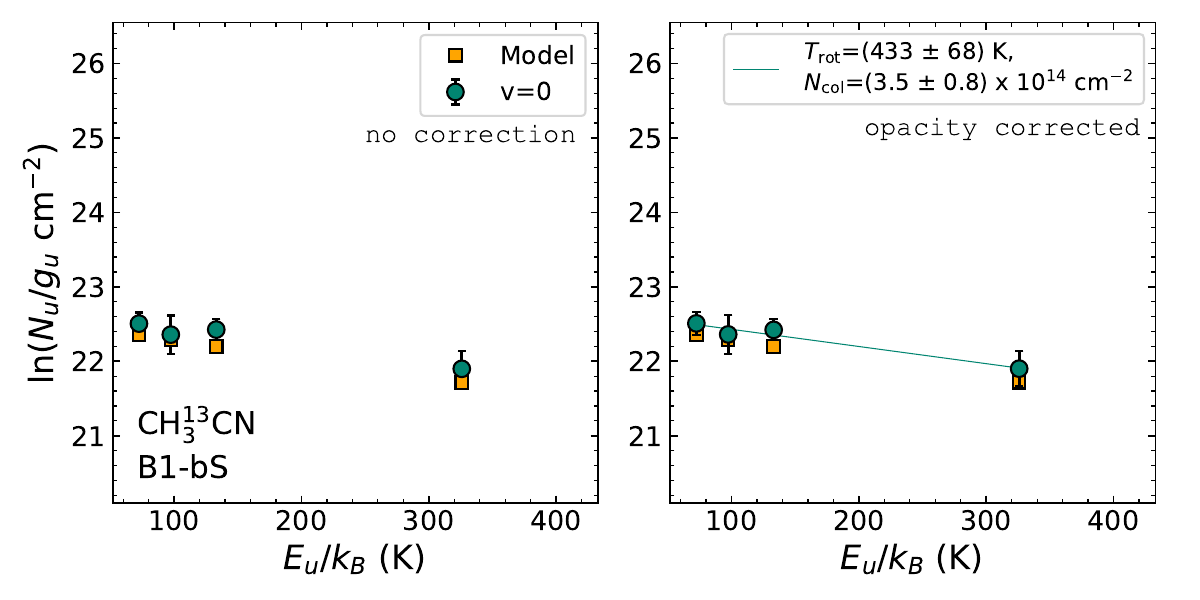}
    \caption{Same as \ref{pd:iras4b}, but for B1-bS. }
    \label{pd:b1bs}
\end{figure}

\begin{figure}[h]
    \centering
    \includegraphics[width=0.48\textwidth]{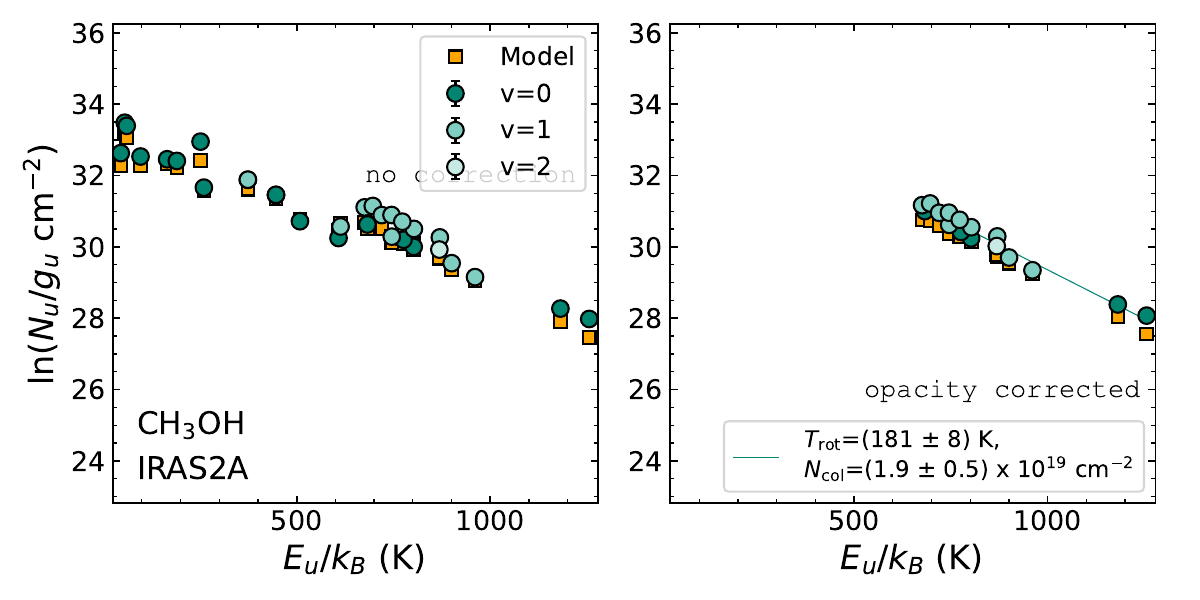}
    \includegraphics[width=0.49\textwidth]{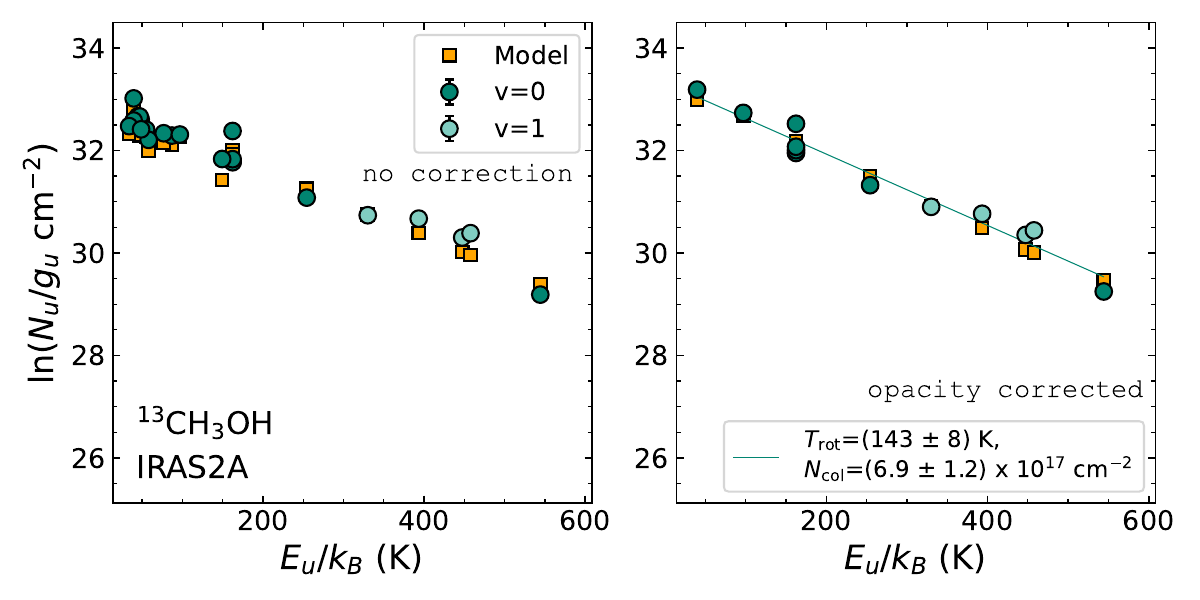}
    \includegraphics[width=0.49\textwidth]{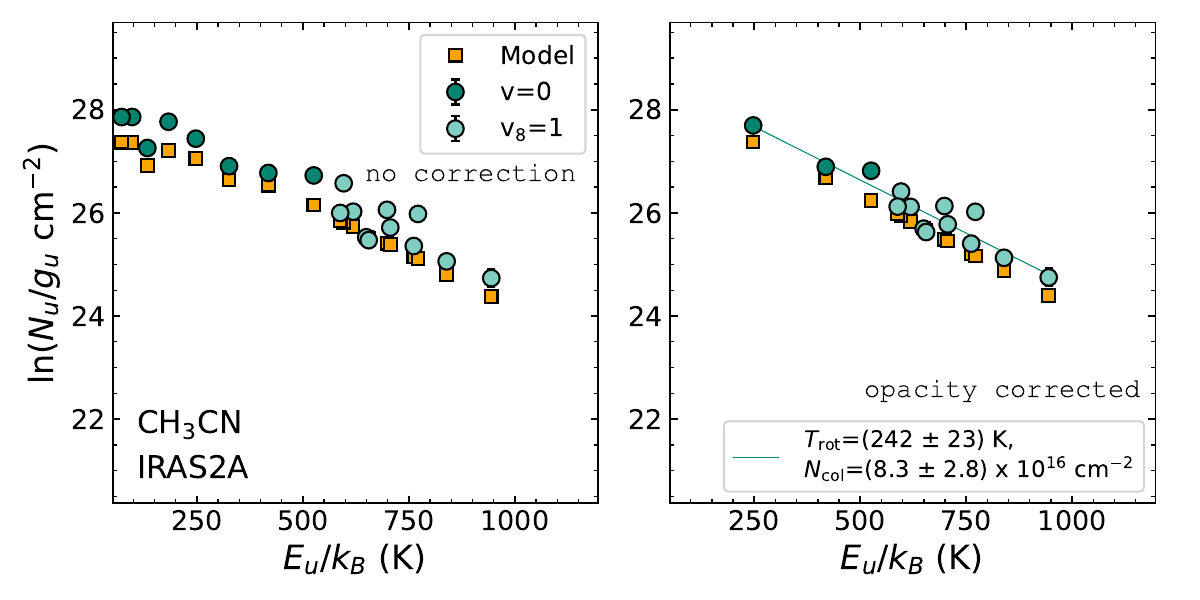}
    \includegraphics[width=0.49\textwidth]{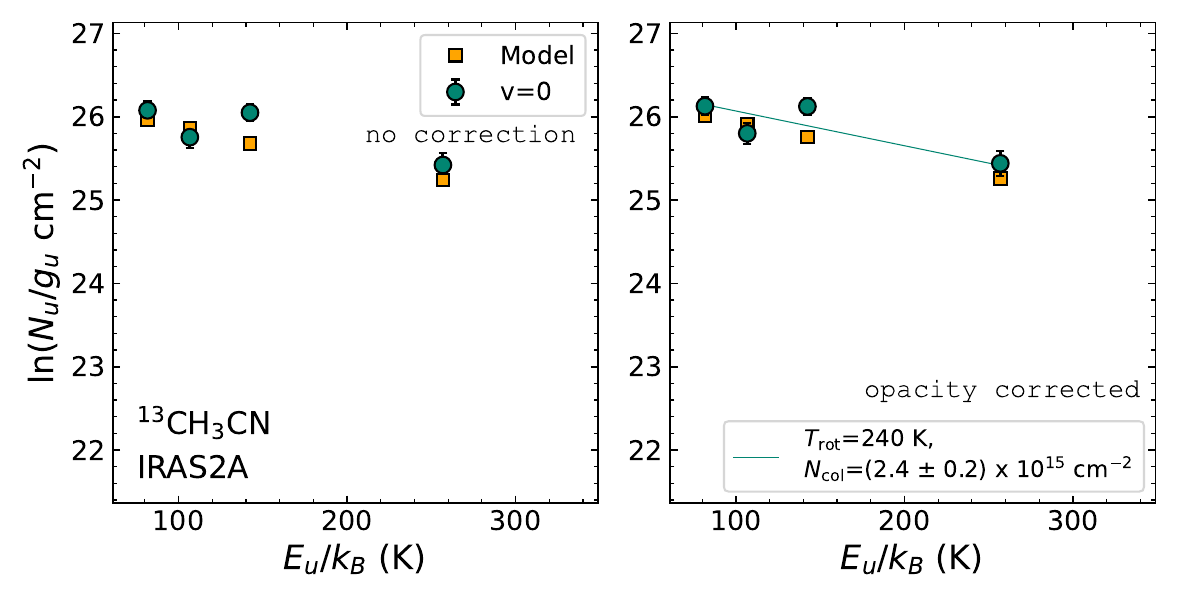}
    \includegraphics[width=0.49\textwidth]{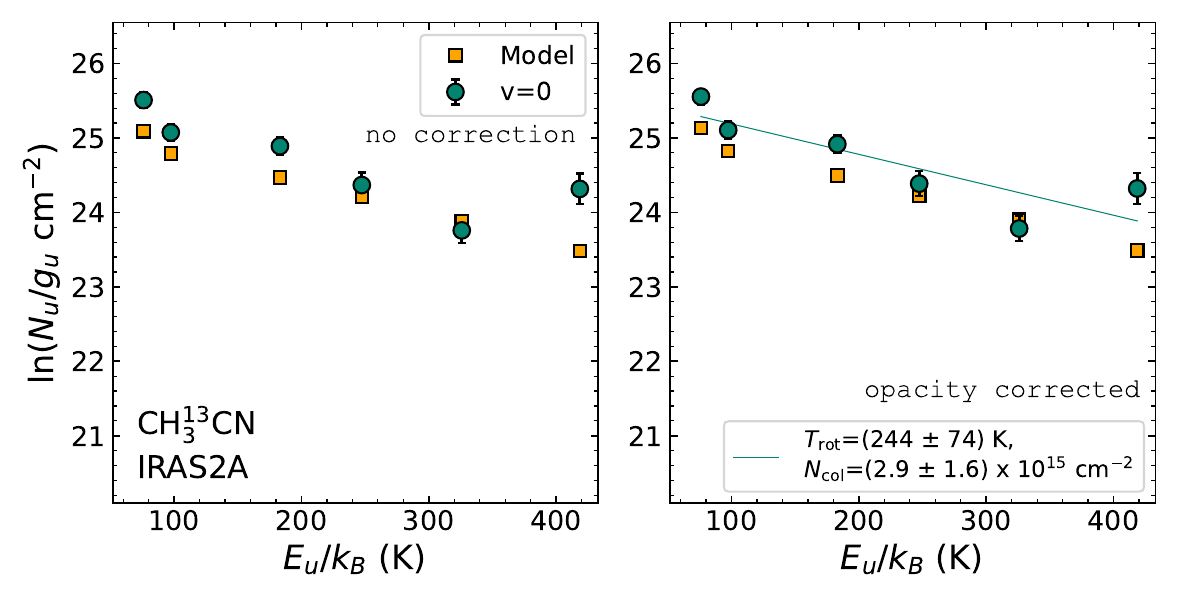}
    \caption{Same as \ref{pd:iras4b}, but for IRAS\,2A. }
    \label{pd:iras2a}
\end{figure}

\begin{figure}[h]
    \centering
    \includegraphics[width=0.48\textwidth]{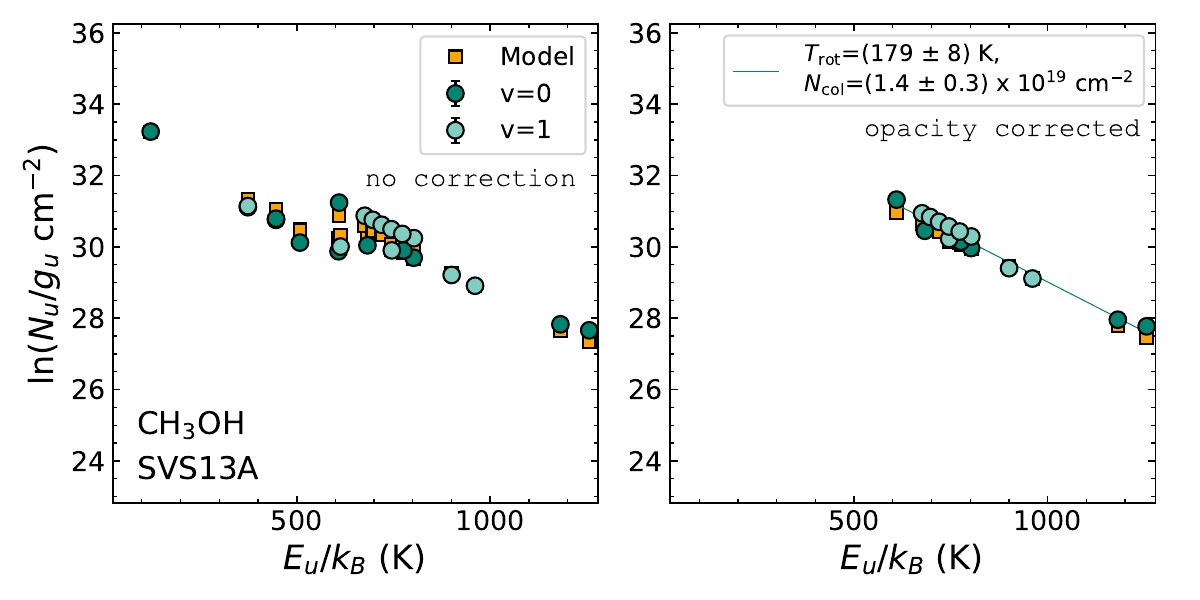}
    \includegraphics[width=0.49\textwidth]{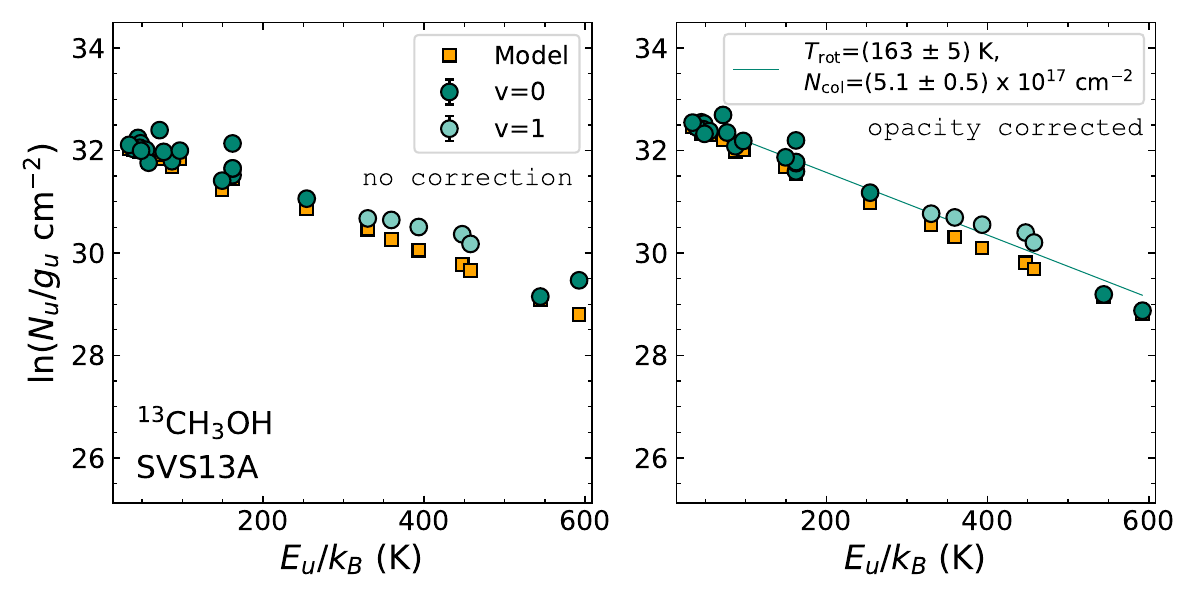}
    \includegraphics[width=0.49\textwidth]{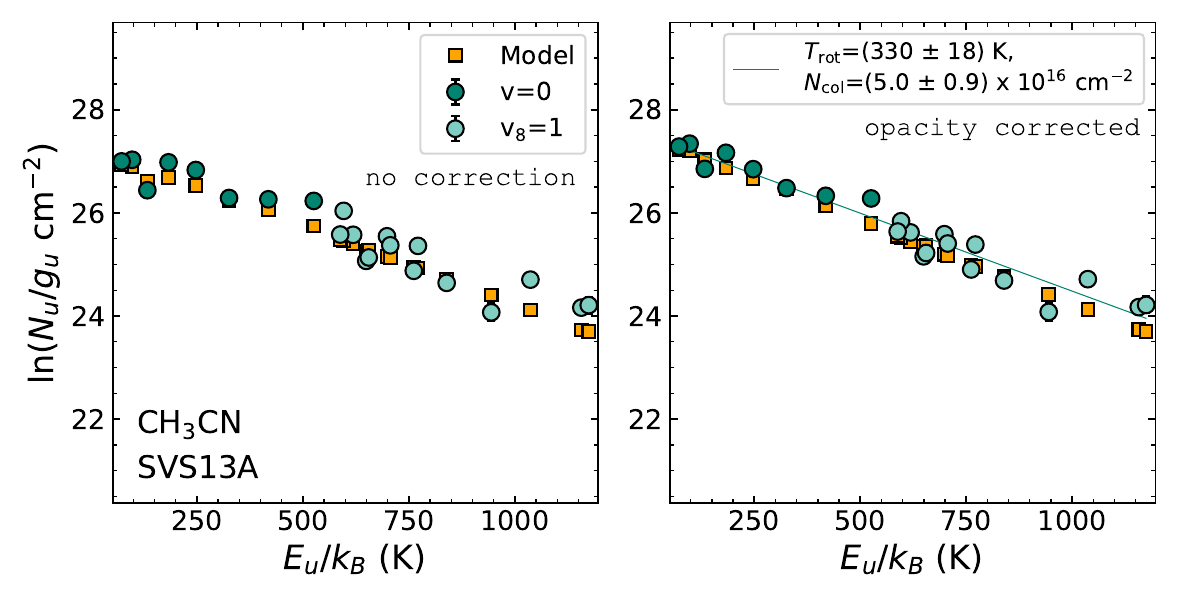}
    \includegraphics[width=0.49\textwidth]{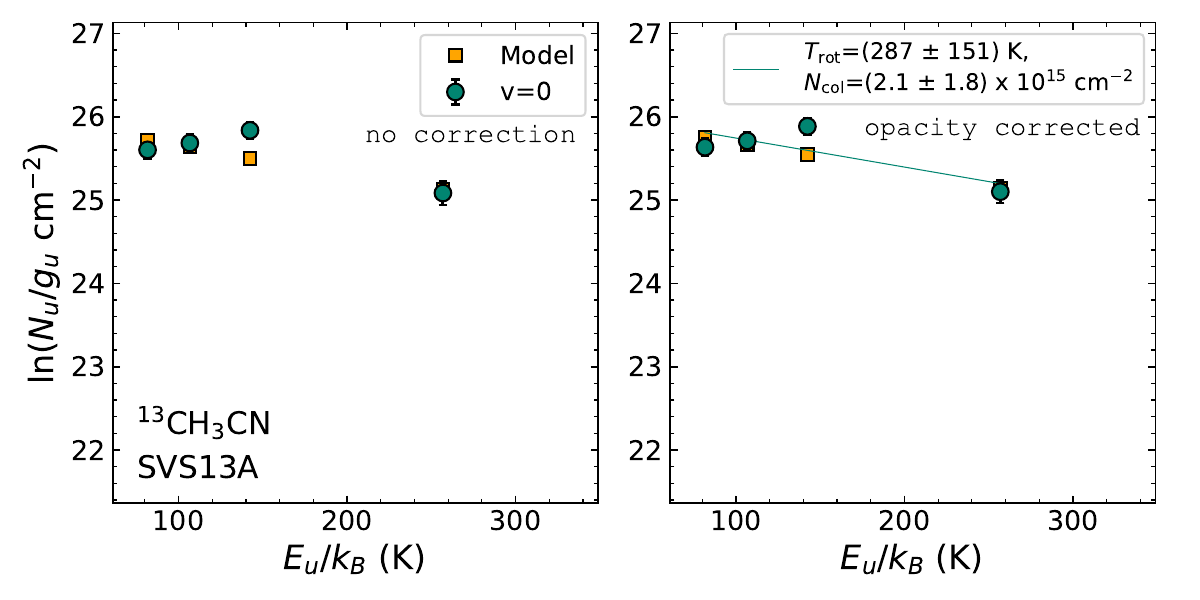}
    \includegraphics[width=0.49\textwidth]{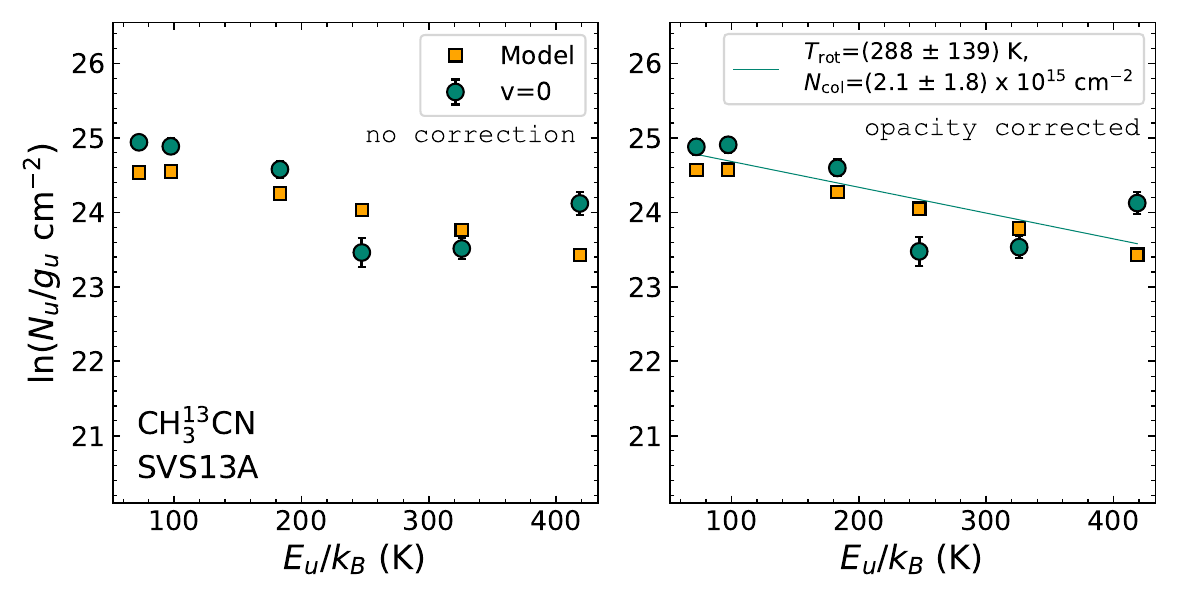}
    \caption{Same as \ref{pd:iras4b}, but for SVS13A. }
    \label{pd:svs13a}
\end{figure}

\begin{figure}[h]
    \centering
    \includegraphics[width=0.48\textwidth]{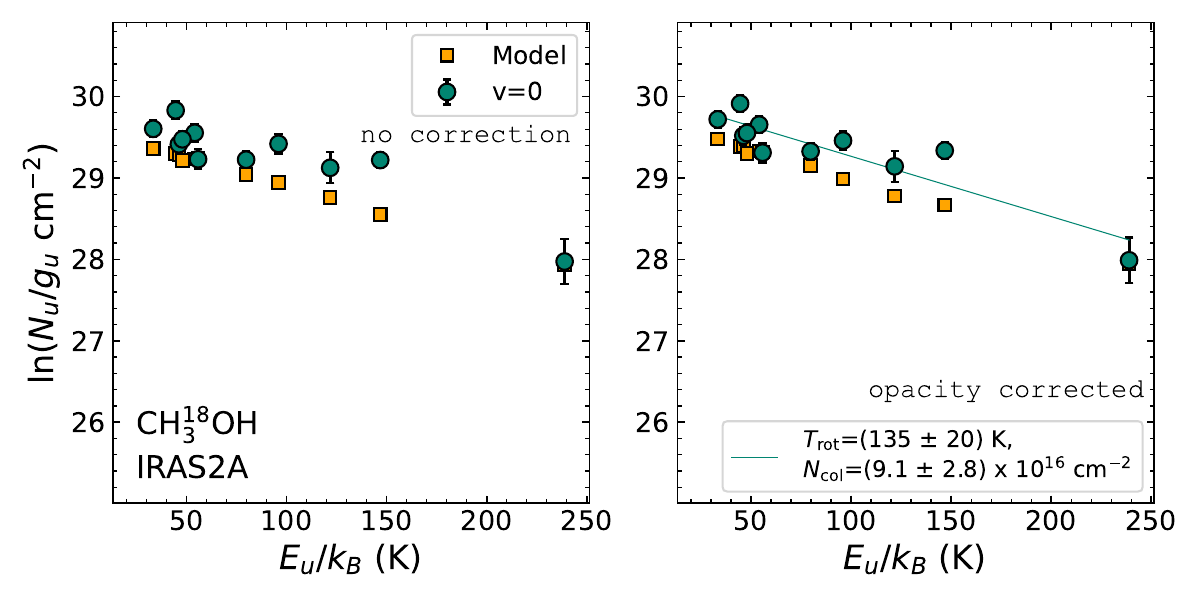}
    \includegraphics[width=0.49\textwidth]{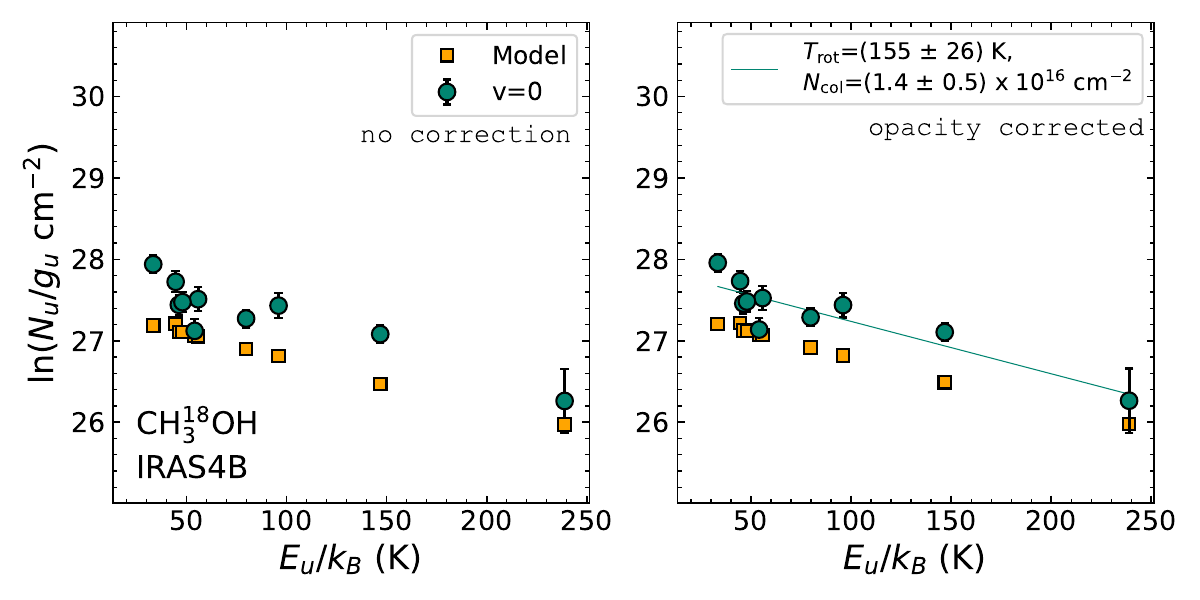}
    \includegraphics[width=0.49\textwidth]{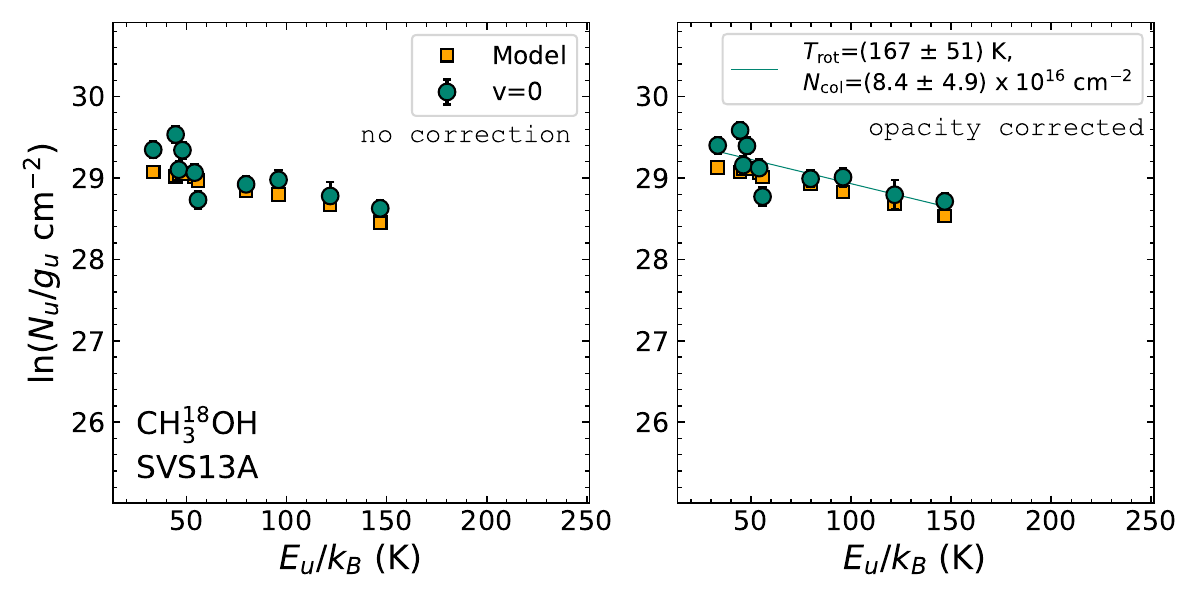}
    \includegraphics[width=0.49\textwidth]{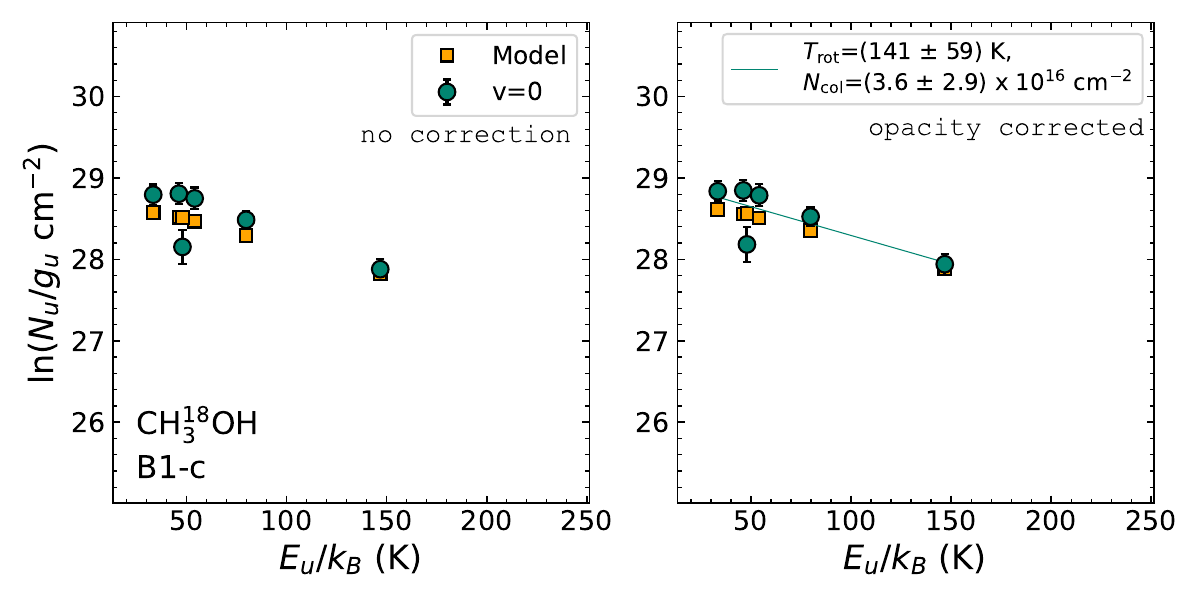}
    \caption{Same as \ref{pd:iras4b},  but for $\rm CH_3^{18}OH$ towards IRAS\,2A, IRAS\,4B, SVS13A, and B1-c.}
    \label{pd:18o}
\end{figure}

In Figs.\,\ref{pd:iras4b}--\ref{pd:18o} all derived population diagrams are displayed.

\clearpage
\twocolumn

\section{Details of the chemical and physical models}\label{a:model}

\begin{figure}[t]
    \centering
    \includegraphics[width=0.45\textwidth]{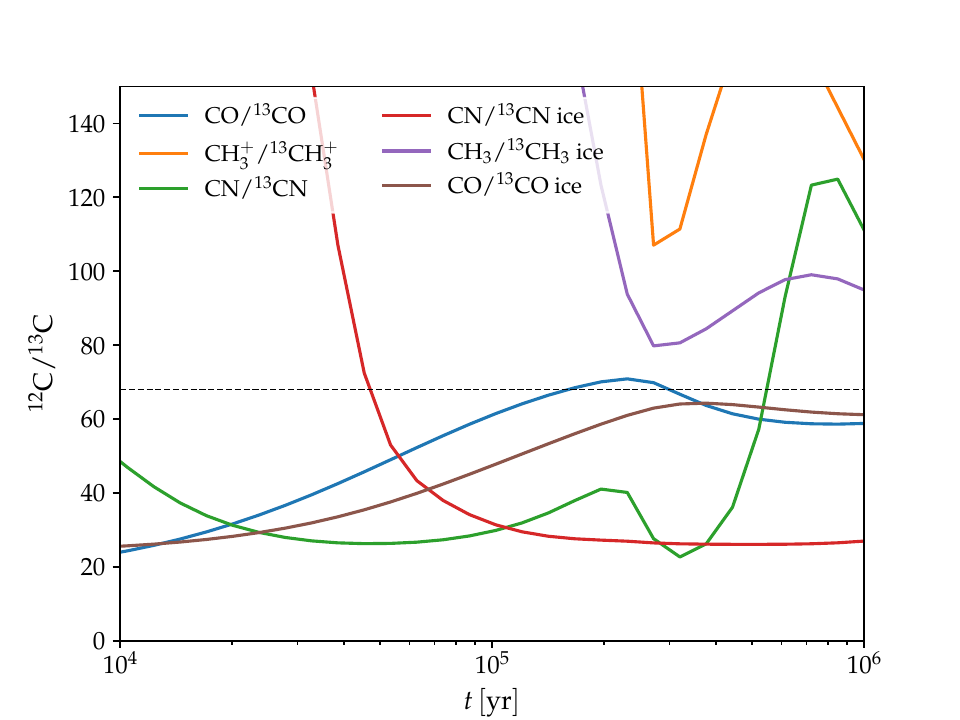}
    \caption{Model predictions for selected gas-phase and ice \ratio ratios as a function of time in the first simulation stage, representing evolution in dark cloud conditions.} 
    \label{fig:darkCloudEvolution}
\end{figure}

As noted in Sect.\,\ref{ss:models}, we have performed the chemical simulations in two stages. In the first (dark cloud) stage, we have run a simple point (i.e. zero-dimensional) model to represent chemical evolution in a dark cloud, assuming canonical physical conditions $n({\rm H_2}) = 10^4 \, \rm cm^{-3}$, $T_{\rm gas} = T_{\rm dust} = 10 \, \rm K$, $A_{\rm V} = 10 \, \rm mag$. The chemistry was let to evolve over a period of $10^6$\,yr in these conditions. Figure~\ref{fig:darkCloudEvolution} shows the time-evolution of the \ratio ratios of selected $\rm CH_3OH$ and $\rm CH_3CN$ precursors in the gas phase and in the ice. It is evident that the ratios not only depend strongly on the time but also on the species in question; similar behavior is shown in our previous works \citep{Colzi2020,Sipila23}. As a result, the initial conditions of the chemistry in the prestellar stage is very sensitive to how long the chemical evolution has proceeded in the dark cloud stage. For example, if we chose a duration of $2 \times 10^5$\,yr for the dark cloud stage, then in the ice the \ratio ratio of CO would (slightly) exceed 68, and we would expect a similar ratio for $\rm CH_3OH$ as it is a product of CO hydrogenation; assuming that no other processes impact the methanol \ratio ratio in the protostellar stage, a very short dark cloud stage would be implied. On the other hand, this would also lead to the two $\rm ^{13}C$ variants of $\rm CH_3CN$ having very different \ratio ratios in the protostellar stage (assuming that $\rm CH_3CN$ forms through $\rm CH_3 + CN$). Figure~\ref{fig:darkCloudEvolution} also shows a dichotomy in the results in that some species are enriched in $\rm ^{12}C$ while others are enriched in $\rm ^{13}C$; at $10^6$\,yr, the two most abundant carbon carriers are $\rm H_2CO$ and $\rm CH_4$ ice (not shown). The former has a \ratio ratio close to that of CO (62), while the latter is anti-fractionated with a \ratio ratio of 134.

\begin{figure}[t]
    \centering
    \includegraphics[width=0.45\textwidth]{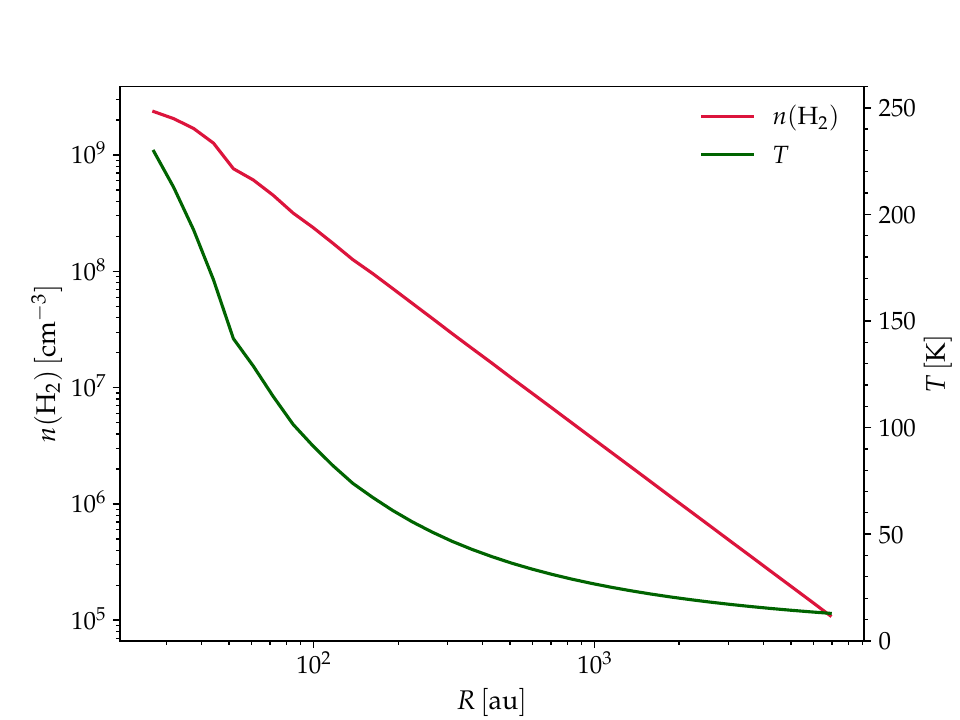}
    \caption{$\rm H_2$ density (left-hand y-axis) and kinetic temperature ($T_{\rm gas} = T_{\rm dust}$; right-hand y-axis) as a function of distance from the center in the IRAS16293-2422 model (from \citealt{Crimier10}).} 
    \label{fig:physicalModel}
\end{figure}

For the protostellar stage we employ the physical model by \citet{Crimier10} which gives the density and temperature profiles as a function of distance from the center of the core. These profiles are displayed in Fig.\,\ref{fig:physicalModel}. We assume $T_{\rm gas} = T_{\rm dust}$, which is reasonable given the high volume density and hence efficient gas-dust thermal coupling. We note that this physical model is not appropriate for describing the very innermost regions of the core (inside a few hundred au) where substructure is known to be present, but is applicable in the present context where the observations were not carried out using high angular resolution.

The chemical simulations in the protostellar stage are carried out by selecting a set of points along the physical model profiles and running the chemical simulation separately for each of the selected points, taking the initial abundances obtained from the preceding dark cloud simulation. This very simple model setup hence neglects any physical evolution between the dark cloud and protostellar core stages, and assumes that the initial chemical conditions are equal across the protostellar core. While such an approach is certainly not realistic, it does allow us to obtain rough estimates of the gas-phase and ice \ratio abundance ratios across the protostellar core. Naturally, the simulation results are also in this case sensitive to the choice of evolutionary time when the abundances are extracted from the model output. We chose here an early time of $10^4$\,yr to reflect the fact that the physical model is not describing well the very inner regions of the core where substructure has already formed.

\end{appendix}

\end{document}